\documentclass[11pt]{article}

\usepackage[url=false,doi=false,isbn=false,style=numeric,natbib=true]{biblatex}
\renewbibmacro{in:}{} 
\AtEveryBibitem{%
   \clearfield{month}%
}
\bibliography{library}

\usepackage{setspace}
\usepackage{a4wide}
\usepackage{authblk}
\usepackage{afterpage}
\usepackage{appendix}
\usepackage{amsmath} 
\usepackage{amssymb} 
\usepackage{subfigure} 
\usepackage{graphicx}
\graphicspath{{figures/}}
\usepackage[small]{caption}
\usepackage{rotating}
\usepackage{tabto}
\usepackage{booktabs}
\usepackage{enumerate}
\usepackage{bm}
\usepackage[usenames,dvipsnames]{xcolor}
\usepackage{float} 

\DeclareMathOperator*{\argmin}{arg\,min}
\DeclareMathOperator*{\argmax}{arg\,max}

\floatstyle{ruled}
\newfloat{algorithm}{htbp}{lop}
\floatname{algorithm}{Box}

\usepackage[colorlinks=true,pdfborder={0 0 0.5},citecolor	=RoyalBlue]{hyperref}

\title{Pathways-driven Sparse Regression Identifies Pathways and Genes Associated with High-density Lipoprotein Cholesterol in Two Asian Cohorts}
\author[1,2,*]{Matt Silver}
\author[3]{Peng Chen}
\author[4]{Ruoying Li}
\author[3,5,6]{Ching-Yu Cheng}
\author[5,6]{Tien-Yin Wong}
\author[3,4]{E-Shyong Tai}
\author[3,7,8,9,10]{Yik-Ying Teo}
\author[1]{Giovanni Montana}

\affil[1]{Statistics Section, Department of Mathematics, Imperial College London, UK}
\affil[2]{MRC International Nutrition Group, London School of Hygiene and Tropical Medicine, UK}
\affil[3]{Saw Swee Hock School of Public Health, National University of Singapore}
\affil[4]{Yong Loo Lin School of Medicine, National University of Singapore}
\affil[5]{Department of Ophthalmology, National University of Singapore}
\affil[6]{Singapore Eye Research Institute, Singapore National Eye Center}
\affil[7]{NUS Graduate School for Integrative Science and Engineering, National University of Singapore}
\affil[8]{Life Sciences Institute, National University of Singapore}
\affil[9]{Genome Institute of Singapore, Agency for Science, Technology and Research}
\affil[10]{Department of Statistics and Applied Probability, National University of Singapore}
\affil[*]{email: matt.silver@lshtm.ac.uk}
\date{} 

\begin{document}

\maketitle

\begin{abstract}
Standard approaches to analysing data in genome-wide association studies (GWAS) ignore any potential functional relationships between genetic markers.  In contrast gene pathways analysis uses prior information on functional structure within the genome to identify pathways associated with a trait of interest.  In a second step, important single nucleotide polymorphisms (SNPs) or genes may be identified within associated pathways.  The pathways approach is motivated by the fact that many genes do not act alone, but instead have effects that are likely to be mediated through their interaction in gene pathways.  Where this is the case, pathways approaches may reveal aspects of a trait's genetic architecture that would otherwise be missed when considering SNPs in isolation.  Most pathways methods begin by testing SNPs one at a time, and so fail to capitalise on the potential advantages inherent in a multi-SNP, joint modelling approach.  Here we describe a dual-level, sparse regression model for the simultaneous identification of pathways, genes and SNPs associated with a quantitative trait.  Our method takes account of various factors specific to the joint modelling of pathways with genome-wide data, including widespread correlation between genetic predictors, and the fact that variants may overlap multiple pathways. We use a resampling strategy that exploits finite sample variability to provide robust rankings for pathways, SNPs and genes. We test our method through simulation, and use it to perform pathways-driven SNP selection in a search for pathways, genes and SNPs associated with variation in serum high-density lipoprotein cholesterol (HDLC) levels in two separate GWAS cohorts of Asian adults.  By comparing results from both cohorts we identify a number of candidate pathways including those associated with cardiomyopathy, and T cell receptor and PPAR signalling.  Highlighted genes include those associated with the L-type calcium channel, adenylate cyclase, integrin, laminin, MAPK signalling and immune function.  \emph{Software implementing the methods described here, together with sample data is available at \url{http://www2.imperial.ac.uk/~gmontana/psrrr.htm}}.
\end{abstract}


\vspace{1cm}

\section*{Introduction}\label{sec:Intro}

Much attention continues to be focused on the problem of identifying SNPs and genes influencing a quantitative or dichotomous trait in genome wide scans \citep{McCarthy2008}.  Despite this, in many instances gene variants identified in GWAS have so far uncovered only a relatively small part of the known heritability of most common diseases \citep{Visscher2012}.  Possible explanations include the presence of multiple SNPs with small effects, or of rare variants, which may be hard to detect using conventional approaches \citep{Visscher2012,Manolio2009,Goldstein2009}.

One potentially powerful approach to uncovering the genetic etiology of disease is motivated by the observation that in many cases disease states are likely to be driven by multiple genetic variants of small to moderate effect, mediated through their interaction in molecular networks or pathways, rather than by the effects of a few, highly penetrant mutations \citep{Schadt2009}. Where this assumption holds, the hope is that by considering the joint effects of variants acting in concert, pathways GWAS methods will reveal aspects of a disease's genetic architecture that would otherwise be missed when considering variants individually \citep{Wang2010,Fridley2011}.  In this section we describe a sparse regression method utilising prior information on gene pathways to identify putative causal pathways, along with the constituent SNPs and genes that may be driving pathways association.

Sparse modelling approaches are becoming increasingly popular for the analysis of genome wide datasets \citep{Shi2011,Cho2010,Ayers2010,Wu2009}.  Sparse regression models enable the joint modelling of large numbers of SNP predictors, and perform `model selection' by highlighting small numbers of variants influencing the trait of interest.  These models work by penalising or constraining the size of estimated regression coefficients.  An interesting feature of these methods is that different sparsity patterns, that is different sets of genetic predictors having specified properties, can be obtained by varying the nature of this constraint.  For example, the lasso \citep{Tibshirani1996} selects a subset of variants whose main effects best predict the response.  Where predictors are highly correlated, the lasso tends to select one of a group of correlated predictors at random. In contrast, the elastic net \citep{Zou2005} selects groups of correlated variables.  Model selection may also be driven by external information, unrelated to any statistical properties of the data being analysed.  For example, the fused lasso \citep{Tibshirani2005,Tibshirani2008} uses ordering information, such as the position of genomic features along a chromosome to select `adjacent' features together.  

Prior information on functional relationships between genetic predictors can also been used to drive the selection of groups of variables.  In the present context, information mapping genes and SNPs to functional gene pathways has recently been used in sparse regression models for pathway selection.  \citet{Chen2010} describe a method that uses a combination of lasso and ridge regression to assess the significance of association between a candidate pathway and a dichotomous (case-control) phenotype, and apply this method in a study of colon cancer etiology.  In contrast, \citet{Silver2012a} use group lasso penalised regression to select pathways associated with a multivariate, quantitative brain-imaging phenotype characteristic of structural change in the brains of patients with Alzheimer's disease.

In identifying pathways associated with a trait of interest, a natural follow-up question is to ask which SNPs and/or genes are driving pathway selection? We might further ask a related question: can the use of prior information on putative gene interactions within pathways increase power to identify causal SNPs, compared to alternative methods that disregard such information? One way to answer these questions is by conducting a two-stage analysis, in which we first identify important pathways, and then in a second step search for SNPs within selected pathways \citep{Eleftherohorinou2009,Eleftherohorinou2011}.  There are however a number of problems with this approach.  Firstly, highlighted SNPs are then not necessarily those that were driving pathway selection in the first step of the analysis.  Secondly, the implicit (and reasonable) assumption is that only a small number of SNPs in a pathway are driving pathway selection, so that ideally we would prefer a model that has this assumption built in.  The above considerations point to the use of a `dual-level' sparse regression model that imposes sparsity at both the pathway and SNP level.  Such a model would perform \emph{simultaneous} pathway and SNP selection, with the additional benefit of being simpler to implement.

A suitable sparse regression model enforcing the required dual-level sparsity is the sparse group lasso (SGL) \citep{Simon2012}.  SGL is a comparatively recent development in sparse modelling, and in simulations has been shown to accurately recover dual-level sparsity, in comparison to both the group lasso and lasso \citep{Friedman2010,Simon2012}.  SGL has been used for the identification of rare variants in a case-control study by grouping SNPs into genes \citep{Zhou2010}; for the identification of genomic regions whose copy number variations have an impact on RNA expression levels \citep{Peng2010}; and to model geographical factors driving climate change \citep{Chatterjee2011a}.  SGL can be seen as fitting into a wider class of structured-sparsity inducing models that use prior information on relationships between predictors to enforce different sparsity patterns \citep{Zhao2009,Huang2011,Jenatton2011}.

In the next section (Methods) we outline our method for sparse, pathways-driven SNP selection, and demonstrate through simulation that the incorporation of prior information mapping SNPs to gene pathways can boost the power to detect SNPs associated with a quantitative trait.  In the following section (Results), we describe an application study, in which we investigate pathways, SNPs and genes associated with serum high-density lipoprotein cholesterol (HDLC) levels in two separate cohorts of Asian adults.  HDLC refers to the cholesterol carried by small lipoprotein molecules, so called high density lipoproteins (HDLs). HDLs help remove the cholesterol aggregating in arteries, and are therefore protective against cardiovascular diseases \citep{Toth2005}.  Serum HDLC levels are genetically heritable ($h^2=0.485$) \citep{Namboodiri1985}.  GWAS studies have now uncovered more than 100 HDLC associated loci (see \url{www.genome.gov/gwastudies}, \citet{Hindorff2009}).  However, considering serum lipids as a whole, variants so far  identified account for only 25-30\% of the genetic variance, highlighting the limited power of current methodologies to detect hidden genetic factors \citep{Teslovich2010}.

\section*{Methods}\label{sec:Methods}

This section is organised as follows.  We begin by introducing the sparse group lasso (SGL) model for pathways-driven SNP selection, along with an efficient estimation algorithm, for the case of non-overlapping pathways.  We then describe a simulation study illustrating superior group (pathway) and variant (SNP) selection performance in the case that the true supporting model is group-sparse.  We continue by extending the previous model to the case of overlapping pathways.  In principle, we can then solve this model using the estimation algorithm described for the non-overlapping case.  However, we argue that this approach does not give us the outcome we require.  For this reason we describe a modified estimation algorithm that assumes pathway independence, and demonstrate in a simulation study that this new algorithm is able to identify the correct SNPs and pathways with improved sensitivity and specificity.  We complete this section with a description of a method to reduce bias in SNP and pathway selection, together with a subsampling procedure to rank SNPs and pathways in order of importance.

\subsection*{The sparse group lasso model}\label{sec:SGL_model}

We arrange the observed values for a univariate quantitative trait or phenotype, measured for $N$ unrelated individuals, in an $(N \times 1)$ response vector $\mathbf{y}$.  We assume minor allele counts for $P$ SNPs are recorded for all individuals, and denote by $x_{ij}$ the minor allele count for SNP $j$ on individual $i$.  These are arranged in an $(N\times P)$ genotype design matrix $\mathbf{X}$.  Phenotype and genotype vectors are mean centred, and SNP genotypes are standardised to unit variance, so that $\sum_i x^2_{ij} = 1$, for $j = 1, \ldots, P$.  

We assume that all $P$ SNPs may be mapped to $L$ groups or pathways, $\mathcal{G}_l \subset\{1, \ldots, P\}$, $l=1, \ldots, L$, and begin by considering the case where pathways are disjoint or non-overlapping, so that $\mathcal{G}_l \cap \mathcal{G}_{l'} = \emptyset$ for any $l \ne l'$.  We denote the vector of SNP regression coefficients by $\boldsymbol{\beta}=(\beta_1, \ldots, \beta_P)$, and additionally denote the matrix containing all SNPs mapped to pathway $\mathcal{G}_l$ by $\mathbf{X}_{l} = (\mathbf{x}_{l_1},\mathbf{x}_{l_2},\dots, \mathbf{x}_{P_l})$, where $\mathbf{x}_j = (x_{1j}, x_{2j}, \ldots, x_{Nj})'$, is the column vector of observed SNP minor allele counts for SNP $j$, and $P_l$ is the number of SNPs in $\mathcal{G}_l$.  We denote the corresponding vector of SNP coefficients by $\boldsymbol{\beta}_l = (\beta_{l_1},\beta_{l_2},\dots,\beta_{P_l})$.

\begin{figure}[htbp]
\begin{center}
	\includegraphics[trim = 30mm 65mm 30mm 65mm, clip, scale=0.5]{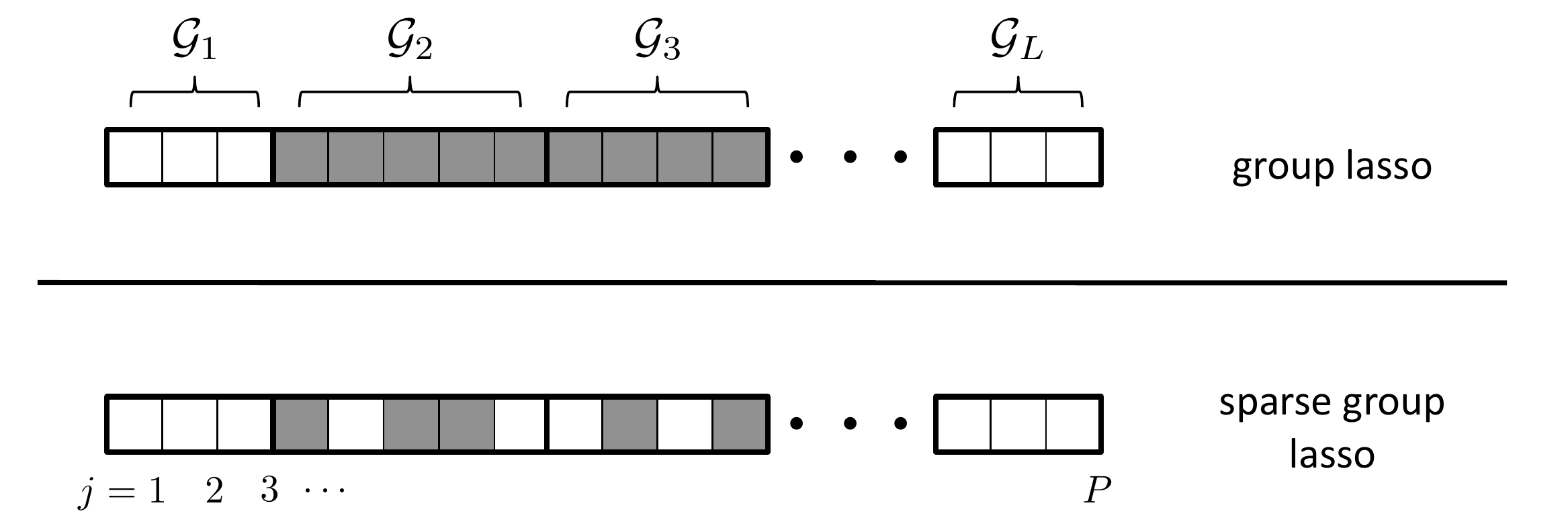}
\caption[Sparsity patterns enforced by the group lasso and sparse group lasso]{Sparsity patterns enforced by the group lasso and sparse group lasso.  The set $\mathcal{S} \subset \{1, \ldots, P\}$ of causal SNPs influencing the phenotype are represented by boxes that are shaded grey.  Causal SNPs are assumed to occur within a set $\mathcal{C} \subset \{1, \ldots, L\}$ of causal pathways, $\mathcal{G}_1, \ldots, \mathcal{G}_L$.  Here $\mathcal{C} = \{2, 3\}$.  The group lasso enforces sparsity at the group or pathway level only, whereas the sparse group lasso additionally enforces sparsity at the SNP level.}
\label{fig:GL_SGL_sparsity_patterns}
\end{center}
\end{figure}

In general, where $P$ is large, we expect only a small proportion of SNPs to be `causal', in the sense that they exhibit phenotypic effects.  A key assumption in pathways analysis is that these causal SNPs will tend to be enriched within a small set, $\mathcal{C} \subset \{1, \ldots, L\}$, of causal pathways, with $|\mathcal{C}| \ll L$, where $|\mathcal{C}|$ denotes the size (cardinality) of $\mathcal{C}$.  We denote the set of causal SNPs mapping to pathway $\mathcal{G}_l$ by $\mathcal{S}_l$, and make the further assumption that most SNPs in a causal pathway are non-causal, so that $|\mathcal{S}_l| < P_l$, where $|\mathcal{S}_l|$ denotes the size (cardinality) of $\mathcal{S}_l$.  A suitable sparse regression model imposing the required, dual-level sparsity pattern is the sparse group lasso (SGL).  We illustrate the resulting causal SNP sparsity pattern in Figure \ref{fig:GL_SGL_sparsity_patterns}, and compare it to that generated by the group lasso (GL), a group-sparse model that we used previously in a sparse regression method to identify gene pathways \citep{Silver2012,Silver2012a}.  

With the SGL \citep{Simon2012}, sparse estimates for the SNP coefficient vector, $\boldsymbol{\beta}$ are given by
\begin{equation}
	\boldsymbol{\hat{\beta}}^{SGL} = 
		\argmin_{\boldsymbol{\beta}} 
		\Big \{
			\frac{1}{2} ||\mathbf{y}-\mathbf{X}\boldsymbol{\beta}||_2^2 + (1-\alpha) \lambda \sum_{l=1}^{L}w_l||\boldsymbol{\beta}_l||_2 + \alpha 					\lambda || \boldsymbol{\beta} ||_1
		\Big \}
	\label{eq:SGL_objF}
\end{equation}
where $\lambda$ $(\lambda>0)$ and  $\alpha$ $(|\alpha| \le 1)$ are parameters controlling sparsity, and $w_l$ is a pathway weighting parameter that may vary across pathways.  \eqref{eq:SGL_objF} corresponds to an ordinary least squares (OLS) optimisation, but with two additional constraints on the coefficient vector, $\boldsymbol{\beta}$, that tend to shrink the size of $\boldsymbol{\beta}$, relative to OLS estimates.  One constraint imposes a group lasso-type penalty on the size ($\ell_2$ norm) of $\boldsymbol{\beta}_l, l = 1, \ldots, L$.  Depending on the values of $\lambda$, $\alpha$ and $w_l$, this penalty has the effect of setting multiple pathway SNP coefficient vectors, $\hat{\boldsymbol{\beta}}_l = \mathbf{0}$, thereby enforcing sparsity at the pathway level.  Pathways with non-zero coefficient vectors form the set $\hat{\mathcal{C}}$ of `selected' pathways, so that
\begin{equation*}
	\hat{\mathcal{C}} (\lambda,\alpha) = \{l: \hat{\boldsymbol{\beta}}_l \ne \mathbf{0}	\}.
\end{equation*}
A second constraint imposes a lasso-type penalty on the size ($\ell_1$ norm) of $\boldsymbol{\beta}$.  Depending on the values of $\lambda$ and  $\alpha$, for a selected pathway $l \in \hat{\mathcal{C}}$, this penalty has the effect of setting multiple SNP coefficient vectors, $\hat{\beta}_j = 0, j \subset \mathcal{G}_l$, thereby enforcing sparsity at the SNP level within selected pathways.  SNPs with non-zero coefficient vectors then form the set $\hat{\mathcal{S}_l}$ of selected SNPs in pathway $l$, so that
\begin{equation*}
	\hat{\mathcal{S}_l} (\lambda,\alpha) = \{j: \hat{\beta}_j \ne 0, j \in \mathcal{G}_l\}.
\end{equation*}
The set of all selected SNPs is given by
\begin{equation*}
	\mathcal{\hat{S}} = \bigcup_{l \in \mathcal{\hat{C}}} \mathcal{\hat{S}}_l.
\end{equation*}
The sparsity parameter $\lambda$ controls the degree of sparsity in $\boldsymbol{\beta}$, such that the number of pathways and SNPs selected by the model increases as $\lambda$ is reduced from a maximal value $\lambda_{max}$, above which $\hat{\boldsymbol{\beta}} = \mathbf{0}$.  The parameter $\alpha$ controls how the sparsity constraint is distributed between the two penalties.  When $\alpha=0$, \eqref{eq:SGL_objF} reduces to the group lasso, so that sparsity is imposed only at the pathway level, and all SNPs within a selected pathway have non-zero coefficients.  When $0 < \alpha <1$, solutions exhibit dual-level sparsity, such that as $\alpha$ approaches 0 from above,  greater sparsity at the group level is encouraged over sparsity at the SNP level.  When $\alpha=1$, \eqref{eq:SGL_objF}  reverts to the lasso, so that pathway information is ignored.   

\subsection*{Model estimation}\label{sec:SGL_model_estimation}

For the estimation of $\hat{\boldsymbol{\beta}}^{SGL}$ we proceed by noting that the optimisation \eqref{eq:SGL_objF} is convex, and (in the case of non-overlapping groups) that the penalty is block-separable, so that we can obtain a solution using block, or group-wise coordinate gradient descent (BCGD) \citep{Tseng2009}.  A detailed derivation of the estimation algorithm is given in the accompanying supplementary information, Section \ref{SGL estimation algorithm}.

From \eqref{eq:SGL_bl_equals_zero_criterion} and \eqref{eq:SGL_SNP_notSel}, the criterion for selecting a pathway $l$ is given by
\begin{equation}
	|| S(\mathbf{X}_l'\hat{\mathbf{r}}_l, \alpha \lambda) ||_2 > (1-\alpha) \lambda w_l,
	\label{eq:SGL_pathway_sel_criterion}
\end{equation}
and the criterion for selecting SNP $j$ in selected pathway $l$ by
\begin{equation}
		||X'_j \mathbf{\hat{r}}_{l,j}||_1 > \alpha \lambda,
	\label{eq:SGL_snp_sel_criterion}		
\end{equation}
where $\mathbf{\hat{r}}_l = \mathbf{\hat{r}}_l - \sum_{m \ne l} \mathbf{X}_l \hat{\boldsymbol{\beta}}_l $ and $\mathbf{\hat{r}}_{l,j} = \mathbf{\hat{r}}_l - \sum_{k \ne j} X_k \hat{\beta}_k $ are respectively the pathway and SNP partial residuals, obtained by regressing out the current estimated effects of all other pathways and SNPs respectively.  The complete algorithm for SGL estimation using BCGD is presented in Box \ref{box:SGL_estimation_algorithm}.

\begin{algorithm}[h]
\NumTabs{25}
\begin{enumerate}
	\item
	initialise $\boldsymbol{\beta} \leftarrow \mathbf{0}$.  
	\item
	\textbf{repeat:} [pathway loop]\\
	\tab\tab for pathway $l = 1, 2, \ldots, L$: \\
	\tab\tab \tab if $|| S(\mathbf{X}_l'\hat{\mathbf{r}}_l, \alpha \lambda) ||_2 \le (1-\alpha) \lambda w_l$\\
	\tab \tab\tab\tab $\boldsymbol{\beta}_l \leftarrow \mathbf{0}$\\
	\tab \tab \tab else\\
	\tab \tab\tab \tab\textbf{repeat:} [SNP loop] \\
	\tab \tab\tab\tab \tab for $j = l_1,\dots,l_{P_l}$:\\
	\tab \tab\tab\tab\tab \tab if $\beta_j = 0:$\\
	\tab \tab\tab\tab\tab\tab \tab Newton update $\beta_j^{**} \leftarrow \beta_j$ using \eqref{eq:SGL_Newton_update_betaj_zero} and \eqref{eq:SGN_Newton_update}\\
	\tab\tab\tab\tab\tab \tab else:\\
	\tab \tab\tab\tab\tab\tab\tab Newton update $\beta_j^{**} \leftarrow \beta_j$ using \eqref{eq:SGL_partial_j} and  \eqref{eq:SGN_Newton_update}\\
	\tab \tab\tab\tab\tab\tab if $f(\boldsymbol{\beta}_l^{**}) > f(\boldsymbol{\beta}_l)$:\\
	\tab \tab\tab\tab\tab\tab\tab $\beta_j^{**} \leftarrow \frac{\beta_j^{**} + \beta_j}{2}$\\
	\tab\tab\tab\tab\tab\tab $\beta_j \leftarrow \beta_j^{**}$\\
	\tab\tab\tab\tab \textbf{until} convergence of $\boldsymbol{\beta}_l$ [SNP loop]\\
	\textbf{until} convergence of $\boldsymbol{\beta}$ [pathway loop]
	\item
	$\boldsymbol{\hat{\beta}}^{SGL} \leftarrow \boldsymbol{\beta}$
\end{enumerate}

\caption{SGL-BCGD estimation algorithm}
\label{box:SGL_estimation_algorithm}
\end{algorithm}

\subsection*{SGL simulation study 1} \label{sec:SGL_non-overlap_sim_study}
We test the hypothesis that where causal SNPs are enriched in a given pathway, pathway-driven SNP selection using SGL will outperform simple lasso selection that disregards pathway information in a simple simulation study.  We simulate $P=2500$ genetic markers for $N=400$ individuals.  Marker frequencies for each SNP are sampled independently from a multinomial distribution following a Hardy Weinberg equilibrium frequency distribution.  SNP minor allele frequencies are sampled from a uniform distribution $\mathcal{U}[0.1,0.5]$.  SNPs are distributed equally between 50 non-overlapping pathways, each containing 50 SNPs.  

We then test each competing method over 500 Monte Carlo (MC) simulations.  At each simulation, a baseline univariate phenotype is sampled from $\mathcal{N}(10, 1)$.  To generate genetic effects, we randomly select 5 SNPs from a single, randomly selected pathway $\mathcal{G}_l$, to form the set $\mathcal{S} \subset \mathcal{G}_l$ of causal SNPs.  Genetic effects are then generated as described in Section \ref{sim_study_1}.

\begin{figure}
	\begin{center}
	\subfigure[]{\includegraphics[trim = 10mm 0mm 10mm 10mm, clip, scale=0.4]{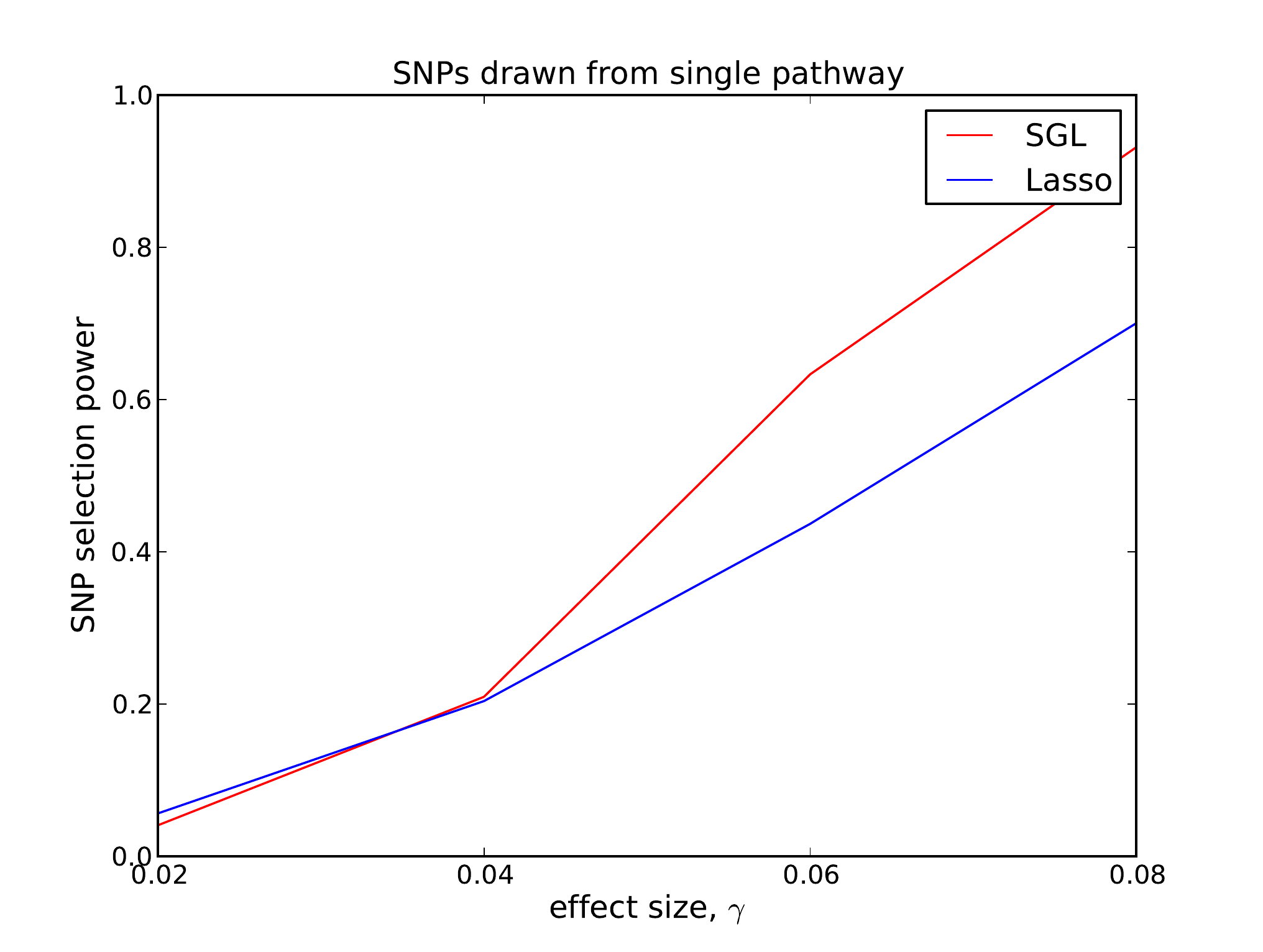}}
	\subfigure[]{\includegraphics[trim = 10mm 0mm 10mm 10mm, clip, scale=0.4]{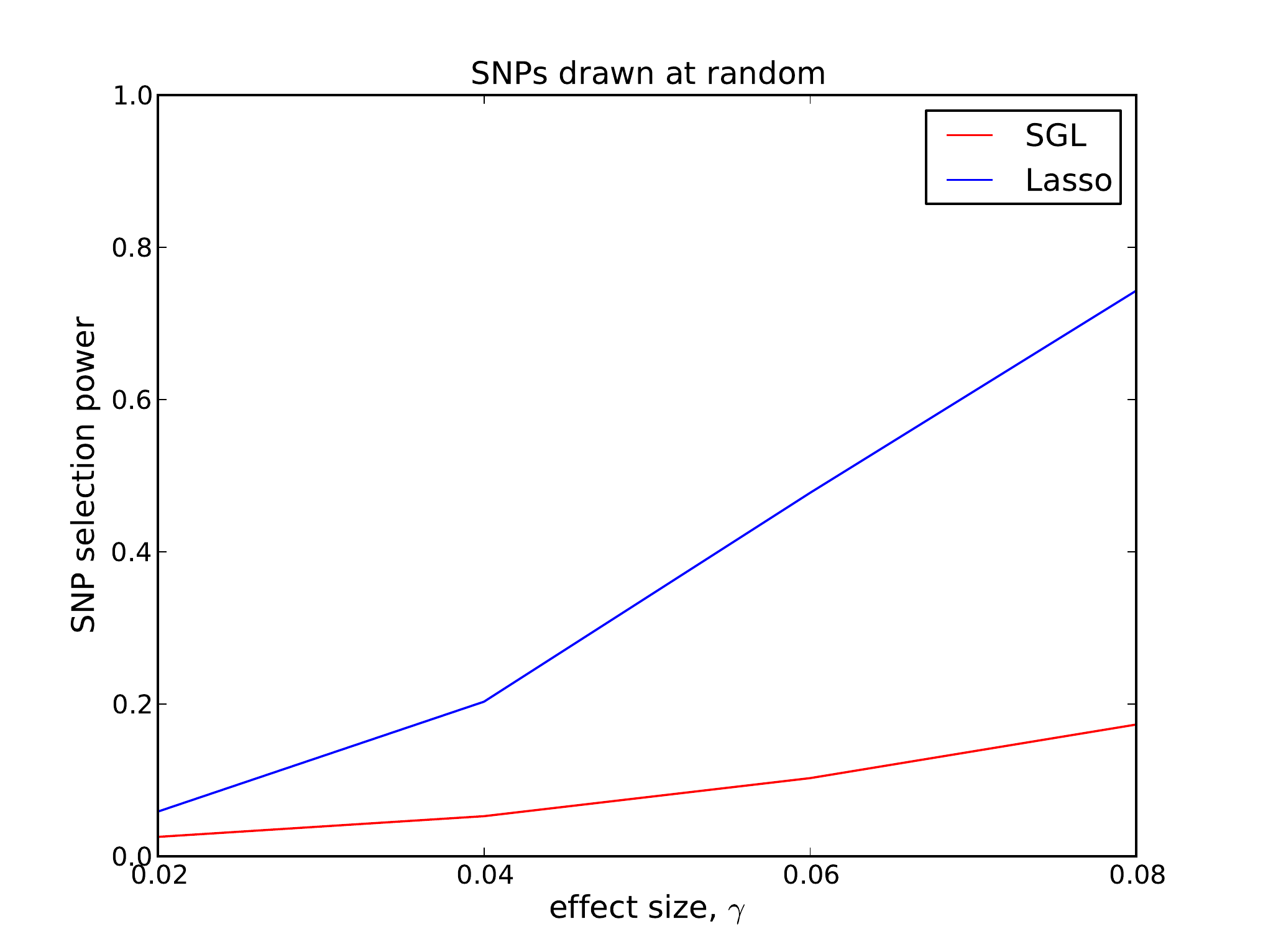}}
	\caption[SGL vs Lasso.  Comparison of power to detect 5 causal SNPs]{SGL vs Lasso.  Comparison of power to detect 5 causal SNPs.  Each data point represents mean power over 500 MC simulations.    \emph{Left:} Causal SNPs drawn from single causal pathway.  \emph{Right: }Causal SNPs drawn at random.}
	\label{fig:SGL_vs_lasso_non-overlapping_power}
	\end{center}
\end{figure}

To enable a fair comparison between the two methods (SGL and lasso), we ensure that both methods select the same number of SNPs at each simulation.  We do this by first obtaining the SGL solution, $\hat{\mathcal{S}}^{SGL}$, with $\lambda = 0.85 \lambda_{max}$ and $\alpha = 0.8$, which ensures sparsity at both the pathway and SNP level.  We use a uniform pathway weighting vector $\boldsymbol{w} = \mathbf{1}$.  We then compute the lasso solution using coordinate descent over a range of values for the lasso regularisation penalty, $\lambda$, and choose the set 
\begin{equation*}
	\hat{\mathcal{S}}^{lasso}(\lambda') \quad \mbox{such that} \quad |\hat{\mathcal{S}}^{lasso}(\lambda')| = |\hat{\mathcal{S}}^{SGL}|
\end{equation*}
where $|\hat{\mathcal{S}}^{SGL}|$ is the number of SNPs previously selected by SGL, and $|\hat{\mathcal{S}}^{lasso}(\lambda')|$ is the number of SNPs selected by the lasso with $\lambda = \lambda'$.  We measure performance as the mean power to detect all 5 causal SNPs over 500 MC simulations, and test a range of genetic effect sizes ($\gamma$) (see \ref{sim_study_1}).  In a follow up study, we compare the performance of the two methods in a scenario in which pathways information is uninformative.  For this we repeat the previous simulations, but with 5 causal SNPs drawn at random from all 2500 SNPs, irrespective of pathway membership.  Results are presented in Figure \ref{fig:SGL_vs_lasso_non-overlapping_power}.

Referring to Figure \ref{fig:SGL_vs_lasso_non-overlapping_power}, we see that where causal SNPs are concentrated in a single causal pathway (Figure \ref{fig:SGL_vs_lasso_non-overlapping_power} - left), SGL demonstrates greater power (and equivalently specificity, since the total number of selected SNPs is constant), compared with the lasso, above a particular effect size threshold (here $\gamma \approx 0.04$).  Where pathway information is not important, that is causal SNPs are not enriched in any particular pathway (Figure \ref{fig:SGL_vs_lasso_non-overlapping_power} - right), SGL performs poorly.

\begin{figure}
	\begin{center}
	\subfigure{\includegraphics[trim = 25mm 0mm 25mm 0mm, clip, scale=0.48]{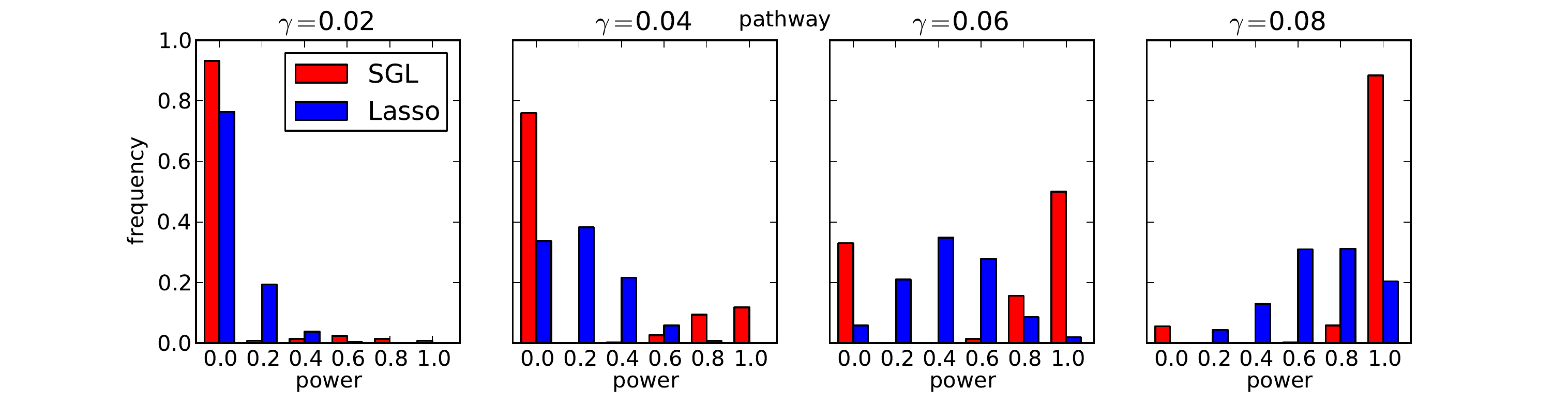}}
	\subfigure{\includegraphics[trim = 25mm 0mm 25mm 0mm, clip, scale=0.48]{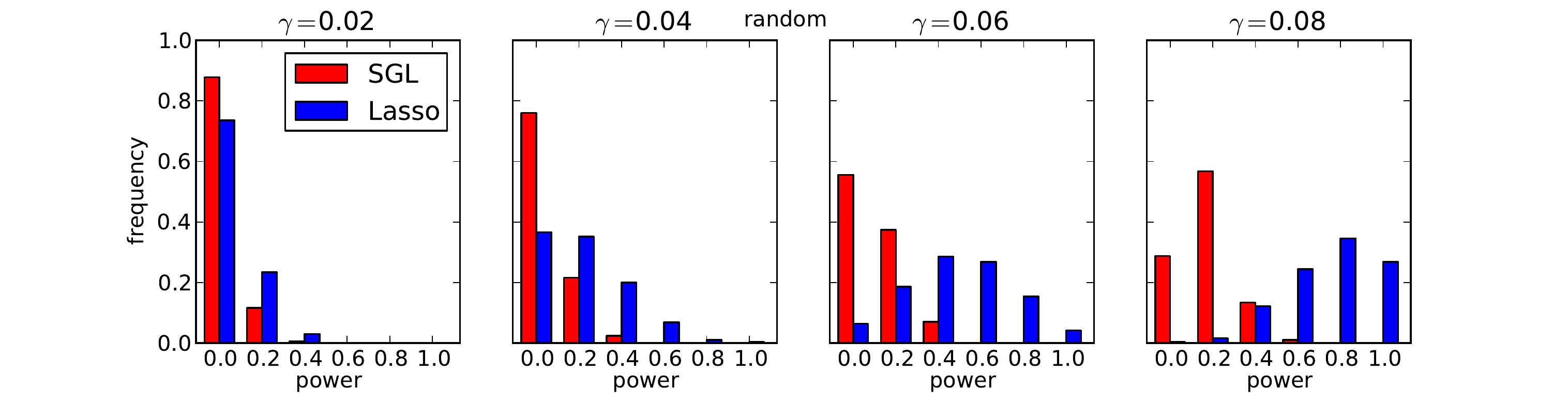}}
	\caption[SGL vs Lasso.  Distribution over 500 MC simulations of power to detect 5 causal SNPs]{SGL vs Lasso.  Distribution over 500 MC simulations of power to detect 5 causal SNPs.  Each plot represents the power distribution at a single data point in Figure \ref{fig:SGL_vs_lasso_non-overlapping_power}.  The power distribution is discrete, since each method can identify $0,1,2,3,4$ or 5 causal SNPs, with corresponding power $0,0.2,0.4,0.6,0.8$ or 1.0.  \emph{Top row:} Causal SNPs drawn from single causal pathway.  \emph{Bottom row:} Causal SNPs drawn at random.}
	\label{fig:SGL_ditribution}
	\end{center}
\end{figure}

To gain a deeper understanding of what is happening here, we also consider the power distributions across all 500 MC simulations corresponding to each point in the plots of Figure \ref{fig:SGL_vs_lasso_non-overlapping_power}.  These are illustrated in Figure \ref{fig:SGL_ditribution}.  The top row of plots illustrates the case where causal SNPs are drawn from a single causal pathway.  Here we see that there is a marked difference between the two distributions (SGL vs lasso).  The lasso shows a smooth distribution in power, with mean power increasing with effect size.  In contrast, with SGL the distribution is almost bimodal, with power typically either 0 or 1, depending on whether or not the correct causal pathway is selected.  This serves as an illustration of the advantage of pathway-driven SNP selection for the detection of causal SNPs in the case that pathways are important.  As previously found by \citet{Zhou2010} in the context of rare variants and gene selection, the joint modelling of SNPs within groups gives rise to a relaxation of the penalty on individual SNPs within selected groups, relative to the lasso.  This can enable the detection of SNPs with small effect size or low MAF that are missed by the lasso, which disregards pathways information and treats all SNPs equally.  Finally, where causal SNPs are not enriched in a causal pathway (bottom row of Figure \ref{fig:SGL_ditribution}), as expected SGL performs poorly.  In this case SGL will only select a SNP where the combined effects of constituent SNPs in a pathway are large enough to drive pathway selection.

\subsection*{The problem of overlapping pathways}\label{sec:pathways_sparse_group_lasso_with_overlaps}

The assumption that pathways are disjoint does not hold in practice, since genes and SNPs may map to multiple pathways (see Figure \ref{fig:pathMappingSchematic}).  This means that typically $\mathcal{G}_l \cap \mathcal{G}_{l'} \ne \emptyset$ for some $l \ne l'$.  In the context of pathways-driven SNP selection using SGL, this has two important implications.  Firstly, the optimisation \eqref{eq:SGL_objF} is no longer separable into groups (pathways), so that convergence using coordinate descent is no longer guaranteed \citep{Tseng2009}.  Secondly, we wish to be able to select pathways independently, and the SGL model as previously described does not allow this .  For example consider the case of an overlapping gene, that is a gene that maps to more than one pathway.  If a SNP mapping to this gene is selected in one pathway, then it must be selected in each and every pathway containing the mapped gene, so that all pathways mapping to the gene are selected.  We instead want to admit the possibility that the joint SNP effects in one pathway may be sufficient to allow pathway selection, while the joint effects in another pathway containing some of the same SNPs do not pass the threshold for pathway selection.

A solution to both these problems is obtained by duplicating SNP predictors in $\mathbf{X}$, so that SNPs belonging to more than one pathway can enter the model separately \citep{Jacob2009,Silver2012}.  The process works as follows.  An expanded design matrix is formed from the column-wise concatenation of the $L, (N \times P_l)$ sub-matrices, $\mathbf{X}_l$, to form the expanded design matrix $\mathbf{X}^* = [\mathbf{X}_1, \mathbf{X}_2, \dots , \mathbf{X}_L]$ of size $(N \times P^*)$, where $P^* = \sum_l P_l$.  The corresponding $P^* \times 1$ parameter vector, $\boldsymbol{\beta}^*$, is formed by joining the $L, ( P_l \times 1)$ pathway parameter vectors, $\boldsymbol{\beta}_l^*$, so that
$
	\boldsymbol{\beta}^* = [{\boldsymbol{\beta}_1^*}', {\boldsymbol{\beta}_2^*}', \dots,{\boldsymbol{\beta}_L^*}']'
$.
Pathway mappings with SNP indices in the expanded variable space are reflected in updated groups $\mathcal{G}_1^*, \ldots, \mathcal{G}_L^*$.  The SGL estimator \eqref{eq:SGL_objF}, adapted to account for overlapping groups, is then given by
\begin{equation}
	\boldsymbol{\hat{\beta}}^{SGL*} = 
		\argmin_{\boldsymbol{\beta}} 
		\Big \{
			\frac{1}{2} ||\mathbf{y}-\mathbf{X}^*\boldsymbol{\beta}^*||_2^2 + (1-\alpha) \lambda \sum_{l=1}^{L}w_l||\boldsymbol{\beta}^*_l||_2 + \alpha 					\lambda || \boldsymbol{\beta}^* ||_1
		\Big \}.
	\label{eq:SGL_overlaps_objF}
\end{equation}
With this overlap expansion, the model is then able to perform pathway and SNP selection in the way that we require, and the corresponding optimisation problem is amenable to solution using the BCGD estimation algorithm described in Box \ref{box:SGL_estimation_algorithm}.  However, for the purpose of pathways-driven SNP selection, the application of this algorithm presents a problem. This arises from the replication of overlapping SNP predictors in each group, $\mathbf{X}_l^*$, that they occur.  

Consider for example the simple situation where there are two pathways, $\mathcal{G}_k^*, \mathcal{G}_l^*$, containing sets of causal SNPs $\mathcal{S}^*_k \subseteq \mathcal{G}_k^*$ and $\mathcal{S}^*_l \subseteq \mathcal{G}_l^*$ respectively.  Here the $^*$ indicates that SNP indices refer to the expanded variable space.  We begin by assuming that $\mathcal{S}^*_k$ and $\mathcal{S}^*_l$ contain the same SNPs, so that in the \emph{unexpanded} variable space, $\mathcal{S}_k = \mathcal{S}_l$.

We then proceed with BCGD by first estimating $\boldsymbol{\beta}_k^*$.  We assume that the correct SNPs are selected, so that $\{ \hat{\beta}_j^* \ne 0: j \in \mathcal{S}^*_k \}$, and $\hat{\beta}_j^* = 0$ otherwise.  For the estimation of $\boldsymbol{\beta}^*_l$, the estimated effect $\sum_{j \in \mathcal{S}^*_k}X_j^* \hat{\beta}_j^*$, of these overlapping causal SNPs is removed from the regression, through its incorporation in the block residual $\mathbf{\hat{r}}^*_l = \mathbf{y} - \sum_{j \in \mathcal{S}^*_k}X_j^* \hat{\beta}_j^*$.  Since no other causal SNPs exist in pathway $\mathcal{G}_l^*$, $\mathbf{X}_l^{*'}\mathbf{\hat{r}}_l^* = \mathbf{0}$, so that the criterion for pathway selection, $|| S(\mathbf{X}^{*'}_l\hat{\mathbf{r}}^*_l, \alpha \lambda) ||_2 > (1-\alpha) \lambda w_l$ \eqref{eq:SGL_pathway_sel_criterion} is not met.  That is $\mathcal{G}_l^*$ is not selected.

\begin{figure}[tbp]
\begin{center}
	\includegraphics[trim = 50mm 80mm 50mm 80mm, clip, scale=0.7]{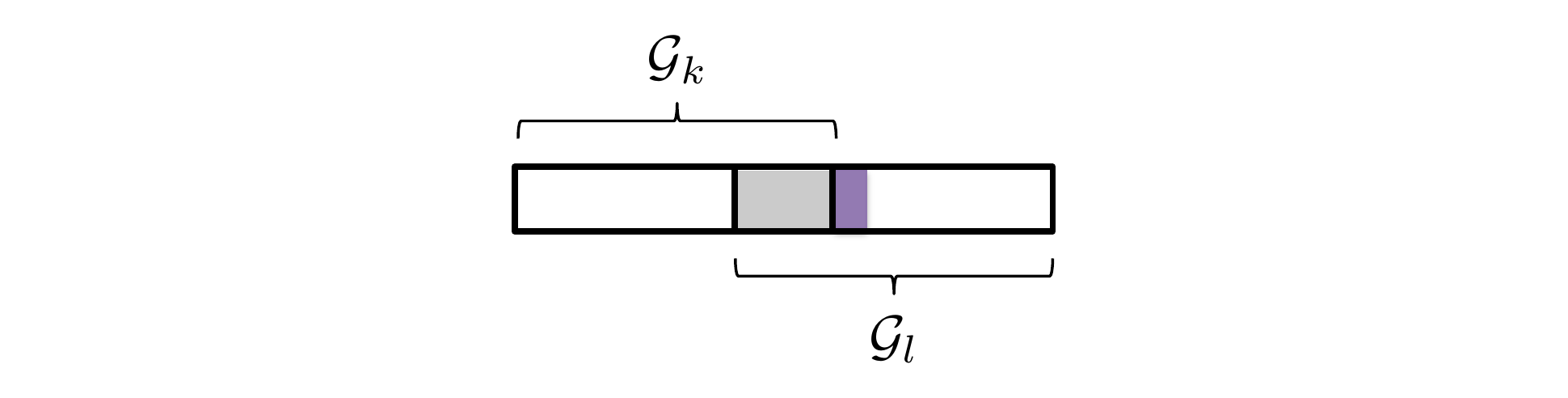}
\caption[Two pathways with partially overlapping causal SNPs]{Two pathways with partially overlapping causal SNPs.  Causal SNPs (marked in grey) in the set $\mathcal{S}_k$ overlap both pathways, so that $\mathcal{S}_k = \mathcal{G}_k \cap \mathcal{G}_l$.   Additional causal SNPs, $\mathcal{S}_l \cap \setminus \mathcal{S}_k$, (marked in purple) occur in pathway $l$ only.}
\label{fig:SGL_partial_overlap}
\end{center}
\end{figure}

Now consider the case where additional, non-overlapping causal SNPs, possibly with smaller effects, occur in $\mathcal{G}_l^*$, so that in the unexpanded variable space, $\mathcal{S}_k \subset \mathcal{S}_l$.  In other words, causal SNPs are \emph{partially overlapping} (see Fig \ref{fig:SGL_partial_overlap})\footnote{This is the situation for example where multiple causal genes overlap both pathways, but one or more additional causal genes occur in $\mathcal{G}_l$.}.  During BCGD pathway $\mathcal{G}_l^*$ is then less likely to be selected by the model, than would be the case if there were no overlapping SNPs, since once again the effects of overlapping causal SNPs, $\mathcal{S}_k \cap \mathcal{S}_l = \mathcal{S}_k$, are removed.  

For pathways-driven SNP selection, we will argue that we instead require that SNPs are selected in each and every pathway whose joint SNP effects pass a revised pathway selection threshold, irrespective of overlaps between pathways.  This is equivalent to the previous pathway selection criterion \eqref{eq:SGL_pathway_sel_criterion}, but with the additional assumption that pathways are independent, in the sense that they do not compete in the model estimation process.  We describe a revised estimation algorithm under the assumption of pathway independence below.

We justify the strong assumption of pathway independence with the following argument.  In reality, we expect that multiple pathways may simultaneously influence the phenotype, and we also expect that many such pathways will overlap, for example through their containing one or more `hub' genes, that overlap multiple pathways \citep{Kim2011,Lehner2006}.  By considering each pathway independently, we aim to maximise the sensitivity of our method to detect these variants and pathways.  In contrast, without the independence assumption, a competitive estimation algorithm will tend to pick out one from each set of similar, overlapping pathways, and miss potentially causal pathways and variants as a consequence.  We illustrate this idea in the simulation study in the following section.  One potential concern is that by not allowing pathways to compete against each other, specificity may be reduced, since too many pathways and SNPs may be selected.  We discuss the issue of specificity further in the context of results from the simulation study.  

A detailed derivation of the SGL model estimation algorithm under the independence assumption is given in supplementary information, Section \ref{SGL with overlaps}.  The main results are that the pathway \eqref{eq:SGL_pathway_sel_criterion} and SNP \eqref{eq:SGL_snp_sel_criterion} selection criteria become
\begin{align}
	|| S(\mathbf{X}_l'\mathbf{y}, \alpha \lambda) ||_2 	&> (1-\alpha) \lambda w_l, \quad \mbox{and} \notag \\
	||X'_j \mathbf{y} ||_1							&> \alpha \lambda
	\label{eq:SGL_overlaps_bl_equals_zero_criterion}
\end{align}
respectively.  The key difference is that partial derivatives $\hat{\mathbf{r}}_l$ and $\hat{\mathbf{r}}_{l,j}$ are replaced by $\mathbf{y}$, that is each pathway is regressed against the phenotype vector $\mathbf{y}$.  This means that there is no block coordinate descent stage in the estimation, so that the revised algorithm utilises only coordinate gradient descent within each selected pathway.  For this reason we use the acronym SGL-CGD for the revised algorithm, and SGL-BCGD for the previous algorithm using block coordinate gradient descent.  The new algorithm is described in Box \ref{box:SGL-CGD_estimation_algorithm}.  

Finally, we note that for SNP selection we are interested only in the set $\mathcal{\hat{S}}$ of selected SNPs in the unexpanded variable space, and not the set $\mathcal{S}^* = \{j^*: \beta_j^* \ne 0, j^* \in \{1, \ldots, P^* \}\}$.  Since, under the independence assumption, the estimation of each $\boldsymbol{\beta}_l^*$ does not depend on the other estimates, $\boldsymbol{\beta}_k^*, k \ne l$, we do not need to record separate coefficient estimates for each pathway in which a SNP is selected.  Instead we need only record the set $\mathcal{\hat{S}}_l, l \in \mathcal{\hat{C}}$ of SNPs selected in each selected pathway.  This has a useful practical implication, since we can avoid the need for an expansion of $\mathbf{X}$ or $\boldsymbol{\beta}$, and simply form the complete set of selected SNPs as
\begin{equation*}
	\mathcal{\hat{S}} = \bigcup_{l \in \mathcal{\hat{C}}} \mathcal{\hat{S}}_l.
\end{equation*}

\begin{algorithm}[h]
\begin{spacing}{1.25}
\NumTabs{25}
\begin{enumerate}
	\item
	initialise $\hat{\boldsymbol{\beta}}^* \leftarrow \mathbf{0}$.  
	\item
	for pathway $l = 1, 2, \ldots, L$: \\
	\tab\tab if $|| S(\mathbf{X}^{*'}_l\mathbf{y}, \alpha \lambda) ||_2 \le (1-\alpha) \lambda w_l$\\
	\tab\tab\tab $\boldsymbol{\hat{\beta}}^*_l \leftarrow \mathbf{0}$\\
	\tab\tab else\\
	\tab\tab\tab \textbf{repeat:} [CGD (SNP) loop] \\
	\tab\tab\tab\tab for $j = l_1,\dots,l_{P_l}$:\\
	\tab\tab\tab\tab\tab if $\hat{\beta}^*_j = 0:$\\
	\tab\tab\tab\tab\tab\tab Newton update $\hat{\beta}_j^{**} \leftarrow \hat{\beta}_j^*$ using \eqref{eq:SGL_directional_derivative_revised} and \eqref{eq:SGN_Newton_update}\\
	\tab\tab\tab\tab\tab else:\\
	\tab\tab\tab\tab\tab\tab Newton update $\hat{\beta}_j^{**} \leftarrow \hat{\beta}_j^*$ using \eqref{eq:SGL_partial_j_revised} and  \eqref{eq:SGN_Newton_update}\\
	\tab\tab\tab\tab\tab if $f(\boldsymbol{\beta}_l^{**}) > f(\boldsymbol{\beta}_l^*)$:\\
	\tab\tab\tab\tab\tab\tab $\hat{\beta}_j^{**} \leftarrow \frac{\hat{\beta}_j^{**} +  \hat{\beta}_j^*}{2}$\\
	\tab\tab\tab\tab\tab $\hat{\beta}_j^* \leftarrow \hat{\beta}_j^{**}$\\
	\tab\tab\tab \textbf{until convergence}
	\item
	$\boldsymbol{\hat{\beta}}^{SGL} \leftarrow \boldsymbol{\beta}^*$

\end{enumerate}
\end{spacing}

\caption{SGL-CGD estimation algorithm for overlapping pathways}
\label{box:SGL-CGD_estimation_algorithm}
\end{algorithm}

\subsection*{SGL simulation study 2} \label{sec:SGL_overlap_sim_study}
We now explore some of the issues raised in the preceding section, specifically the potential impact on pathway and SNP selection power and specificity of treating the pathways as independent in the SGL estimation algorithm.  We do this in a simulation study in which we simulate overlapping pathways.  The simulation scheme is specifically designed to highlight differences in pathway and SNP selection with the independence assumption (using the SGL-CGD estimation algorithm in Box \ref{box:SGL-CGD_estimation_algorithm}) and without it (using the standard SGL estimation algorithm in Box \ref{box:SGL_estimation_algorithm}).

\begin{figure}
	\begin{center}
	\subfigure[]{\includegraphics[trim = 20mm 75mm 20mm 75mm, clip, scale=0.5]{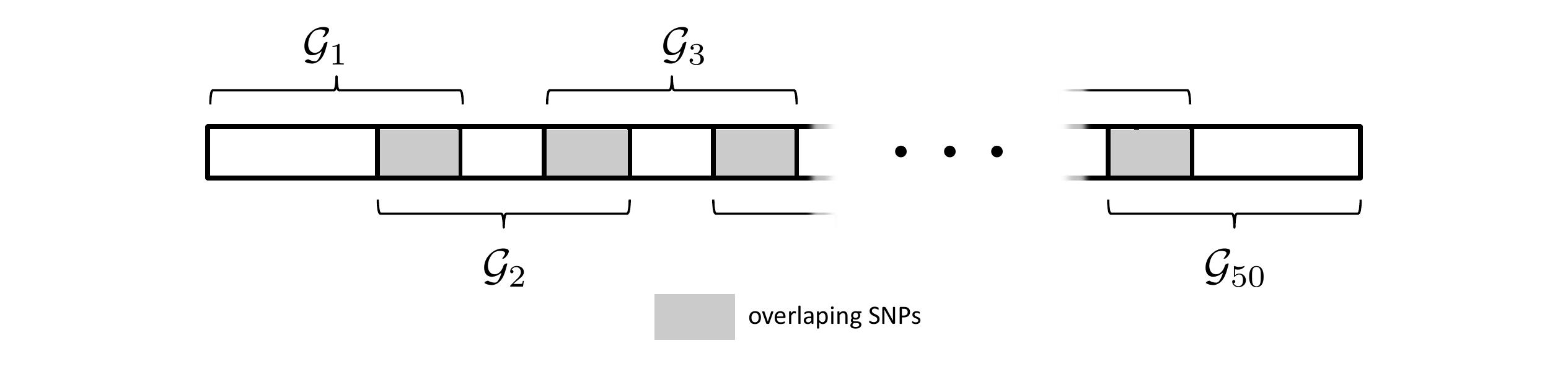}}
	\subfigure[]{\includegraphics[trim = 20mm 85mm 20mm 75mm, clip, scale=0.5]{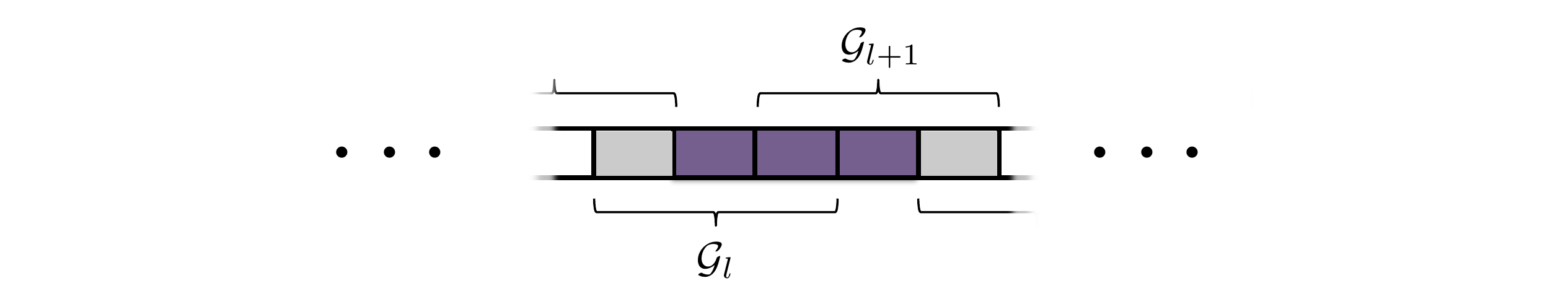}}
	\caption[SGL Simulation Study with overlapping pathways]{SGL Simulation Study with overlapping pathways.  (a): Illustration of pathway overlap scheme.  The are 30 SNPs in each pathway.  Pathways $\mathcal{G}_l, (l = 1, \ldots, 50)$ overlap each adjacent pathway by 10 SNPs. (b): Causal SNPs from adjacent pathways, $l, l+1$ are randomly selected from the region marked in purple, ensuring that SNPs in $\mathcal{S}$ overlap a maximum of two pathways.}
	\label{fig:SNP_overlap_sim_scheme}
	\end{center}
\end{figure}

SNPs with variable MAF are simulated using the same procedure described in the previous simulation study, but this time SNPs are mapped to 50 \emph{overlapping} pathways, each containing 30 SNPs.  Each pathway overlaps any adjacent (by pathway index) pathway by 10 SNPs.  This overlap scheme is illustrated in Figure \ref{fig:SNP_overlap_sim_scheme} (a).

As before we consider a range of overall genetic effect sizes, $\gamma$.  A total of 2000 MC simulations are conducted for each effect size.  At  MC simulation $z$, we randomly select two adjacent pathways, $\mathcal{G}_l, \mathcal{G}_{l+1}$ where $l \in \{1, \ldots, 49\}$.  From these two pathways we randomly select 10 SNPs according to the scheme illustrated in Figure \ref{fig:SNP_overlap_sim_scheme} (b).  This ensures that causal SNPs overlap a minimum of 1, and a maximum of 2 pathways, with $\mathcal{S}_z \subset (\mathcal{G}_l \cap \setminus \mathcal{G}_{l-1}) \cup (\mathcal{G}_{l+1} \cap \setminus \mathcal{G}_{l+2})$.  The true set of causal pathways, $\mathcal{C}$, is then given by $\{l\}$, $\{l+1\}$ or $\{l,l+1\}$ (although simulations where $|\mathcal{C}| = 1$ will be extremely rare).  Genetic effects on the phenotype are generated as described previously (Section \ref{sim_study_1}).  

SNP coefficients are estimated for each algorithm, SGL-BCGD and SGL-CGD, using the same regularisation with $\lambda = 0.85 \lambda_{max}$ and $\alpha = 0.85$ for both.

The average number of pathways and SNPs selected by SGL-BCGD and SGL-CGD across all 2000 MC simulations is reported in Table \ref{tab:SGL_simStudy2_nSelVars}.  As expected, for both models, the number of selected variables (pathways or SNPs) increases with decreasing effect size, as the number of pathways close to the selection threshold set by $\lambda_{max}$ increases.

\begin{table}[htdp]
\caption{Simulation Study 2: Mean number of pathways and SNPs selected by each model at each effect size, $\gamma$, across 2000 MC simulations.}
\begin{center}
\begin{tabular}{rr|cccccc}
&&\multicolumn{6}{c}{$\gamma$}\\
&&0.02&0.04&0.06&0.08&0.1&0.12\\
\hline
pathways & SGL-CGD &  5.8 & 5.9 & 5.4 & 4.8 & 3.9 & 3.2 \\
& SGL-BCGD &  5.8 & 5.9 & 5.4 & 4.8 & 3.9 & 3.2 \\
\hline
SNPs & SGL-CGD &  26.6 & 27.0 & 24.8 & 22.2 & 18.5 & 15.3 \\
& SGL-BCGD &  28.8 & 29.3 & 26.7 & 23.6 & 19.4 & 15.8 \\
\end{tabular}
\end{center}
\label{tab:SGL_simStudy2_nSelVars}
\end{table}%

For each model, at MC simulation $z$ we record the pathway and SNP selection power, $|\mathcal{\hat{C}}_z \cap \mathcal{C}_z| / |\mathcal{C}_z|$ and $|\mathcal{\hat{S}}_z \cap \mathcal{S}_z| / |\mathcal{S}_z|$ respectively.  Since the number of selected variables can vary slightly between the two models, we also record false positive rates (FPR) for pathway and SNP selection as $|\mathcal{\hat{C}}_z \cap \setminus \mathcal{C}_z| / |\mathcal{\hat{C}}_z|$ and $|\mathcal{\hat{S}}_z \cap \setminus \mathcal{S}_z| / |\mathcal{\hat{S}}_z|$ respectively.  

\begin{figure}[htbp]
\begin{center}
	\includegraphics[trim = 40mm 20mm 25mm 20mm, clip, scale=0.3]{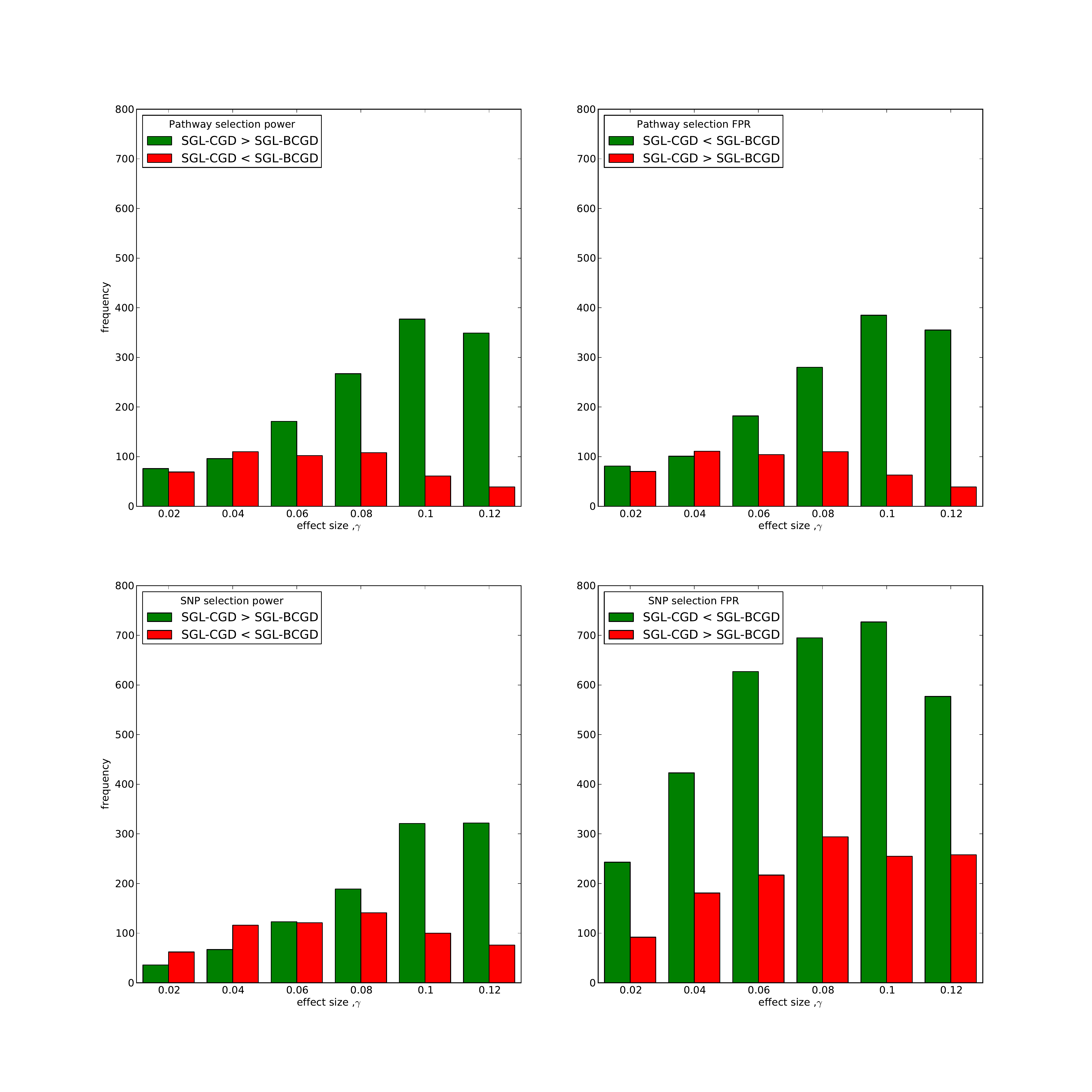}
\caption[SGL-CGD vs SGL-BCGD performance, measured across 2000 MC simulations]{SGL-CGD vs SGL-BCGD performance, measured across 2000 MC simulations. \emph{Top row: } Pathway selection performance.  (Left) green bars indicate the number of MC simulations where SGL-CGD has greater pathway selection power than SGL.  Red bars indicate where SGL-BCGD has greater power than SGL-CGD.  (Right) green bars indicate the number of MC simulations where SGL-CGD has a lower FPR than SGL.  Red bars indicate the opposite.  \emph{Bottom row: }  As above, but for SNP selection performance.}
\label{fig:SGL-CGD_vs_SGL_performance}
\end{center}
\end{figure}

The large possible variation in causal SNP distributions, causal SNP MAFs etc.~make a comparison of mean power and FPR between the two methods somewhat unsatisfactory.  For example, depending on effect size, a large number of simulations can have either very high, or very low pathway and SNP selection power, masking subtle differences in performance between the two methods.  Since we are specifically interested in establishing the relative performance of the two methods, we instead illustrate the number of simulations at which one method outperforms the other across all 2000 MC simulations, and show this in Figure \ref{fig:SGL-CGD_vs_SGL_performance}.  In this figure, the number of simulations in which SGL-CGD outperforms SGL, i.e. where SGL-CGD power $>$ SGL-BCGD power, or SGL-CGD FPR $<$ SGL-BCGD FPR, are shown in green.  Conversely, the number of simulations where SGL-BCGD outperforms SGL-CGD are shown in red.

We first consider pathway selection performance (top row of Figure \ref{fig:SGL-CGD_vs_SGL_performance}).  For both methods, the same number of pathways are selected on average, across all effect sizes (Table \ref{tab:SGL_simStudy2_nSelVars}).  At low effect sizes, there is no difference in performance between the two methods for the large majority of MC simulations, and where there is a difference, the two methods are evenly balanced.  As with SGL Simulation Study 1, this is the region (with $\gamma \le 0.04$) where pathway selection fairs no better than chance.  With $\gamma > 0.04$, SGL-CGD consistently outperforms SGL, both in terms of pathway selection sensitivity and control of false positives (measured by FPR).  

To understand why, we turn to SNP selection performance (bottom row of Figure \ref{fig:SGL-CGD_vs_SGL_performance}).  At small effect sizes ($\gamma \le 0.04$), in the small minority of simulations where the correct pathways are identified, SGL-BCGD tends to demonstrate greater power than SGL-CGD (Figure \ref{fig:SGL-CGD_vs_SGL_performance} bottom left).  However, this is at the expense of lower specificity (Figure \ref{fig:SGL-CGD_vs_SGL_performance} bottom right).  These difference are due to the slightly larger number of SNPs selected by SGL-BCGD (see Table \ref{tab:SGL_simStudy2_nSelVars}), which in turn is due to the `screening out' of previously selected SNPs from the adjacent causal pathway during BCGD, as described previously.  This results in the selection of a larger number of SNPs when any two overlapping pathways are selected by the model.  In the case where two causal pathways are selected, SNP selection power is then likely to be higher, although at the expense of a greater number of false positives.

When pathway effects are just on the margin of detectability ($\gamma = 0.06$), SGL-CGD is more often able to select both causal pathways, although this doesn't translate into increased SNP selection power.  This is most likely because at this effect size neither model can detect SNPs with low MAF, so that SGL-CGD is detecting the same (overlapping) SNPs in both causal pathways.  Note that once again SGL-BCGD typically has a higher FPR than SGL-CGD, since more SNPs are selected from non-causal pathways.

As the effect size increases, the number of simulations in which SGL-CGD outperforms SGL-BCGD for SNP selection power grows, paralleling the former method's enhanced pathway selection power.  This is again a demonstration of the screening effect with SGL-BCGD described previously.  This means that SGL-CGD is more often able to select both causal pathways, and to select additional causal SNPs that are missed by SGL.  These additional SNPs are likely to be those with lower MAF, for example, that are harder to detect with SGL, once the effect of overlapping SNPs are screened out during estimation using BCGD.  Interestingly, as before SGL-CGD continues to exhibit lower false positive rates than SGL.  This suggests that, with the simulated data considered here, the independence assumption offers better control of false positives by enabling the selection of causal SNPs in each and every pathway to which they are mapped.  In contrast, where causal SNPs are successively screened out during the estimation using BCGD, too many SNPs with spurious effects are selected.  

The relative advantage of SGL-CGD over SGL-BCGD on all performance measures starts to decrease around $\gamma = 0.1$, as SGL-BCGD becomes better able to detect all causal pathways and SNPs, irrespective of the screening effect.

\subsection*{Pathway and SNP selection bias}\label{sec:bias}

One issue that must be addressed is the problem of selection bias, by which we mean the tendency of SGL to favour the selection of particular pathways or SNPs under the null, where no SNPs influence the phenotype.  Possible biasing factors include variations in pathway size, that is the number of SNPs mapping to a pathway, or varying patterns of SNP-SNP correlations and gene sizes.  Common strategies for bias reduction include the use of dimensionality reduction techniques and permutation methods \citep{Wang2009,Holmans2009,Zhao2011,Chen2011b}.  

In earlier work we described an adaptive weight-tuning strategy, designed to reduce selection bias in a group lasso-based pathway selection method \citep{Silver2012}.  This works by tuning the pathway weight vector, $\mathbf{w} = (w_1, w_2, \ldots, w_L)$, so as to ensure that pathways are selected with equal probability under the null.  This strategy can be readily extended to the case of dual-level sparsity with the SGL.

Our procedure rests on the observation that for pathway selection to be unbiased, each pathway must have an equal chance of being selected.  For a given $\alpha$, and with $\lambda$ tuned to ensure that a single pathway is selected, pathway selection probabilities are then described by a uniform distribution, $\Pi_l = 1/L$, for $l = 1,\ldots,L$.  We proceed by calculating an empirical pathway selection frequency distribution, $\Pi^*(\mathbf{w})$, by determining which pathway will first be selected by the model as $\lambda$ is reduced from its maximal value, $\lambda_{max}$, over multiple permutations of the response, $\mathbf{y}$.  This process is described in detail in Supplementary Information \ref{weight_tuning}.

Our iterative weight tuning procedure then works by applying successive adjustments to the pathway weight vector, $\boldsymbol{w}$, so as to reduce the difference, $d_l = \Pi^*_l(\boldsymbol{w}) - \Pi_l$, between the unbiased and empirical (biased) distributions for each pathway.  At iteration $\tau$, we compute the empirical pathway selection probability distribution $\Pi^*(\mathbf{w}^{(\tau)})$, determine $d_l$ for each pathway, and then apply the following weight adjustment
\begin{equation*}
	w_l^{(\tau+1)} 	= w_l^{(\tau)} \left[1 - \text{sign} (d_l) (\eta - 1) L^2 d_l^2 \right] 	\qquad 0 <  \eta < 1, \quad l = 1, \ldots, L.
	\label{eq:adjustWeights}	
\end{equation*}
The parameter $\eta$ controls the maximum amount by which each $w_l$ can be reduced in a single iteration, in the case that pathway $l$ is selected with zero frequency.  The square in the weight adjustment factor ensures that large values of $|d_l|$ result in relatively large adjustments to $w_l$.  Iterations continue until convergence, where $\sum_{l=1}^L |d_l| < \epsilon$.  

Note that when multiple pathways are selected by the model, the expected pathway selection frequency distribution under the null will not be uniform.  This is because pathways overlap, so that selection frequencies will reflect the complex distribution of overlapping genes, as indeed will unbiased empirical selection frequencies.  We have shown previously in extensive simulations that this adaptive weight-tuning procedure gives rise to substantial gains in sensitivity and specificity with regard to pathway selection \citep{Silver2012}.

\subsection*{Pathway, SNP and gene ranking}\label{sec:SGL-CGD_ranking}

With most variable selection methods, a choice for the regularisation parameter, $\lambda$, must be made, since this determines the number of variables selected by the model.  Common strategies include the use of cross validation to choose a $\lambda$ value that minimises the prediction error between training and test datasets \citep{Hastie}.  One drawback of this approach is that it focuses on optimising the \emph{size} of the set, $\hat{\mathcal{C}}$, of selected pathways (more generally, selected variables) that minimises the cross validated prediction error.   Since the variables in $\hat{\mathcal{C}}$ will vary across each fold of the cross validation, this procedure is not in general a good means of establishing the importance of a unique set of variables, and  can give rise to the selection of too many variables \citep{Vounou2011,Meinshausen2010}.  For the lasso, alternative approaches, based on data subsampling or bootstrapping have been shown to improve model consistency, in the sense that the correct model is selected with a high probability \citep{Bach2008b,Meinshausen2010,Chatterjee2011}.  These methods work by recording selected variables across multiple subsamples of the data, and forming the final set of selected variables either as the intersection of variables selected at each model fit, or by assessing variable selection frequencies.  Examples of the use of such approaches can be found in a number of recent gene mapping studies involving model selection using either the lasso or elastic net \citep{Cho2010,Eleftherohorinou2011,Motyer2011,Vounou2011}.  Motivated by these ideas, we adopt a resampling strategy in which we calculate pathway, gene and SNP selection frequencies by repeatedly fitting the model over $B$ subsamples of the data, at fixed values for $\alpha$ and $\lambda$.  Each random subsample of size $N/2$ is drawn without replacement.   Our motivation here is to exploit knowledge of finite sample variability obtained by subsampling, to achieve better estimates of a variable's importance.  With this approach, which in some respects resembles the `pointwise stability selection' strategy of \citet{Meinshausen2010}, selection frequencies provide a direct measure of confidence in the selected pathways in a finite sample.  This resampling strategy also allows us to rank pathways, genes and SNPs in order of their strength of association with the phenotype, so that we expect the true set of causal variables to achieve a high ranking, whereas non-causal variables will be ranked low.   

For pathway ranking, we denote the set of selected pathways at subsample $b$ by
\begin{equation*}
	\hat{\mathcal{C}}^{(b)} = \{ l: \boldsymbol{\hat{\beta}}_l^{(b)} \ne \mathbf{0} \}	\quad b = 1, \ldots, B,
\end{equation*}
where $\boldsymbol{\hat{\beta}}_l^{(b)}$ is the estimated SNP coefficient vector for pathway $l$ at subsample $b$.  The selection probability for pathway $l$ measured across all $B$ subsamples is then
\begin{equation*}
	\pi^{path}_l = \frac{1}{B} \sum_{b=1}^B I_l^{(b)}	\quad l = 1, \ldots, L
\end{equation*}
where the indicator function, $I_l^{(b)} = 1$ if $l \in \hat{\mathcal{C}}^{(b)}$, and 0 otherwise.  Pathways are ranked in order of their selection probabilities, $\pi^{path}_{l_1} \ge, \ldots, \ge \pi^{path}_{l_L}$.

For SNP and gene ranking, we denote the set of SNPs selected at subsample $b$ (in the unexpanded variable space) by $\hat{\mathcal{S}}^{(b)}$, and further denote the set of selected genes to which the SNPs in $\mathcal{\hat{S}}^{(b)}$ are mapped by $\hat{\phi}^{(b)} \subset \Phi$, where $\Phi = \{1, \ldots, G \}$ is the set of gene indices corresponding to all $G$ mapped genes.  Using the same strategy as for pathway ranking, we obtain an expression for the selection probability of SNP $j$ across $B$ subsamples as
\begin{equation*}
	\pi^{SNP}_{j} = \frac{1}{B} \sum_{b=1}^B J_{j}^{(b)}	
\end{equation*}
where the indicator function, $J_{j}^{(b)} = 1$ if $j \in \mathcal{\hat{S}}^{(b)}$, and 0 otherwise.  A similar expression for the selection probability for gene $g$  is
\begin{equation*}
	\pi^{gene}_g = \frac{1}{B} \sum_{b=1}^B K_{g}^{(b)}
\end{equation*}
where the indicator function, $K_{g}^{(b)} = 1$ if $g \in \hat{\phi}^{(b)}$, and 0 otherwise.  SNPs and genes are then ranked in order of their respective selection frequencies. 

\section*{Results}\label{sec:Results}

\subsection*{Subjects, genotypes and phenotypes}\label{sec:P-SGLAW_study_genotypes}

The analysis is carried out using data from two separate cohorts of Asian adults.  These datasets have previously been used to search for novel variants associated with type 2 diabetes mellitus (T2D) in Asian populations.  The first (discovery) cohort is from the Singapore Prospective Study Program, hereafter referred to as `SP2', and the second (replication) dataset is from the Singapore Malay Eye Study or `SiMES'.  Detailed information on both datasets can be found in \citet{Sim2011}, but we briefly outline some salient features here.

Both datasets comprise whole genome data for T2D cases and controls, genotyped on the Illumina HumanHap 610 Quad array.  For the present study we use controls only, since variation in lipid levels between cases and controls can be greater than the variation within controls alone. The use of both cases and controls in our analysis might then lead to a confounded analysis, where any associations could be linked to T2D status or some other spurious factor.

The SP2 dataset consists entirely of ethnic Chinese, and shows no evidence of population stratification.  The SiMES dataset comprises ethnic Malays, and shows some evidence of cryptic relatedness between samples.  For this reason, the first two principal components of a PCA for population structure are used as covariates in our analysis of this dataset.  Again full details of the stratification analysis can be found in \citet{Sim2011} and associated supplementary information.

A summary of information pertaining to genotypes for each dataset, both before and after imputation and pathway mapping, is given in Table \ref{tab:lipid_study}, along with a list of phenotypes and covariates.

\begin{table}[htdp]
\small
\caption{Genotype and phenotype information corresponding to the SP2 and SiMES datasets used in the study.}
\begin{center}
\begin{tabular}{lcc}
&SP2&Simes\\
\hline
Sample size&$N=1,040$&$N=1,099$\\\\
\textbf{Genotypes}\\
\emph{Before imputation}\\
SNPs available for analysis$^($\footnotemark[1]$^)$&$542,297$&$557,824$\\
SNPs with missing genotypes$^($\footnotemark[2]$^)$&$152,372$&$282,549$\\
\emph{Post imputation}\\
SNPs available for analysis$^($\footnotemark[3]$^)$&$492,639$&$515,503$\\\\
\textbf{Phenotypes/covariates}\\
quantitative trait (phenotype)$^($\footnotemark[4]$^)$&HDLC&HDLC\\
covariates&gender, age, age$^2$,&gender, age, age$^2$,\\ &BMI$^($\footnotemark[5]$^)$& BMI, PC1, PC2$^($\footnotemark[6]$^)$\\
\end{tabular}
\end{center}
\small
$^($\footnotemark[1]$^)$after first round of quality control \citep{Sim2011} and removal of monomorphic SNPs\\
$^($\footnotemark[2]$^)$maximum 5\% missing rate per SNP\\
$^($\footnotemark[3]$^)$after imputation and removal of SNPs with MAF$<0.01$\\
$^($\footnotemark[4]$^)$mg/dL\\
$^($\footnotemark[5]$^)$body mass index (kg/m$^2$)\\
$^($\footnotemark[6]$^)$principal components relating to cryptic relatedness
\label{tab:lipid_study}
\end{table}

\subsection*{Genotype imputation}\label{sec:P-SGLAW_study_genotype_imputation}
After the initial round of quality control, genotypes for both datasets have a maximum SNP missingness of 5\%.  Since our method cannot handle missing values, we perform `missing holes' SNP imputation, so that all missing SNP calls are estimated against a reference panel of known haplotypes.  

SNP imputation proceeds in two stages.  First, imputation requires accurate estimation of haplotypes from diploid genotypes (phasing).  This is performed using SHAPEIT v1 (\url{http://www.shapeit.fr}).  This uses a hidden Markov model to infer haplotypes from sample genotypes using a map of known recombination rates across the genome \citep{Delaneau2012}.  The recombination map must correspond to genotype coordinates in the dataset to be imputed, so we use recombination data from HapMap phase II, corresponding to genome build NCBI b36 (\url{http://hapmap.ncbi.nlm.nih.gov/downloads/recombination/2008-03_rel22_B36/}).

Following the primary phasing stage, SNP imputation is performed using IMPUTE v2.2.2 (\url{http://mathgen.stats.ox.ac.uk/impute/impute_v2.html}).  IMPUTE uses a reference panel of known haplotypes to infer unobserved genotypes, given a set of observed sample haplotypes \citep{Howie2011}.  The latest version (IMPUTE 2) uses an updated, efficient algorithm, so that a custom reference panel can be used for each study haplotype, and for each region of the genome, enabling the full range of reference information provided by HapMap3 \citep{The1000GenomesProjectConsortium2011} to be used.  Following IMPUTE 2 guidelines, we use HapMap3 reference data corresponding to NCBI b36 (\url{http://mathgen.stats.ox.ac.uk/impute/data_download_hapmap3_r2.html}) which includes haplotype data for 1,011 individuals from Africa, Asia, Europe and the Americas.  SNPs are imputed in 5MB chunks, using an effective population size (\emph{Ne}) of 15,000, and a buffer of 250kb to avoid edge effects, again as recommended for IMPUTE 2.

The phasing and imputation process is complex and computationally intensive.  For this reason we implement a pipeline in Python, with phasing and imputation for each chromosome conducted in parallel across multiple nodes in a computing cluster.  This enables full genome imputation that would otherwise take days, to be completed in a matter of hours.

\subsection*{Pathway mapping}\label{sec:P-SGLAW_study_pathway_mapping}

Pathways GWAS methods rely on prior information mapping SNPs to functional networks or pathways.  Since pathways are typically defined as groups of interacting genes, SNP to pathway mapping is a two-part process, requiring the mapping of genes to pathways, and of SNPs to genes.  A consistent strategy for this mapping process has however yet to be established, a situation compounded by a lack of agreement on what constitutes a pathway in the first place \citep{Cantor2010}.  

The number and size of databases devoted to classifying genes into pathways is growing rapidly, as is the range and diversity of gene interactions considered (see for example \url{http://www.pathguide.org/}).  Databases such as those provided by KEGG (\url{http://www.genome.jp/kegg/pathway.html}), Reactome (\url{http://www.reactome.org/}) and Biocarta (\url{http://www.biocarta.com/}) classify pathways across a number of functional domains, for example apoptosis, cell adhesion or lipid metabolism; or crystallise current knowledge on specific disease-related molecular reaction networks.  Strategies for pathways database assembly range from a fully-automated text-mining approach, to that of careful curation by experts.  Inevitably therefore, there is considerable variation between databases, in terms of both gene coverage and consistency \citep{Soh2010}, so that the choice of database(s) will itself influence results in pathways GWAS.  

\begin{figure}[htbp]
\begin{center}
	\includegraphics[trim = 50mm 70mm 40mm 70mm, clip, scale=0.7]{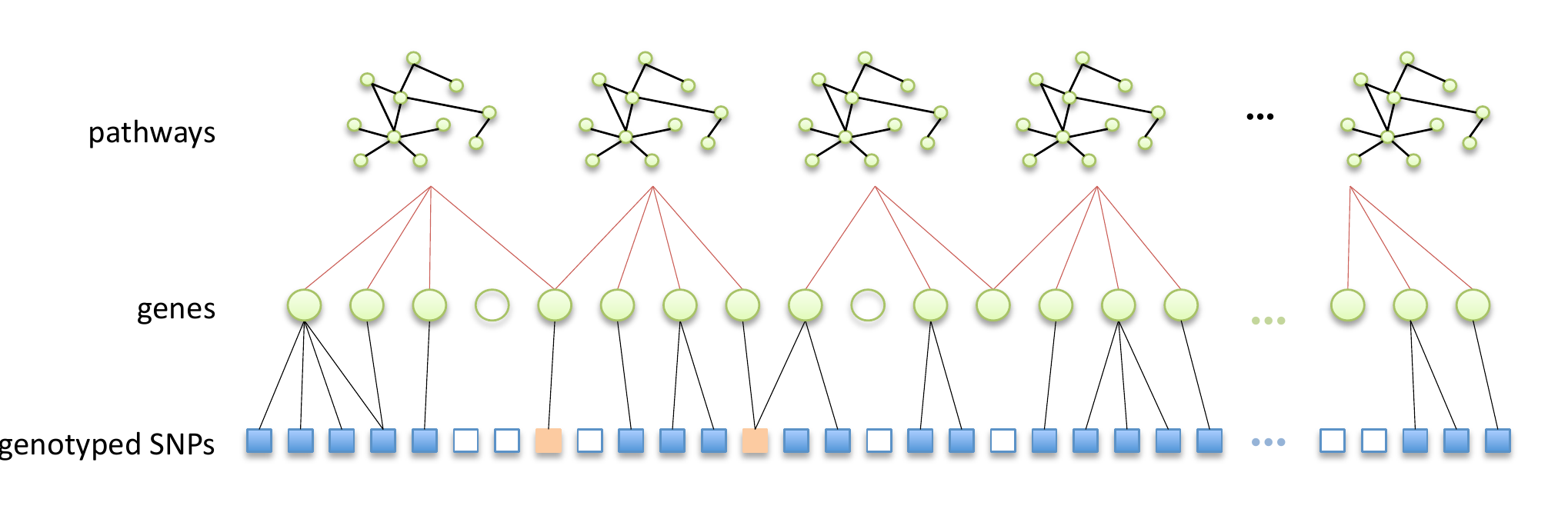}
\caption[Schematic illustration of the SNP to pathway mapping process]{Schematic illustration of the SNP to pathway mapping process.  (i) Genes (green circles) are mapped to pathways using information on gene-gene interactions (top row), obtained from a gene pathways database.  Many genes do not map to any known pathway (unfilled circles).  Also, some genes may map to more than one pathway.  (ii) Genes that map to a pathway are in turn mapped to genotyped SNPs within a specified distance.  Many SNPs cannot be mapped to a pathway since they do not map to a mapped gene (unfilled squares).  Note SNPs may map to more than one gene.  Some SNPs (orange squares) may map to more than one pathway, either because they map to multiple genes belonging to different pathways, or because they map to a single gene that belongs to multiple pathways.}
\label{fig:pathMappingSchematic}
\end{center}
\end{figure}

The mapping of SNPs to genes adds a further layer of complexity, since although many SNPs may occur within gene boundaries, on a typical GWAS array the vast majority of SNPs will reside in inter-genic regions.  In an attempt to include variants potentially residing in functionally significant regions lying outside gene boundaries, SNPs may be mapped to nearby genes using various distance thresholds.  Various values for SNP to gene mapping distances, measured in thousands of nucleotide base pairs (kb), have been suggested in the literature, ranging from mapping SNPs to genes only if they fall within a specific gene, to the attempt to encompass upstream promoters and enhancers by extending the range to $10, 20$ or even $500$kb and beyond \citep{Wang2009,Eleftherohorinou2009,Cantor2010}.  This process is illustrated schematically in Figure \ref{fig:pathMappingSchematic}.  Notable features of the SNP to pathway mapping process include the fact that genes (and therefore SNPs) may map to more than one pathway, and also that many SNPs and genes do not currently map to any known pathway \citep{Fridley2011}.  

\begin{figure}[htbp]
\begin{center}
	\includegraphics[trim = 60mm 62mm 50mm 62mm, clip, scale=0.45]{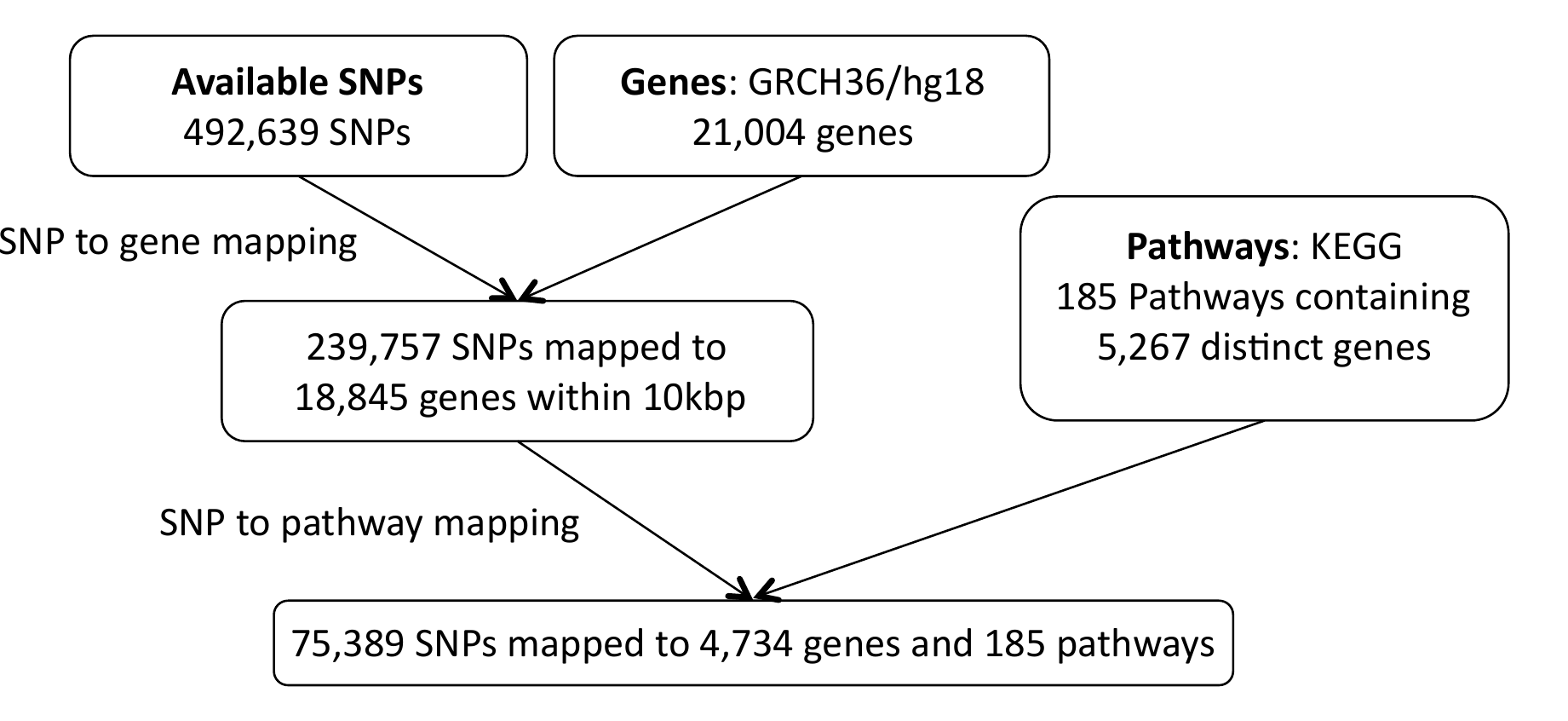}
\caption{\emph{SP2 dataset}.  SNP to pathway mapping.}
\label{fig:SP2_pathway_mapping_flowchart}
\end{center}
\end{figure}

\begin{figure}[htbp]
\begin{center}
	\includegraphics[trim = 60mm 62mm 50mm 62mm, clip, scale=0.45]{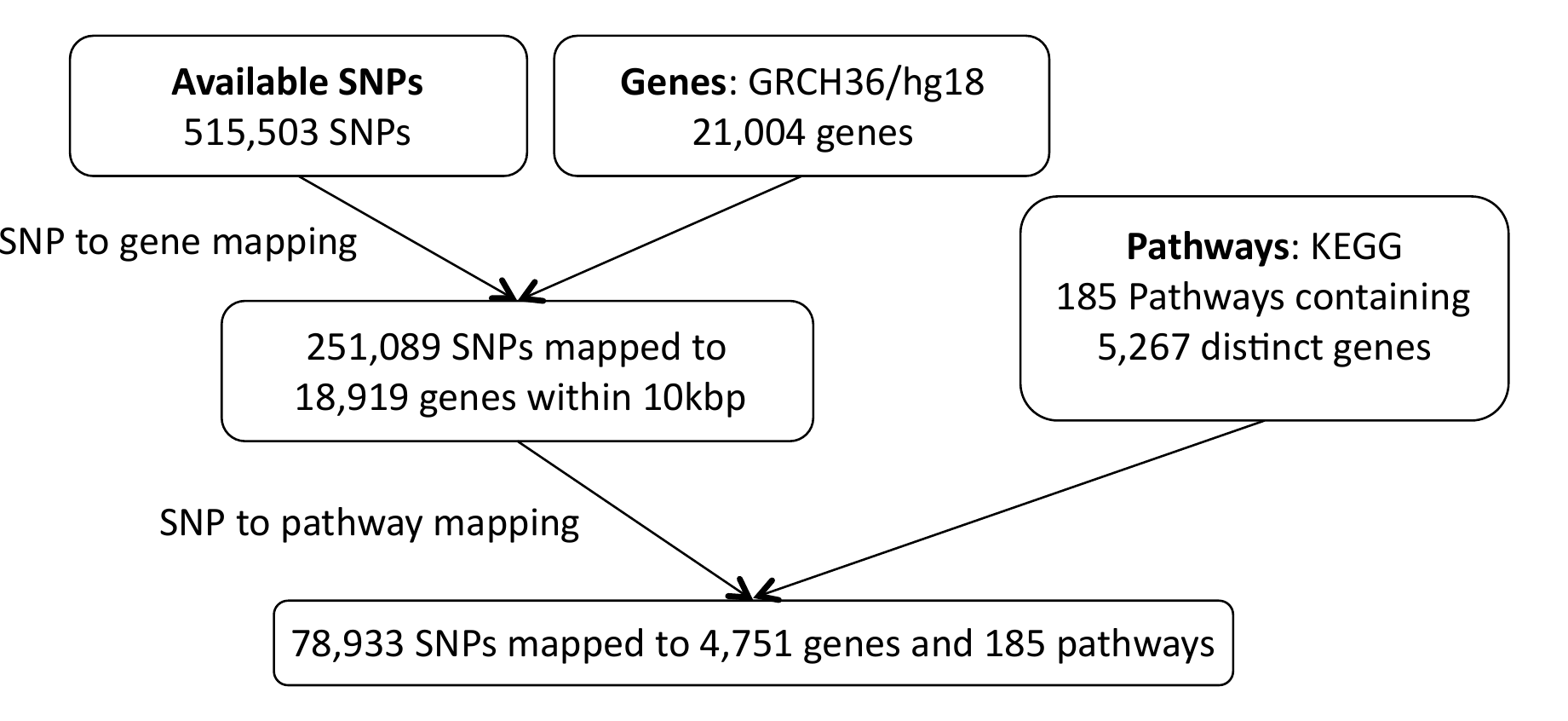}
\caption{\emph{SiMES dataset}.  SNP to pathway mapping.}
\label{fig:SiMES_pathway_mapping_flowchart}
\end{center}
\end{figure}

Following imputation, SNPs for both datasets in the present study are mapped to KEGG canonical pathways from the MSigDB database (\url{http://www.broadinstitute.org/gsea/msigdb/index.jsp}).   We exclude the largest KEGG pathway (by number of mapped SNPs), `Pathways in Cancer', since it is highly redundant in that it contains multiple other pathways as subsets.  Details of the pathway mapping process are given in Figures \ref{fig:SP2_pathway_mapping_flowchart} and \ref{fig:SiMES_pathway_mapping_flowchart}.  

Note that there is a difference in the number of SNPs available for the pathway mapping between the two datasets, and this results in a small discrepancy in the total number of mapped genes (SP2: 4,734 mapped genes; SiMES: 4,751).   However, both datasets map to all 185 KEGG pathways, and  a large majority of mapped genes and SNPs overlap both datasets.  Detailed information on the pathway mapping process for the two datasets is presented in Table \ref{tab:lipid_pathway_mapping_stats}.\\\\
We perform pathways-driven SNP selection on both datasets, using the procedures described in Methods.  We present results for each dataset separately below.

\begin{table}[htdp]
\small
\caption{Comparison of SNP and gene to pathway mappings for the SP2 and SiMES datasets.}
\begin{center}
\begin{tabular}{lcc}
&SP2&SiMES\\
\hline
Total SNPs mapping to pathways&75,389&78,933\\
Total SNPs mapping to pathways in both datasets (intersection)&\multicolumn{2}{c}{74,864}\\
Total mapped genes&4,734&4,751\\
Total genes mapping to pathways in both datasets (intersection)&\multicolumn{2}{c}{4,726}\\
Total mapped pathways&185&185\\
Minimum number of genes mapping to single pathway&11&11\\
Maximum number of genes mapping to single pathway&63&63\\
Minimum number of SNPs mapping to single pathway&66&67\\
Maximum number of SNPs mapping to single pathway&5,759&6,058\\
Minimum number of pathways mapping to a single SNP&1&1\\
Maximum number of pathways mapping to a single SNP&45&45
\end{tabular}
\end{center}
\label{tab:lipid_pathway_mapping_stats}
\end{table}

\newpage
\subsection*{SP2 Analysis}

For the SP2 dataset we consider two separate scenarios for the regularisation parameters $\lambda$ and $\alpha$.  For the two scenarios we set the sparsity parameter, $\lambda=0.95 \lambda_{max}$, but consider two values for $\alpha$, namely $\alpha = 0.95, 0.85$.  We test each scenario over 1000 $N/2$ subsamples.  We also compare the resulting pathway and SNP selection frequency distributions with null distributions, again over 1000 $N/2$ subsamples, but with phenotype labels permuted, so that no SNPs can influence the phenotype.

The parameter $\alpha$ controls how the regularisation penalty is distributed between the $\ell_2$ (pathway) and $\ell_1$ (SNP) norms of the coefficient vector.  Each scenario therefore entails different numbers of selected pathways and SNPs, and this information is presented in Table \ref{tab:SP2_nsel_vars}.  

\begin{table}[htdp]
\caption[Separate combinations of regularisation parameters, $\lambda$ and $\alpha$ used for analysis of the SP2 dataset]{Separate combinations of regularisation parameters, $\lambda$ and $\alpha$ used for analysis of the SP2 dataset.  For each $\lambda$, $\alpha$ combination, the mean ($\pm$SD) number of selected pathways and SNPs across all $1000$ subsamples is reported.}
\begin{center}
\begin{tabular}{l|cc}
&\multicolumn{2}{c}{$\lambda=0.95 \lambda_{max}$}\\
&$\alpha=0.85$&$\alpha=0.95$\\
\hline
\emph{empirical}\\
selected pathways & $7.9 \pm 6.1$&$4.8 \pm 4.1$\\
selected SNPs & $1551 \pm 1294$ & $160 \pm 185$\\\\
\emph{null}\\
selected pathways & $9.1 \pm 7.2$ & $5.0 \pm 4.55$\\
selected SNPs & $1656 \pm 1401$ & $155 \pm 194$\\
\end{tabular}
\end{center}
\label{tab:SP2_nsel_vars}
\end{table}%

\begin{figure}
	\begin{center}
	\subfigure[]{\includegraphics[trim = 10mm 0mm 0mm 0mm, clip, scale=0.55]{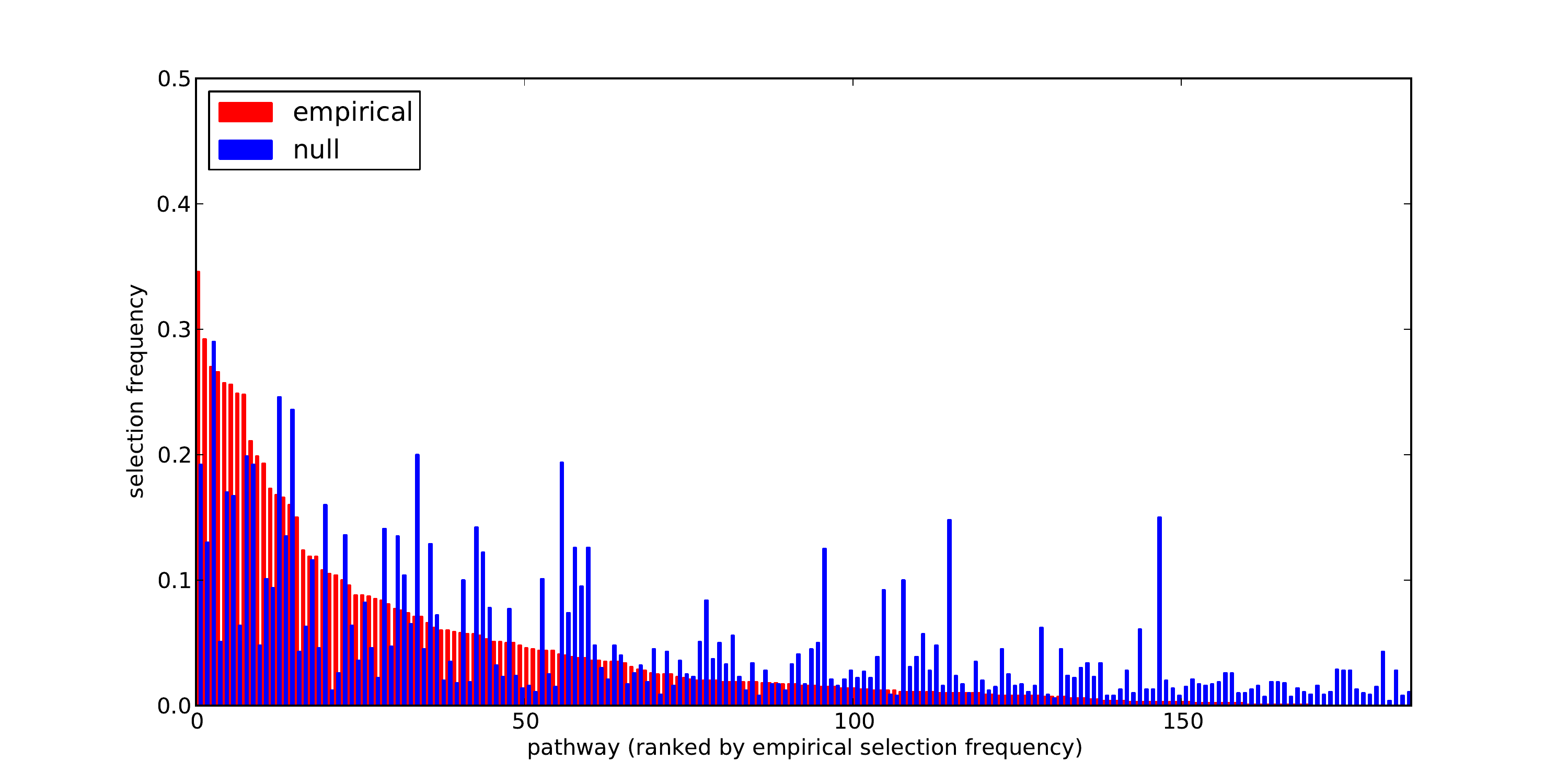}}
	\subfigure[]{\includegraphics[trim = 10mm 0mm 0mm 0mm, clip, scale=0.55]{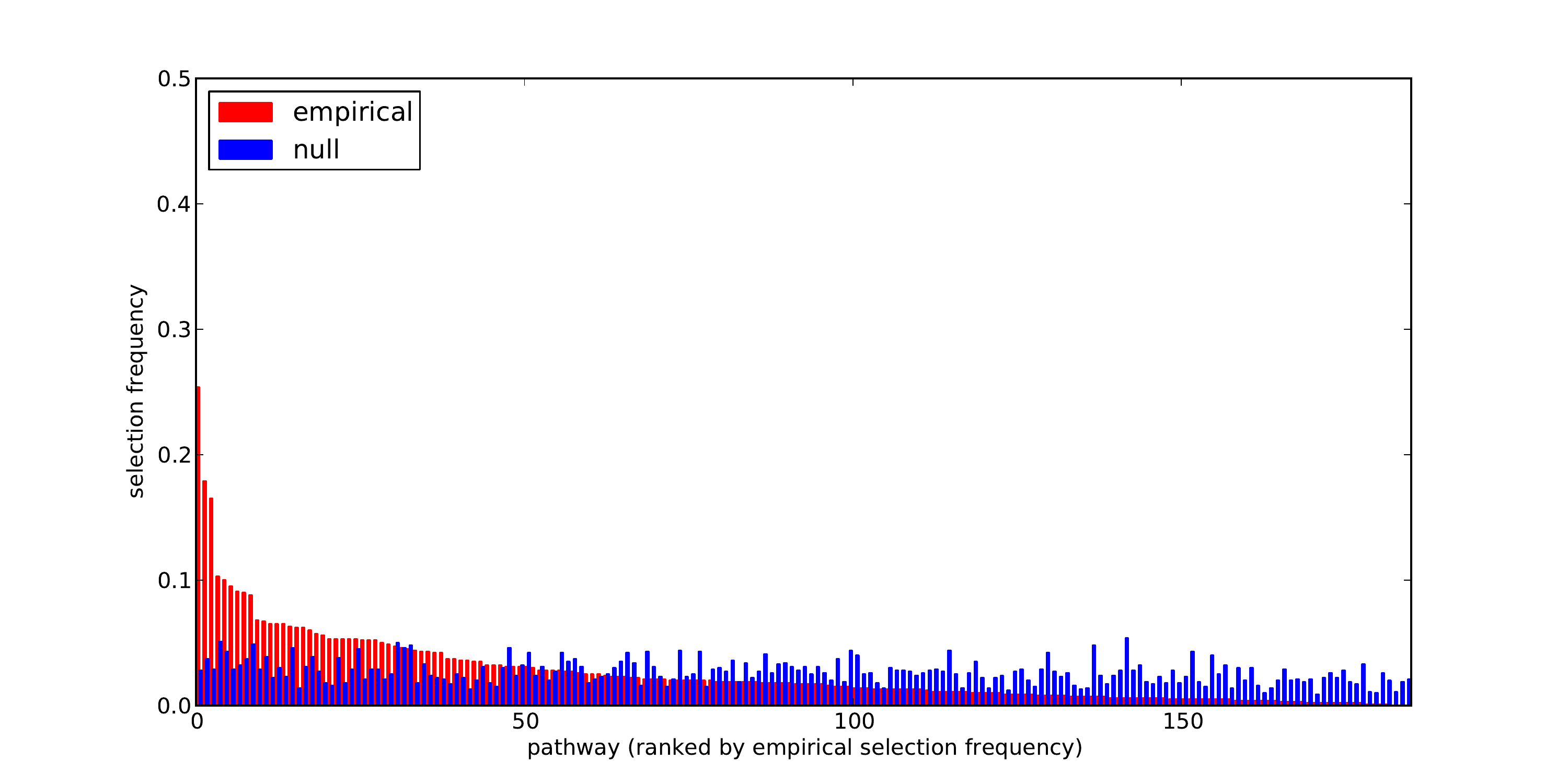}}
	\caption[Empirical and null pathway selection frequency distributions for all 185 KEGG pathways with the SP2 dataset]{Empirical and null pathway selection frequency distributions for all 185 KEGG pathways with the SP2 dataset. For each scenario, pathways are ranked along the $x$-axis in order of their empirical pathway selection frequency, $\pi^{path}_{l_1} >, \ldots, > \pi^{path}_{l_L}$.  (a) $\alpha = 0.85$.  (b) $\alpha = 0.95$. }
	\label{fig:SP2_pathways_distributions}
	\end{center}
\end{figure}

\begin{figure}
	\begin{center}
	\subfigure[]{\includegraphics[trim = 40mm 40mm 35mm 40mm, clip, scale=0.7]{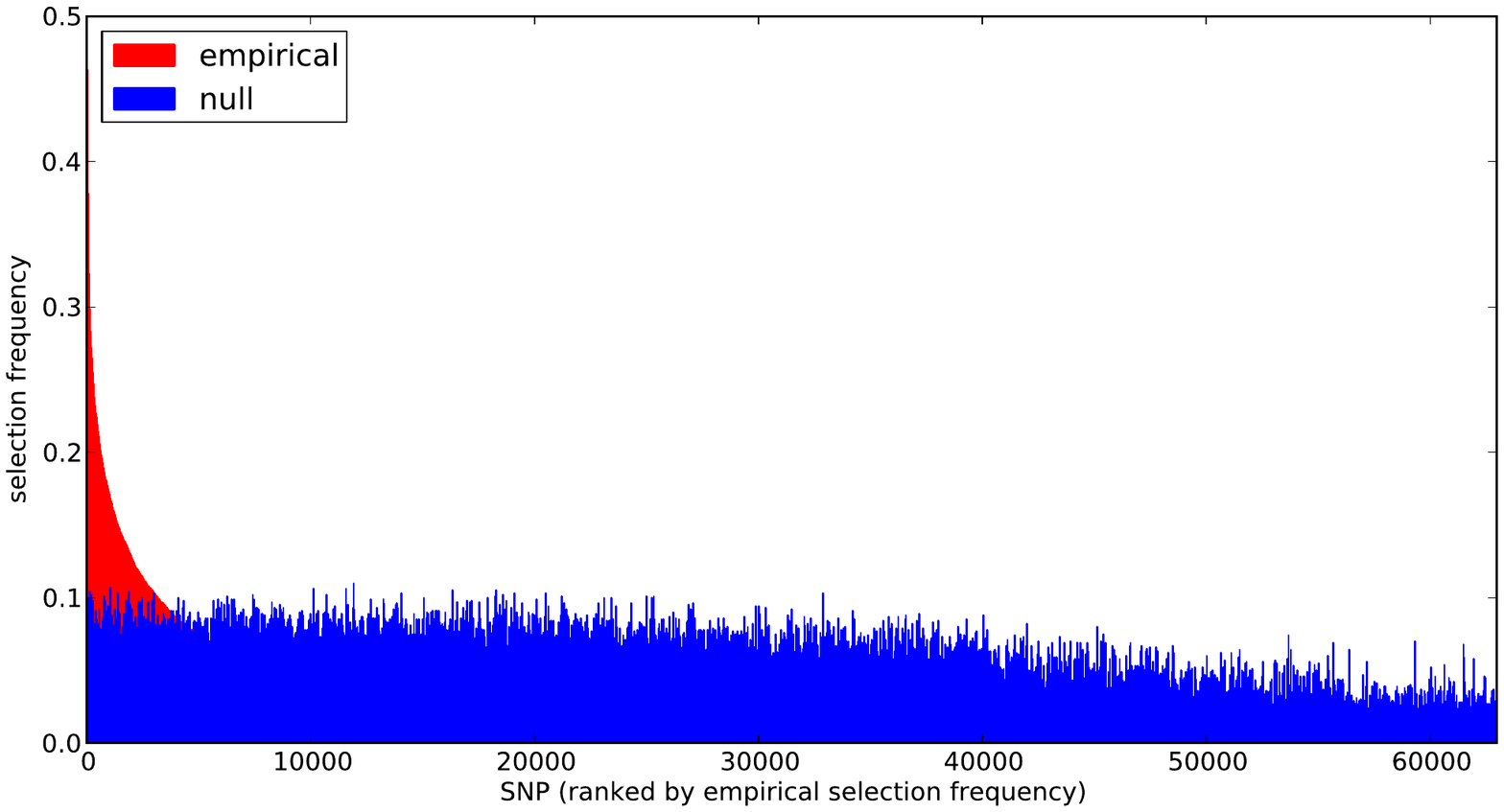}}
	\subfigure[]{\includegraphics[trim = 40mm 40mm 35mm 40mm, clip, scale=0.7]{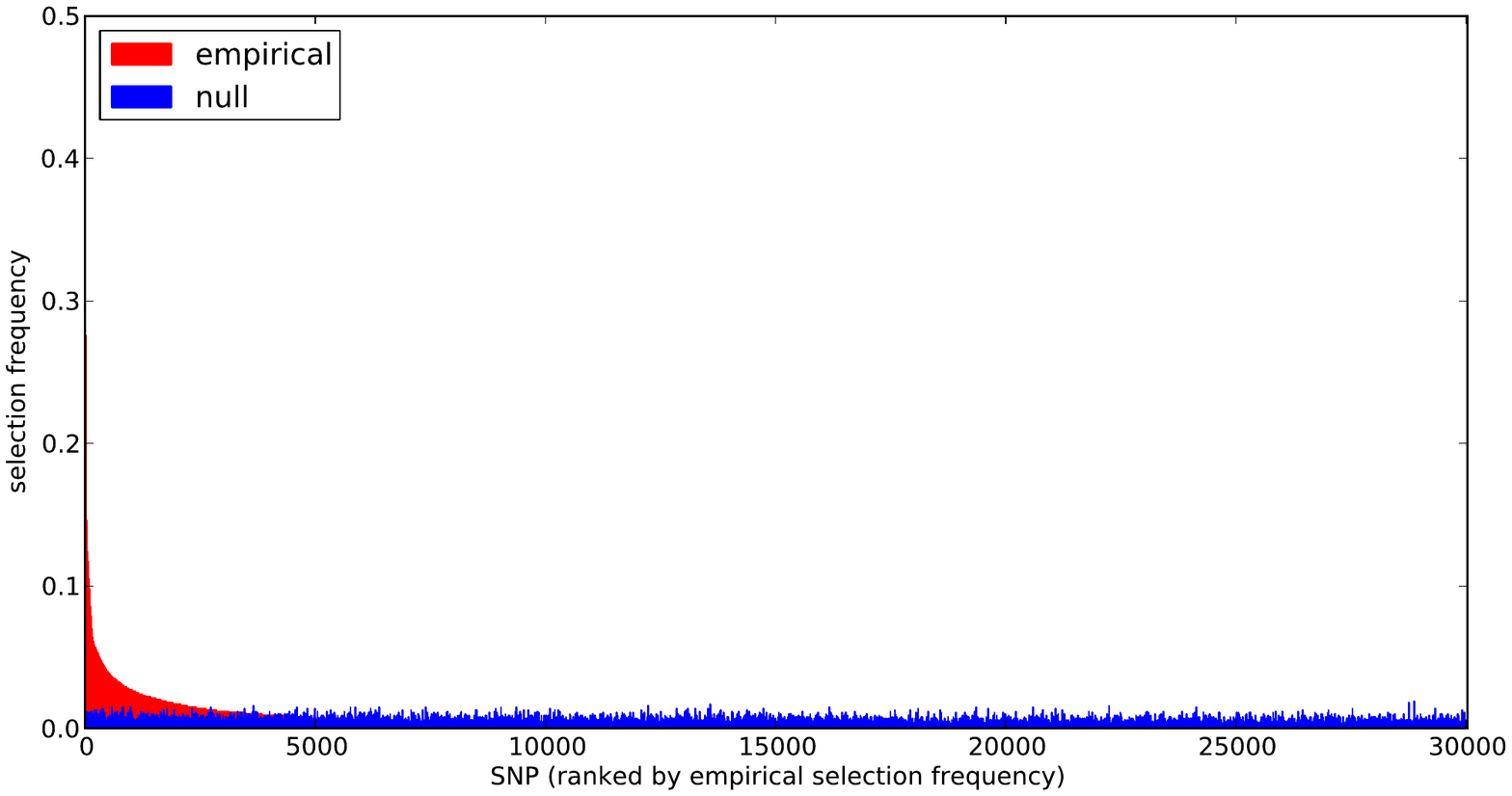}}
	\caption[Empirical and null SNP selection frequency distributions with the SP2 dataset]{Empirical and null SNP selection frequency distributions with the SP2 dataset. For each scenario, SNPs are ranked along the $x$-axis in order of their empirical pathway selection frequency, $\pi^{SNP}_{j_1} > \pi^{SNP}_{j_2} > \ldots $.  (a) $\alpha = 0.85$.  (b) $\alpha = 0.95$. Note fewer SNPs are selected with nonzero empirical selection frequency with $\alpha=0.95$, so that the $x$-axis range in (b) is reduced.}
	\label{fig:SP2_SNP_distributions}
	\end{center}
\end{figure}

Comparisons of empirical and null pathway selection frequency distributions for each scenario are presented in Figure \ref{fig:SP2_pathways_distributions}.  The same comparisons for SNP selection frequencies are presented in Figure \ref{fig:SP2_SNP_distributions}.  In these plots, null distributions (coloured blue) are ordered along the $x$-axis according to their corresponding ranked empirical selection frequencies (marked in red).  This is to help visualise any potential biases that may be influencing variable selection (see below).

To interpret these results, we begin by noting from Table \ref{tab:SP2_nsel_vars} that many more SNPs are selected with $\alpha=0.85$, resulting in higher SNP selection frequencies, compared to those obtained with $\alpha=0.95$ (see Figure \ref{fig:SP2_SNP_distributions}).  This is as expected, since a lower value for $\alpha$ implies a reduced $\ell_1$ penalty on the SNP coefficient vector, resulting in more SNPs being selected.  Perhaps surprisingly, given that the $\ell_2$ group penalty $(1-\alpha)\lambda$ is increased, the number of selected pathways is also greater.  This must reflect the reduced $\ell_1$ penalty, which allows a greater number of SNPs to contribute to a putative selected pathway's coefficient vector.  This in turn increases the number of pathways that pass the threshold for selection.

\begin{figure}
	\begin{center}
	\subfigure[]{\includegraphics[trim = 0mm 0mm 0mm 0mm, clip, scale=0.35]{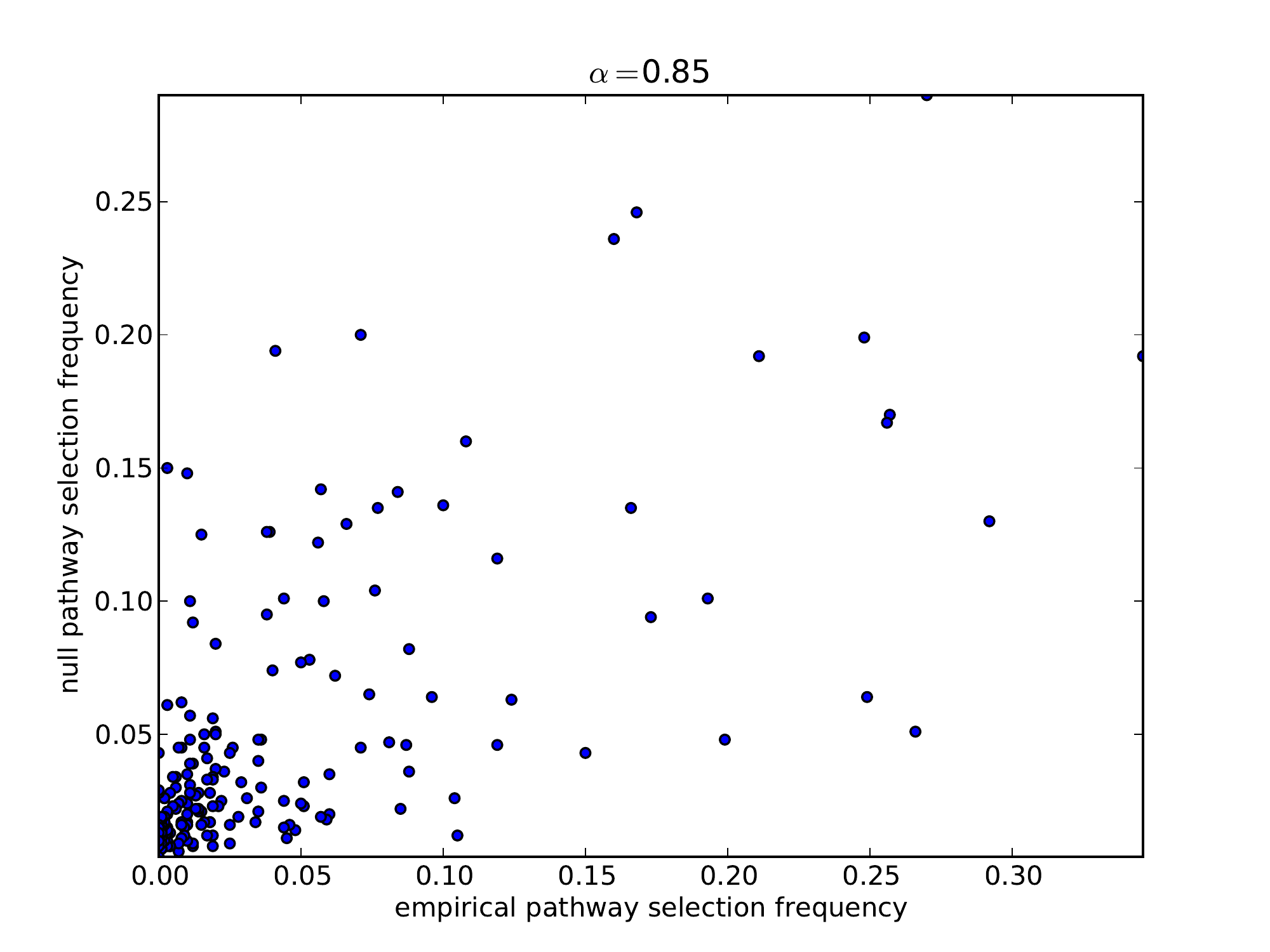}}
	\subfigure[]{\includegraphics[trim = 0mm 0mm 0mm 0mm, clip, scale=0.35]{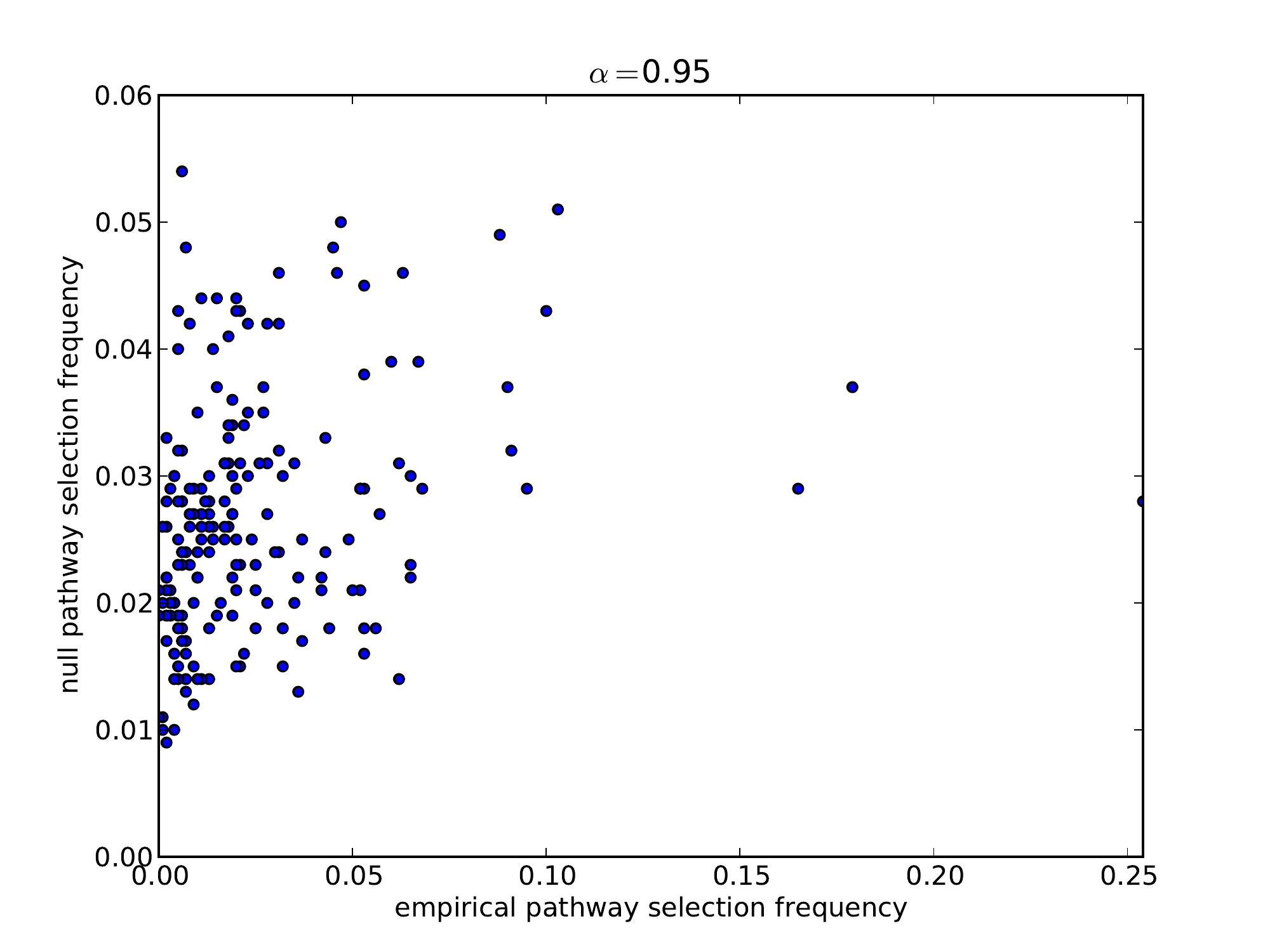}}
	\subfigure[]{\includegraphics[trim = 0mm 0mm 0mm 0mm, clip, scale=0.35]{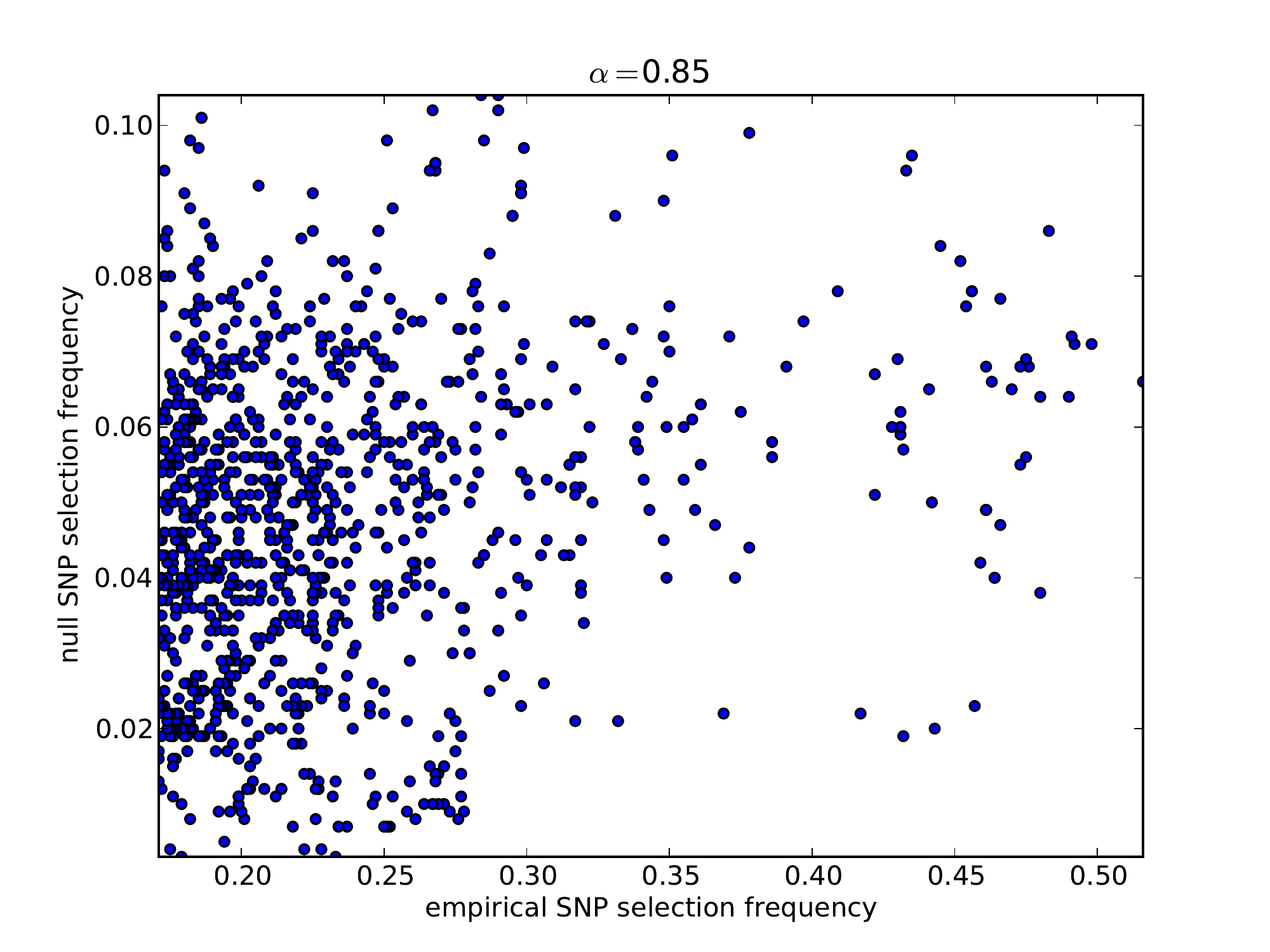}}
	\subfigure[]{\includegraphics[trim = 0mm 0mm 0mm 0mm, clip, scale=0.35]{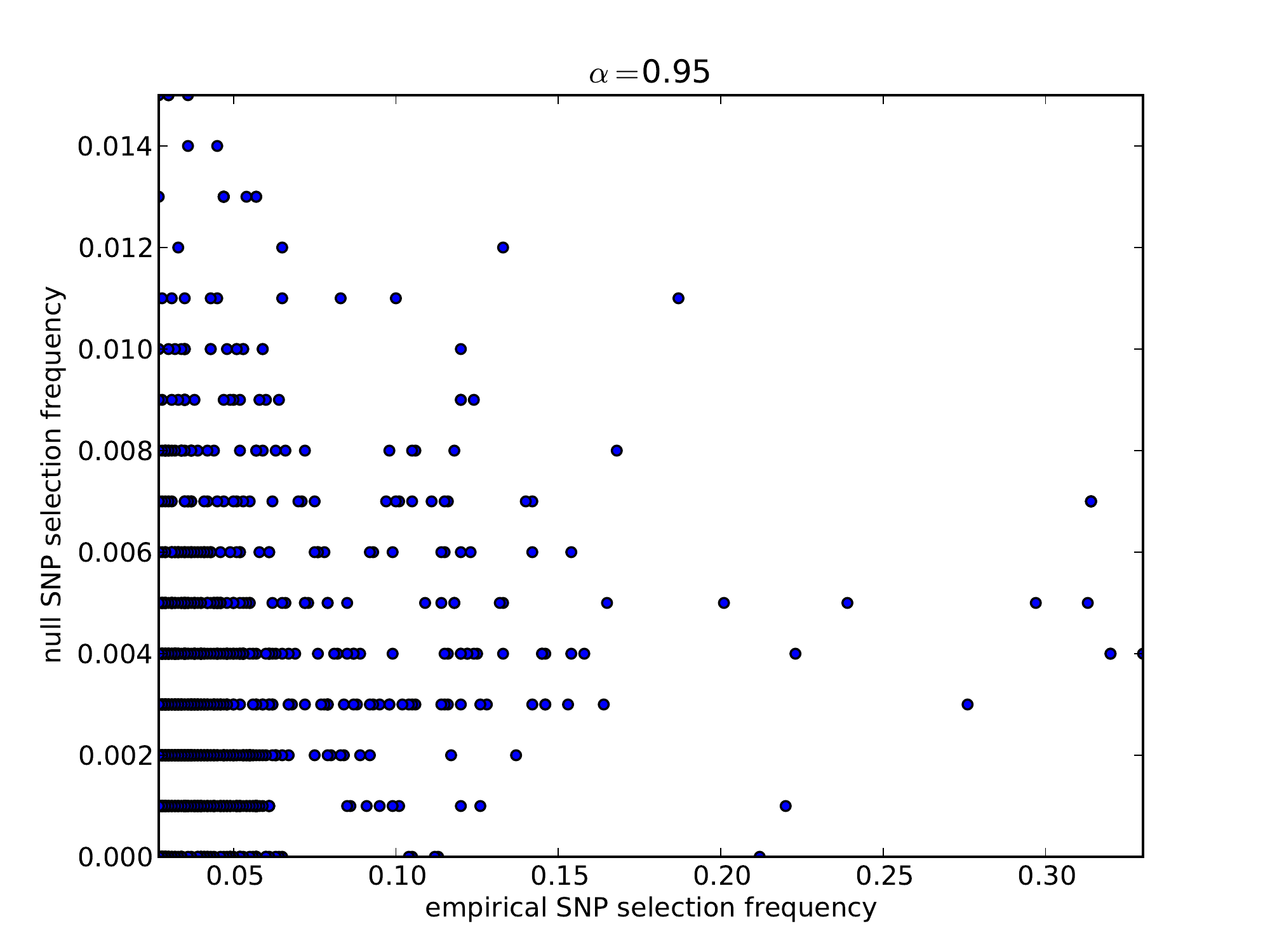}}
	\caption[SP2 dataset: Scatter plots comparing empirical and null selection frequencies]{SP2 dataset: Scatter plots comparing empirical and null selection frequencies presented in Figures \ref{fig:SP2_pathways_distributions} and \ref{fig:SP2_SNP_distributions}. \emph{(a)} and \emph{(b)}: Pathway selection frequencies with $\alpha=0.85, 0.95$ respectively.  \emph{(c)} and \emph{(d)}: SNP selection frequencies for the same $\alpha$ values.  For clarity, SNP selection frequencies are plotted for the top 1000 SNPs (by empirical selection frequency) only.  Corresponding correlation coefficients (for all ranked SNPs) are presented in Table \ref{tab:SP2_pearson_coeffs}.  Note that pathway and SNP selection frequencies are much higher at the lower $\alpha$ value (left hand plots), since many more variables are selected (see Table \ref{tab:SP2_nsel_vars}.)}
	\label{fig:SP2_scatter_plots}
	\end{center}
\end{figure}

This raises the question of what might be considered to be an optimal choice for the regularisation-distributional parameter $\alpha$, since different assumptions about the number of SNPs potentially influencing the phenotype may affect the resulting pathway and SNP rankings.  To answer this, we turn our attention to the pathway and SNP selection frequency distributions for each $\alpha$ value in Figures \ref{fig:SP2_pathways_distributions} and \ref{fig:SP2_SNP_distributions}.  At the lower value of $\alpha=0.85$ (top plots in Figures \ref{fig:SP2_pathways_distributions} and \ref{fig:SP2_SNP_distributions}), empirical pathway and SNP selection frequency distributions appear to be biased, in the sense that there is a suggestion that pathways and SNPs with the highest empirical selection frequencies also tend to be selected with a higher frequency under the null, where there is no association between genotype and phenotype.  This relationship appears to be diminished with $\alpha=0.95$, when fewer SNPs are selected by the model.  We investigate this further by plotting empirical vs.~null selection frequencies as a sequence of scatter plots in Figure \ref{fig:SP2_scatter_plots}, and we report Pearson correlation coefficients and p-values for these in Table \ref{tab:SP2_pearson_coeffs}.

\begin{table}[htdp]
\caption[SP2 dataset: Pearson correlation coefficients $(r)$and p-values ]{SP2 dataset: Pearson correlation coefficients $(r)$and p-values for the data plotted in Figure \ref{fig:SP2_scatter_plots}.  $n$ denotes the number of predictors considered.  For SNPs, coefficients describe correlations for all predictors selected with nonzero empirical selection frequencies only, since a large number of SNPs are not selected by the model at any subsample.}
\begin{center}
\begin{tabular}{cccc|ccc}
&\multicolumn{3}{c|}{$\alpha=0.85$}&\multicolumn{3}{c}{$\alpha=0.95$}\\
&$n$&$r$&p-value&$n$&$r$&p-value\\
\hline
pathways&185&$0.66$&$1.3\times10^{-24}$&$185$&$0.26$&$2.9\times10^{-4}$\\
SNPs&$62,965$&$0.37$&$0$&$30,027$&$0.11$&$1.2\times10^{-84}$\\
\end{tabular}
\end{center}
\label{tab:SP2_pearson_coeffs}
\end{table}%

These provide further evidence of increased correlation between empirical and null selection frequency distributions at the lower $\alpha$ value for both pathways and SNPs, again suggesting increased bias in the empirical results, in the sense that certain pathways and SNPs tend to be selected with a higher frequency, irrespective of whether or not a true signal may be present.  Further qualitative evidence of reduced bias with $\alpha=0.95$ is suggested by the clearer separation of empirical and null distributions at the higher $\alpha$ value in Figures \ref{fig:SP2_pathways_distributions} and \ref{fig:SP2_SNP_distributions}.  For example, the maximum empirical pathway selection frequency is reduced by a factor of 0.29 (0.35 to 0.25) as $\alpha$ is increased from 0.85 to 0.95, whereas the maximum pathway selection frequency under the null is reduced by a factor of 0.81 (0.29 to 0.054).  Similarly for SNPs, the maximum empirical SNP selection frequency is reduced by a factor of 0.37 (0.52 to 0.33), whereas the maximum SNP selection frequency under the null is reduced by a factor of 0.9 (0.11 to 0.011).

The increased bias with $\alpha=0.85$ is most likely due to the selection of too many SNPs, in the sense that many selected SNPs do not exhibit real phenotypic effects.  These extra SNPs effectively add noise to the model, in the form of multiple weak, spurious signals.  This in turn will add bias to the resulting selection frequency distributions, tending to favour, for example, SNPs that overlap multiple pathways, and the pathways that contain them.  As $\alpha$ is increased, we would expect this biasing effect to be reduced, until a point where too few SNPs are selected, when there is then a risk that some of the true signal may be lost.  

Note that the reduced but still significant correlations between empirical and null selection frequency distributions at $\alpha=0.95$ in Table \ref{tab:SP2_pearson_coeffs} are not unexpected.  These may reflect the complex overlap structure between pathways, meaning that pathways (and associated SNPs) with a relatively high degree of overlap with other pathways, due for example to the presence of so called `hub genes', are more likely to harbour true signals, as well as spurious ones \citep{Lehner2006,Carter2004,Jeong2001}.  

Taking all the above into consideration, we choose to report results with $\alpha=0.95$, where there is less evidence of bias due to the selection of too many SNPs.  The top 30 pathways, ranked by selection frequency are presented in Table \ref{tab:SP2_pathway_rankings}, and the top 30 ranked SNPs, together with corresponding genes to which they are mapped are presented in Table \ref{tab:SP2_snp_gene_ranking}.  Versions of these tables extending to lower ranks are provided as supplementary information.


\begin{sidewaystable}[p]
\caption[SP2 dataset: Top 30 pathways, ranked by pathway selection frequency]{SP2 dataset: Top 30 pathways, ranked by pathway selection frequency, $\pi^{path}$.  The final column lists genes in the pathway that are in the top 30 ranked genes selected in the study (see Table \ref{tab:SP2_snp_gene_ranking}).  Pathways falling in the consensus set, $\Psi_{25}^{path}$, obtained by comparing pathway ranking results from both SP2 and SiMES datasets (see Table \ref{tab:lipid_study_consensus_pathway_rankings}), are marked with a $^*$.}
\begin{center}
\scriptsize
\begin{tabular}{rlccl}
Rank&KEGG pathway name&$\pi^{path}$&Size&top 30 ranked genes in pathway\\
&&&(\# SNPs)\\
\hline
1 & Toll Like Receptor Signaling Pathway & 0.254 & 766 & \emph{TIRAP}   \emph{RAC1}   \emph{IFNAR1}   \emph{CD80}   \emph{IL12B}   \emph{PIK3R1}   \\
2 & Jak Stat Signaling Pathway & 0.179 & 1447 & \emph{PIAS2}   \emph{IL5RA}   \emph{TPO}   \emph{IFNAR1}   \emph{IL12B}   \emph{PIK3R1}   \emph{IL2RA}   \\
3 & Ubiquitin Mediated Proteolysis & 0.165 & 1603 & \emph{PIAS2}   \emph{RFWD2}   \emph{PARK2}   \\
4 & $^*$Dilated Cardiomyopathy & 0.103 & 3054 & \emph{ADCY2}   \emph{TGFB3}   \emph{PRKACB}   \emph{RYR2}   \emph{ITGB8}   \emph{ITGA1}   \emph{CACNA2D3}   \emph{LAMA2}   \emph{CACNA1C}   \\
5 & Cytokine Cytokine Receptor Interaction & 0.100 & 2553 & \emph{IL5RA}   \emph{IL12B}   \emph{TGFB3}   \emph{EGFR}   \emph{TPO}   \emph{IFNAR1}   \emph{IL2RA}   \\
6 & Ecm Receptor Interaction & 0.095 & 2271 & \emph{ITGB8}   \emph{ITGA1}   \emph{LAMA2}   \\
7 & Arginine And Proline Metabolism & 0.091 & 432 & \emph{NOS1}   \\
8 & Parkinson's Disease & 0.090 & 1320 & \emph{PARK2}   \\
9 & $^*$ Hypertrophic Cardiomyopathy  & 0.088 & 2819 & \emph{TGFB3}   \emph{RYR2}   \emph{ITGB8}   \emph{ITGA1}   \emph{CACNA2D3}   \emph{LAMA2}   \emph{CACNA1C}   \\
10 & Small Cell Lung Cancer & 0.068 & 1808 & \emph{PIAS2}   \emph{PIK3R1}   \emph{LAMA2}   \\
11 & Natural Killer Cell Mediated Cytotoxicity & 0.067 & 1781 & \emph{KRAS}   \emph{RAC1}   \emph{VAV3}   \emph{VAV2}   \emph{PRKCA}   \emph{IFNAR1}   \emph{PRKCB}   \emph{PIK3R1}   \\
12 & $^*$ T Cell Receptor Signaling Pathway & 0.065 & 1541 & \emph{KRAS}   \emph{VAV3}   \emph{VAV2}   \emph{PIK3R1}   \\
13 & Tgf Beta Signaling Pathway & 0.065 & 947 & \emph{TGFB3}   \\
14 & Olfactory Transduction & 0.065 & 2497 & \emph{PRKACB}   \\
15 & $^*$ Arrhythmogenic Right Ventricular Cardiomyopathy  & 0.063 & 3726 & \emph{RYR2}   \emph{TCF7L1}   \emph{ITGB8}   \emph{ITGA1}   \emph{CACNA2D3}   \emph{LAMA2}   \emph{CACNA1C}   \\
16 & $^*$ Ppar Signaling Pathway & 0.062 & 758 & \\
17 & Taste Transduction & 0.062 & 941 & \emph{PRKACB}   \\
18 & Type I Diabetes Mellitus & 0.060 & 776 & \emph{CD80}   \emph{IL12B}   \\
19 & $^*$ Ribosome & 0.057 & 261 & \\
20 & $^*$ Terpenoid Backbone Biosynthesis & 0.056 & 147 & \\
21 & Neuroactive Ligand Receptor Interaction & 0.053 & 5745 & \emph{GRIN3A}   \\
22 & Regulation Of Actin Cytoskeleton & 0.053 & 3803 & \emph{KRAS}   \emph{RAC1}   \emph{EGFR}   \emph{ITGB8}   \emph{VAV3}   \emph{ITGA1}   \emph{VAV2}   \emph{PIK3R1}   \\
23 & Mismatch Repair & 0.053 & 222 & \\
24 & Cell Adhesion Molecules Cams & 0.053 & 3977 & \emph{ITGB8}   \emph{CD80}   \\
25 & Maturity Onset Diabetes Of The Young & 0.053 & 239 & \\
26 & Butanoate Metabolism & 0.052 & 383 & \\
27 & Purine Metabolism & 0.052 & 3224 & \emph{ADCY2}   \\
28 & P53 Signaling Pathway & 0.052 & 598 & \emph{RFWD2}   \\
29 & Dorso Ventral Axis Formation & 0.050 & 581 & \emph{KRAS}   \emph{EGFR}   \\
30 & Basal Cell Carcinoma & 0.049 & 589 & \emph{TCF7L1}   \\
\end{tabular}
\end{center}
\label{tab:SP2_pathway_rankings}
\end{sidewaystable}

\begin{table}[htdp]
\caption{SP2 dataset: Top 30 SNPs and genes, respectively ranked by SNP and gene selection frequency.  Genes falling in the top 30 ranks of the consensus gene set, $\Psi_{244}^{gene}$, obtained by comparing gene ranking results from both SP2 and SiMES datasets (see Table \ref{tab:lipid_study_consensus_gene_rankings}), are marked with a $^*$.}
\begin{center}
\footnotesize
\begin{tabular}{r l c c | l  c c c c} \\
&  \multicolumn{3}{c|}{SNP RANKING} &  \multicolumn{3}{c}{GENE RANKING}\\
Rank & SNP & $\pi^{SNP}$ & Mapped gene(s) & Gene & $\pi^{gene}$ & \# mapped SNPs\\[3pt]
\hline\noalign{\smallskip}
1 & rs2257167 & 0.33 & \emph{IFNAR1} & \emph{IFNAR1} & 0.33 & 11 \\
2 & rs2254315 & 0.32 & \emph{IFNAR1} & \emph{IL12B} & 0.30 & 9 \\
3 & rs1041868 & 0.32 & \emph{IFNAR1} & \emph{PIAS2} & 0.30 & 7 \\
4 & rs7364085 & 0.31 & \emph{IFNAR1} & \emph{TIRAP} & 0.22 & 5 \\
5 & rs2850021 & 0.31 & \emph{IFNAR1} & \emph{RAC1} & 0.21 & 10 \\
6 & rs2253413 & 0.31 & \emph{IFNAR1} & \emph{LAMA2}$^*$ & 0.19 & 111 \\
7 & rs2243590 & 0.31 & \emph{IFNAR1} & \emph{ADCY2}$^*$ & 0.19 & 94 \\
8 & rs2834204 & 0.31 & \emph{IFNAR1} & \emph{PIK3R1} & 0.19 & 28 \\
9 & rs3181224 & 0.30 & \emph{IL12B} & \emph{PARK2} & 0.19 & 460 \\
10 & rs512868 & 0.28 & \emph{PIAS2} & \emph{IL2RA} & 0.19 & 55 \\
11 & rs641366 & 0.24 & \emph{PIAS2} & \emph{PRKCA}$^*$ & 0.19 & 123 \\
12 & rs2032215 & 0.22 & \emph{PIAS2} & \emph{ITGB8} & 0.18 & 27 \\
13 & rs8177375 & 0.22 & \emph{TIRAP} & \emph{TCF7L1} & 0.18 & 55 \\
14 & rs10893493 & 0.21 & \emph{TIRAP} & \emph{CD80}$^*$ & 0.18 & 21 \\
15 & rs4890341 & 0.20 & \emph{PIAS2} & \emph{GRIN3A} & 0.18 & 60 \\
16 & rs2303361 & 0.19 & \emph{RAC1} & \emph{PRKCB}$^*$ & 0.18 & 83 \\
17 & rs7873495 & 0.17 & \emph{GRIN3A}   \emph{PPP3R2}   & \emph{CACNA1C}$^*$ & 0.17 & 180 \\
18 & rs1323653 & 0.17 & \emph{IL2RA} & \emph{TGFB3} & 0.16 & 7 \\
19 & rs11762117 & 0.16 & \emph{ITGB8} & \emph{PRKACB} & 0.16 & 16 \\
20 & rs3807955 & 0.16 & \emph{ITGB8} & \emph{KRAS}$^*$ & 0.16 & 21 \\
21 & rs10462842 & 0.15 & \emph{ADCY2} & \emph{VAV3} & 0.16 & 97 \\
22 & rs10215885 & 0.15 & \emph{ITGB8} & \emph{IL5RA} & 0.15 & 38 \\
23 & rs3807936 & 0.15 & \emph{ITGB8} & \emph{ITGA1}$^*$ & 0.15 & 77 \\
24 & rs530205 & 0.15 & \emph{PIAS2} & \emph{VAV2}$^*$ & 0.15 & 85 \\
25 & rs3823974 & 0.15 & \emph{ITGB8} & \emph{EGFR}$^*$ & 0.14 & 61 \\
26 & rs2074425 & 0.15 & \emph{ITGB8} & \emph{TPO} & 0.14 & 50 \\
27 & rs6725799 & 0.15 & \emph{TCF7L1} & \emph{CACNA2D3}$^*$ & 0.14 & 283 \\
28 & rs2301727 & 0.15 & \emph{ITGB8} & \emph{RYR2}$^*$ & 0.14 & 214 \\
29 & rs10486391 & 0.14 & \emph{ITGB8} & \emph{NOS1} & 0.14 & 49 \\
30 & rs3779505 & 0.14 & \emph{ITGB8} & \emph{RFWD2} & 0.13 & 31 \\
\end{tabular}
\end{center}
\label{tab:SP2_snp_gene_ranking}
\end{table}

\newpage
\subsection*{SiMES Analysis}

\begin{figure}[h]
	\begin{center}
	\subfigure{\includegraphics[trim = 5mm 6mm 5mm 10mm, clip, scale=0.35]{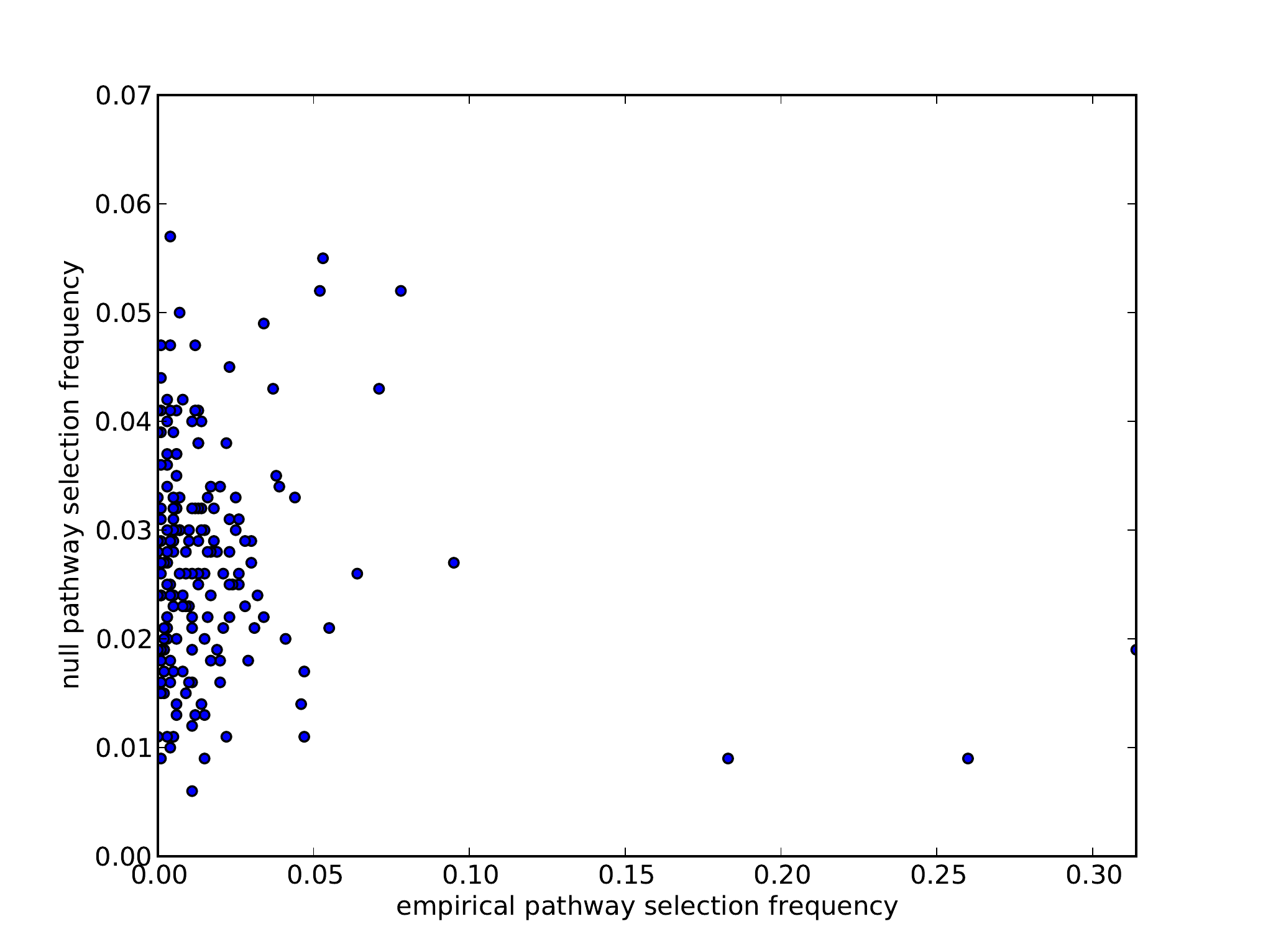}}
	\subfigure{\includegraphics[trim = 5mm 6mm 5mm 10mm, clip, scale=0.35]{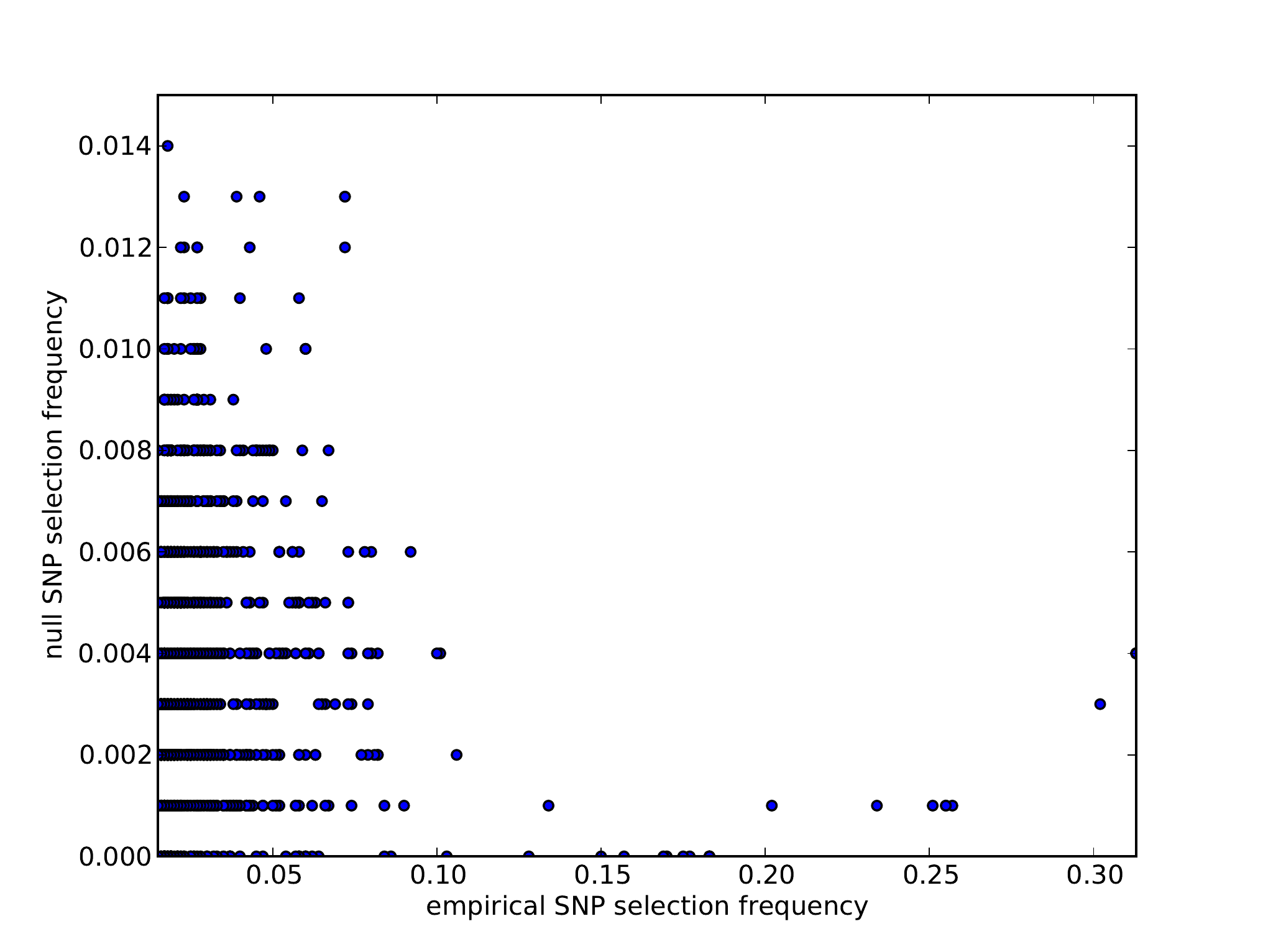}}
	\caption[SiMES dataset: Scatter plots comparing empirical and null selection frequencies ]{SiMES dataset: Scatter plots comparing empirical and null pathway (\emph{left}) and SNP (\emph{right}) selection frequencies presented in Figure \ref{fig:SiMES_selFreq_distributions}.  For clarity, SNP selection frequencies are plotted for the top 1000 SNPs (by empirical selection frequency) only.}
	\label{fig:SiMES_scatter_plots}
	\end{center}
\end{figure}

\begin{figure}[p]
	\begin{center}
	\subfigure{\includegraphics[trim = 15mm 0mm 0mm 10mm, clip, scale=0.555]{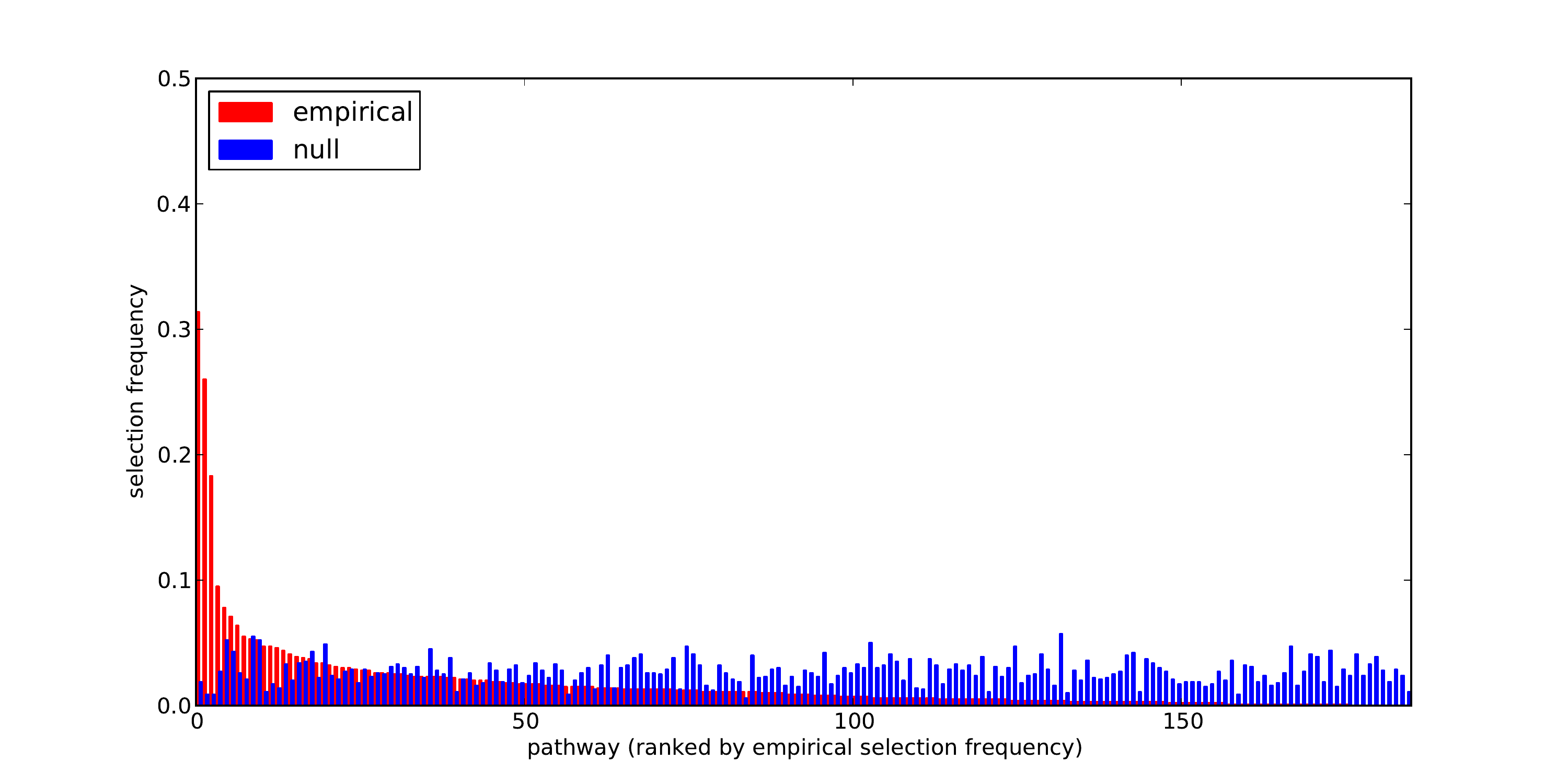}}
	\subfigure{\includegraphics[trim = 30mm 40mm 30mm 40mm, clip, scale=0.63]{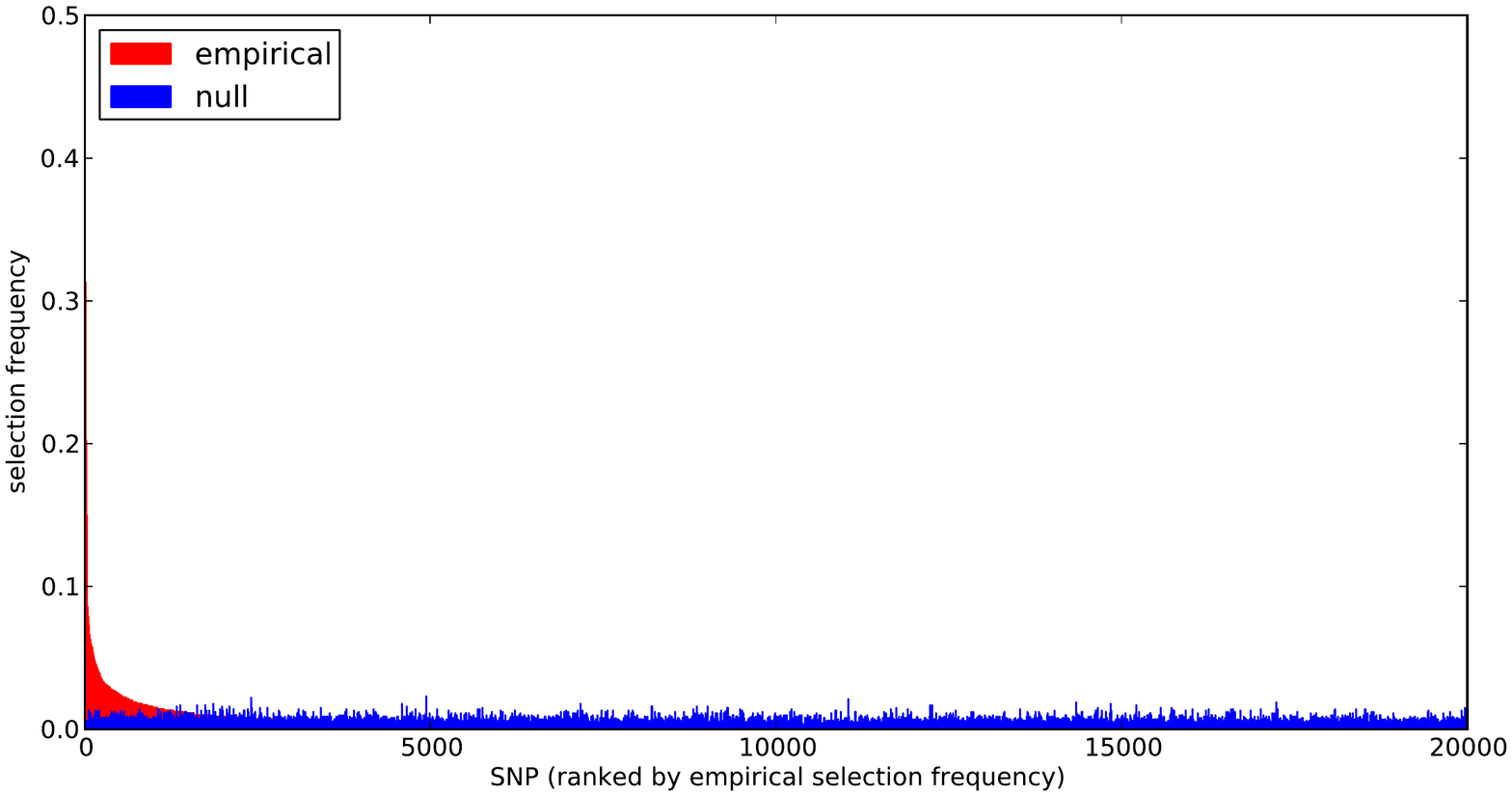}}
	\caption[Selection frequency distributions for the SiMES dataset]{Empirical and null pathway (\emph{top}) and SNP (\emph{bottom}) selection frequency distributions for the SiMES dataset.  $\alpha=0.95$. For both empirical (red) and null (blue) distributions, variables (pathways and SNPs) are ranked along the $x$-axis in order of their empirical selection frequencies.}
	\label{fig:SiMES_selFreq_distributions}
	\end{center}
\end{figure}

For the replication SiMES dataset, we repeat the above analysis design, but consider only the `low bias' scenario where $\lambda=0.95 \lambda_{max}$ and $\alpha=0.95$.  Once again we test each scenario over 1000 $N/2$ subsamples, and compare the resulting pathway and SNP selection frequency distributions with null distributions generated over 1000 $N/2$ subsamples with phenotype labels permuted.  Pathway and SNP selection frequency distributions are presented in Figure \ref{fig:SiMES_selFreq_distributions}.  An investigation of pathway and SNP selection bias is presented in the form of scatter plots illustrating potential correlation between empirical and null selection frequencies in Figure \ref{fig:SiMES_scatter_plots}, with corresponding Pearson correlation coefficients and p-values presented in Table \ref{tab:SiMES_pearson_coeffs}.  The top 30 ranked pathways, and SNPs and genes are presented in Tables \ref{tab:SiMES_pathway_rankings} and \ref{tab:SiMES_snp_gene_ranking} respectively.

\begin{table}[h]
\caption[SiMES dataset: Pearson correlation coefficients $(r)$and p-values ]{SiMES dataset: Pearson correlation coefficients $(r)$and p-values for the data plotted in Figure \ref{fig:SiMES_scatter_plots}.  $n$ denotes the number of predictors considered.    For SNPs, coefficients describe correlations for all predictors selected with nonzero empirical selection frequencies only, since a large number of SNPs are not selected by the model at any subsample.}
\begin{center}
\begin{tabular}{cccc}
&$n$&$r$&p-value\\
\hline
pathways&185&$-0.094$&$0.20$\\
SNPs&$20,006$&$0.058$&$2.63\times10^{-16}$\\
\end{tabular}
\end{center}
\label{tab:SiMES_pearson_coeffs}
\end{table}%

\begin{sidewaystable}[htdp]
\caption[SiMES dataset: Top 30 pathways, ranked by pathway selection frequency]{SiMES dataset: Top 30 pathways, ranked by pathway selection frequency, $\pi^{path}$.  The final column lists genes in the pathway that are in the top 30 ranked genes selected in the study (i.e.~genes in the top 30 gene ranking in Table \ref{tab:SiMES_snp_gene_ranking}).  Pathways falling in the consensus set, $\Psi_{25}^{path}$, obtained by comparing pathway ranking results from both SP2 and SiMES datasets (see Table \ref{tab:lipid_study_consensus_pathway_rankings}), are marked with a $^*$.}
\begin{center}
\scriptsize
\begin{tabular}{rlccl}
Rank&KEGG pathway name&$\pi^{path}$&Size&top 30 ranked genes in pathway\\
&&&(\# SNPs)\\
\hline
1 & Oxidative Phosphorylation & 0.314 & 871 & \emph{PPA2}   \emph{NDUFA4}   \emph{SDHB}   \emph{SDHC}   \emph{ATP6V0A4}   \\
2 & $^*$ Terpenoid Backbone Biosynthesis & 0.260 & 158 & \emph{PDSS2}   \\
3 & Regulation Of Autophagy & 0.183 & 215 & \emph{GABARAPL1}   \\
4 & Glycerolipid Metabolism & 0.095 & 1074 & \emph{ALDH7A1}   \emph{DGKB}   \emph{DGKH}   \emph{ALDH2}   \emph{LIPC}   \\
5 & $^*$Dilated Cardiomyopathy & 0.078 & 3177 & \emph{ADCY2}   \emph{RYR2}   \emph{ITGA11}   \emph{ITGB1}   \emph{SLC8A1}   \emph{ITGA1}   \emph{CACNA2D3}   \emph{LAMA2}   \emph{CACNA1C}   \emph{CACNA1D}   \\
6 & $^*$ Hypertrophic Cardiomyopathy  & 0.071 & 2932 & \emph{PRKAG2}   \emph{RYR2}   \emph{ITGA11}   \emph{ITGB1}   \emph{SLC8A1}   \emph{ITGA1}   \emph{CACNA2D3}   \emph{LAMA2}   \emph{CACNA1C}   \emph{CACNA1D}   \\
7 & $^*$ Ribosome & 0.064 & 270 & \\
8 & Glutathione Metabolism & 0.055 & 389 & \emph{MGST3}   \\
9 & $^*$ Arrhythmogenic Right Ventricular Cardiomyopathy  & 0.053 & 3899 & \emph{RYR2}   \emph{ITGA11}   \emph{ITGB1}   \emph{SLC8A1}   \emph{ITGA1}   \emph{CACNA2D3}   \emph{LAMA2}   \emph{CACNA1C}   \emph{CACNA1D}   \\
10 & $^*$ T Cell Receptor Signaling Pathway & 0.052 & 1624 & \emph{PAK7}   \emph{FYN}   \\
11 & Cardiac Muscle Contraction & 0.047 & 1952 & \emph{RYR2}   \emph{SLC8A1}   \emph{CACNA2D3}   \emph{CACNA1C}   \emph{CACNA1D}   \\
12 & Biosynthesis Of Unsaturated Fatty Acids & 0.047 & 282 & \\
13 & Lysosome & 0.046 & 1322 & \emph{ATP6V0A4}   \\
14 & Apoptosis & 0.044 & 954 & \emph{BCL2}   \\
15 & Pathogenic Escherichia Coli Infection & 0.041 & 538 & \emph{ITGB1}   \emph{FYN}   \\
16 & Metabolism Of Xenobiotics By Cytochrome P450 & 0.039 & 880 & \emph{MGST3}   \\
17 & Drug Metabolism Cytochrome P450 & 0.038 & 910 & \emph{MGST3}   \\
18 & Autoimmune Thyroid Disease & 0.037 & 686 & \\
19 & Focal Adhesion & 0.034 & 4787 & \emph{ITGA11}   \emph{LAMA2}   \emph{BCL2}   \emph{FYN}   \emph{EGFR}   \emph{ITGB1}   \emph{ITGA1}   \emph{PAK7}   \emph{PRKCB}   \emph{IGF1R}   \\
20 & Leishmania Infection & 0.034 & 718 & \emph{PRKCB}   \emph{ITGB1}   \\
21 & $^*$ Ppar Signaling Pathway & 0.032 & 800 & \\
22 & Rna Polymerase & 0.031 & 193 & \\
23 & Lysine Degradation & 0.030 & 423 & \emph{ALDH7A1}   \emph{ALDH2}   \\
24 & Endocytosis & 0.030 & 3436 & \emph{EGFR}   \emph{IGF1R}   \\
25 & Glycosaminoglycan Biosynthesis Chondroitin Sulfate & 0.029 & 727 & \\
26 & Melanoma & 0.028 & 1189 & \emph{EGFR}   \emph{IGF1R}   \\
27 & Nucleotide Excision Repair & 0.028 & 330 & \\
28 & Prostate Cancer & 0.026 & 1419 & \emph{EGFR}   \emph{IGF1R}   \emph{BCL2}   \\
29 & Renal Cell Carcinoma & 0.026 & 1004 & \emph{PAK7}   \\
30 & Glycine Serine And Threonine Metabolism & 0.026 & 268 & \\
\end{tabular}
\end{center}
\label{tab:SiMES_pathway_rankings}
\end{sidewaystable}

\begin{table}[htdp]
\caption{SiMES dataset: Top 30 SNPs and genes, respectively ranked by SNP and gene selection frequency.  Genes falling in the top 30 ranks of the consensus gene set, $\Psi_{244}^{gene}$, obtained by comparing gene ranking results from both SP2 and SiMES datasets (see Table \ref{tab:lipid_study_consensus_gene_rankings}), are marked with a $^*$.}
\begin{center}
\footnotesize
\begin{tabular}{r l c c | l  c c c c} \\
&  \multicolumn{3}{c|}{SNP RANKING} &  \multicolumn{3}{c}{GENE RANKING}\\
Rank & SNP & $\pi^{SNP}$ & Mapped gene(s) & Gene & $\pi^{gene}$ & \# mapped SNPs\\[3pt]
\hline\noalign{\smallskip}
1 & rs2636698 & 0.31 & \emph{PPA2} & \emph{PPA2} & 0.31 & 16 \\
2 & rs2726503 & 0.31 & \emph{PPA2} & \emph{PDSS2} & 0.26 & 59 \\
3 & rs2713829 & 0.31 & \emph{PPA2} & \emph{GABARAPL1} & 0.18 & 11 \\
4 & rs2636726 & 0.31 & \emph{PPA2} & \emph{ATP6V0A4} & 0.15 & 35 \\
5 & rs2713834 & 0.31 & \emph{PPA2} & \emph{ITGB1} & 0.13 & 14 \\
6 & rs2726471 & 0.31 & \emph{PPA2} & \emph{CACNA1C}$^*$ & 0.11 & 186 \\
7 & rs2636713 & 0.31 & \emph{PPA2} & \emph{PRKCB}$^*$ & 0.11 & 84 \\
8 & rs2726516 & 0.31 & \emph{PPA2} & \emph{FYN} & 0.11 & 46 \\
9 & rs2636739 & 0.31 & \emph{PPA2} & \emph{BCL2}$^*$ & 0.10 & 61 \\
10 & rs2713861 & 0.31 & \emph{PPA2} & \emph{PAK7}$^*$ & 0.10 & 127 \\
11 & rs2636751 & 0.31 & \emph{PPA2} & \emph{DGKB} & 0.10 & 233 \\
12 & rs2298733 & 0.30 & \emph{PPA2} & \emph{LAMA2}$^*$ & 0.10 & 118 \\
13 & rs6568474 & 0.26 & \emph{PDSS2} & \emph{NDUFA4} & 0.10 & 7 \\
14 & rs9386622 & 0.26 & \emph{PDSS2} & \emph{DGKH} & 0.10 & 70 \\
15 & rs6924886 & 0.26 & \emph{PDSS2} & \emph{ADCY2}$^*$ & 0.09 & 104 \\
16 & rs759440 & 0.25 & \emph{PDSS2} & \emph{LIPC} & 0.09 & 69 \\
17 & rs11759792 & 0.23 & \emph{PDSS2} & \emph{SLC8A1}$^*$ & 0.09 & 240 \\
18 & rs10457161 & 0.20 & \emph{PDSS2} & \emph{EGFR}$^*$ & 0.09 & 74 \\
19 & rs12821011 & 0.18 & \emph{GABARAPL1} & \emph{PRKAG2} & 0.09 & 118 \\
20 & rs11053685 & 0.18 & \emph{GABARAPL1} & \emph{CACNA1D} & 0.09 & 83 \\
21 & rs4764324 & 0.18 & \emph{GABARAPL1} & \emph{ITGA11}$^*$ & 0.09 & 63 \\
22 & rs4764327 & 0.18 & \emph{GABARAPL1} & \emph{IGF1R}$^*$ & 0.09 & 100 \\
23 & rs9373924 & 0.18 & \emph{PDSS2} & \emph{SDHC} & 0.09 & 9 \\
24 & rs10845074 & 0.18 & \emph{GABARAPL1} & \emph{CACNA2D3}$^*$ & 0.08 & 294 \\
25 & rs9320215 & 0.17 & \emph{PDSS2} & \emph{RYR2}$^*$ & 0.08 & 221 \\
26 & rs10845073 & 0.17 & \emph{GABARAPL1} & \emph{ITGA1}$^*$ & 0.08 & 77 \\
27 & rs4946826 & 0.16 & \emph{PDSS2} & \emph{ALDH7A1} & 0.08 & 23 \\
28 & rs6938393 & 0.15 & \emph{PDSS2} & \emph{MGST3}$^*$ & 0.08 & 40 \\
29 & rs13202332 & 0.13 & \emph{PDSS2} & \emph{ALDH2} & 0.08 & 12 \\
30 & rs9480754 & 0.13 & \emph{PDSS2} & \emph{SDHB} & 0.08 & 13 \\
\end{tabular}
\end{center}
\label{tab:SiMES_snp_gene_ranking}
\end{table}

\newpage
\subsection*{Comparison of ranked pathway lists}\label{sec:PSGLAW_comparison_of_ranked_pathway_lists}

We now consider the problem of comparing rankings (pathways, genes and SNPs) obtained for each dataset.  To do this we require some measure of distance between each pair of lists.  Ideally this measure should place more emphasis on differences between highly-ranked variables, since we expect the association signal, and hence agreement between the ranked lists, to be strongest there.  By the same reasoning, we expect there to be little or no agreement between variables at lower rankings, where selection frequencies are low.  Indeed a consideration of empirical and null selection frequency distributions (Figures \ref{fig:SP2_pathways_distributions} and \ref{fig:SiMES_selFreq_distributions}, top) suggests that only the very top ranked variables are likely to reflect any true signal, so that we would additionally like our distance metric to be able to accommodate consideration of the top-$k$ variables only, with $k < p$, where $p$ is the total number of variables ranked in either dataset.  One complication with top-\emph{k} lists is that they are \emph{partial}, in the sense that unlike complete ($k=p$) lists, a variable may occur in one list, but not the other.

In order to consider this problem, we introduce the following notation.  We denote the complete set of ranked predictors by $\mathcal{L}=\{1, \ldots, p\}$, and begin by assuming that all variables are ranked in both datasets.  We denote the rank of each variable in list 1 by $\tau(i), i = 1, \ldots, p$, so that $\tau(5) = 1$ if variable 5 is ranked first and so on.  The corresponding ranks for list 2 are denoted by $\sigma(i), i = 1, \ldots, p$.  A suitable metric describing the distance between two top-\emph{k} rankings is the \emph{Canberra distance} \citep{Jurman2008a},
\begin{equation}
	Ca(k,\tau,\sigma) = \sum_{i=1}^p \frac{| \min \{\tau(i), k+1\} - \min \{\sigma(i), k+1\} |}{\min \{\tau(i), k+1\} + \min \{\sigma(i), k+1\}}.
	\label{eq:Ca}
\end{equation}
This has the properties that we require, in that the denominator ensures more emphasis is placed on differences in the ranks of highly ranked variables in either dataset.  Furthermore, this distance measure allows comparisons between partial, top-\emph{k} lists, since a variable occurring in one top-\emph{k} list but not the other is assigned a ranking of $k+1$ in the list from which it is missing.  Note also that a variable $i$ that is not in either of the top-$k$ ranks, that is $\tau(i),\sigma(i) > k$, makes no contribution to $Ca(k,\tau,\sigma)$.

In order to gauge the extent to which the distance measure \eqref{eq:Ca} differs from that expected between two random lists, we require a value for the expected Canberra distance between two random lists, which we denote $E[Ca(k,p)]$.  \citet{Jurman2008a} derive an expression for this quantity, and we use this to compute the normalised Canberra distance,
\begin{equation}
	Ca^*(k,\tau,\sigma) = \frac{Ca(k,\tau,\sigma)}{E[Ca(k,p)]}.
	\label{eq:Ca_star}
\end{equation}

Note that this has a lower bound of 0, corresponding to exact agreement between the lists.  For two random lists, the upper bound will generally be close to 1, although it can exceed 1, particularly for small \emph{k}, since the expected value for random lists is not necessarily the highest value.

\begin{figure}[htbp]
\begin{center}
	\includegraphics[trim = 10mm 0mm 10mm 10mm, clip, scale=0.5]{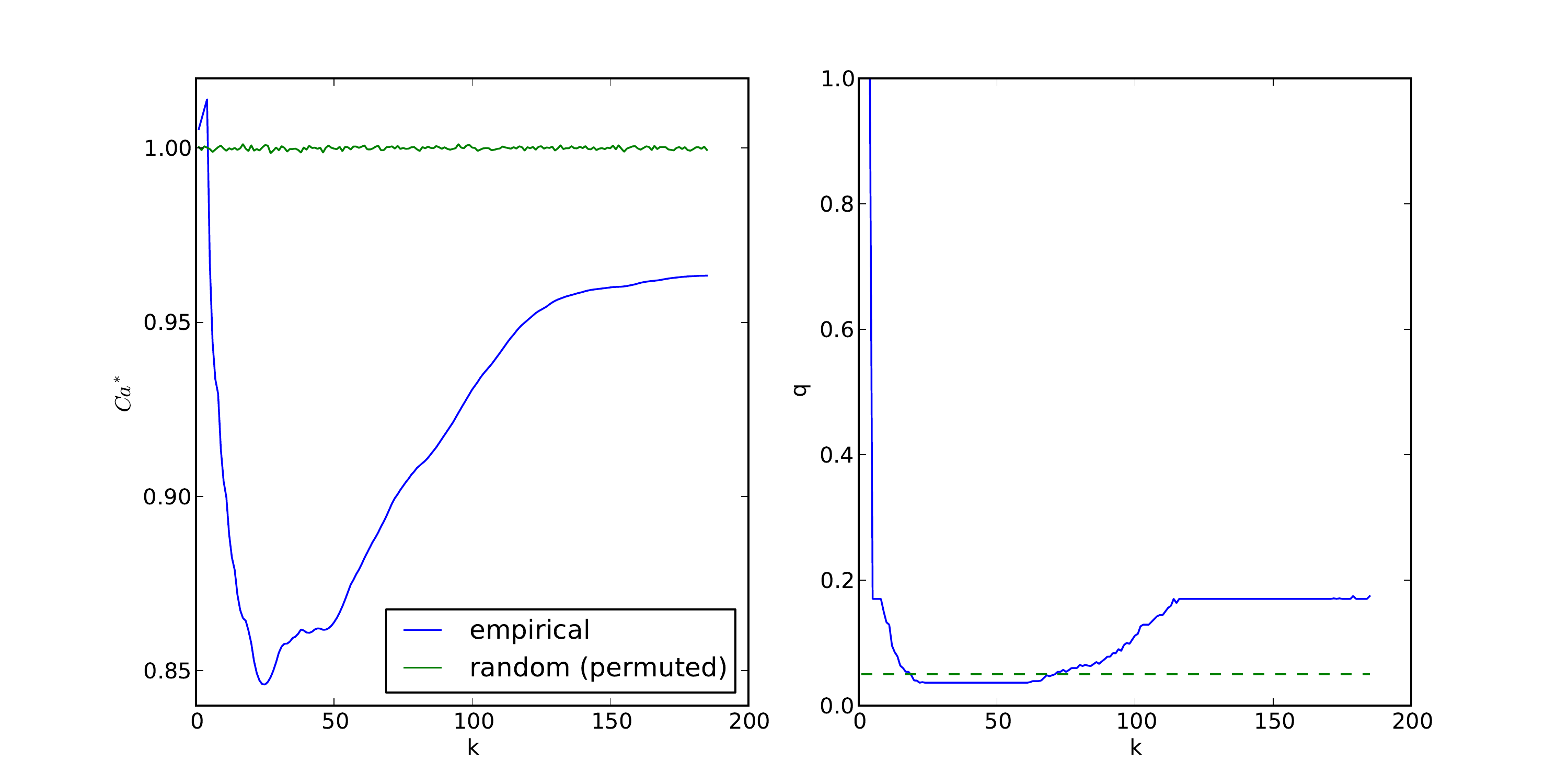}
\caption[Comparison of top-\emph{k} SP2 and SiMES pathway rankings]{Comparison of top-\emph{k} SP2 and SiMES pathway rankings.  \emph{Left hand plot:} Variation of normalised Canberra distance, $Ca^*$ with \emph{k} \eqref{eq:Ca_star} (blue curve).  Corresponding mean values over $Z=10,000$ permutations of SiMES rankings \eqref{eq:Ca_star_null} (green curve).  \emph{Right hand plot:} FDR $q$-values (blue curve).  Dotted green line shows the threshold for FDR control at the 5\% level.}
\label{fig:lipid_study_compare_ranked_pathways}
\end{center}
\end{figure}

We illustrate the variation of the normalised Canberra distance \eqref{eq:Ca_star} between SP2 and SiMES pathway rankings in the left hand plot in Figure \ref{fig:lipid_study_compare_ranked_pathways} (blue curve).  We consider all possible top-$k$ lists, $k=1,\ldots, 185$ since all 185 pathways are ranked in both datasets.  In the same plot, we also show
\begin{equation}
	Ca^*_\pi (k,\tau,\sigma) =
	\frac{1}{Z} \sum_{\pi = 1}^{Z} \frac{Ca(k,\tau,\sigma^\pi)}{E[Ca(k,p)]} \quad k = 1, \ldots, 185
	\label{eq:Ca_star_null}
\end{equation}
obtained by comparing empirical SP2 rankings $(\tau)$ against $Z=10,000$ permutations of the SiMES pathway rankings, $\sigma^\pi, \pi = 1, \ldots, 10,000$ (green curve).  This latter curve confirms that the expected value, $E[Ca(k,p)]$, is indeed a good measure of $Ca$ in the random case where there is no agreement between rankings.

Using the same permuted rankings, $\sigma^\pi$, we next test the null hypothesis that the observed normalised Canberra distance, $Ca^*(k,$ $\tau,\sigma)$, is not significantly different from that between $\tau$ and a random list $\sigma^\pi$, by computing a p-value as
\begin{equation*}
	p^*(k) = 
		\frac{1}{Z} \sum_{\pi = 1}^{Z} I_{Ca^*(k,\tau,\sigma) \le Ca^*(k,\tau,\sigma^\pi)},
\end{equation*}
for $k = 1, \ldots, 185$.  We then obtain FDR q-values using the Benjamini-Hochberg procedure \citep{Benjamini1995a} and illustrate these for each $k$ in the right hand plot of Figure \ref{fig:lipid_study_compare_ranked_pathways}.  FDR is controlled at a nominal 5\% level for $19 \le k \le 71$, indicating that the distance between the top-\emph{k} pathway rankings for both datasets is significantly different from the random ranking case for a wide range of possible values of \emph{k}.  The distance $Ca^*$ between SP2 and SiMES pathway rankings however attains its minimum value when $k=25$ with q$(25) = 0.037$, so that on this measure, the two pathway rankings are in closest agreement when we consider the top 25 pathways in each ranked list only.  Some intuitive understanding of why this might be so can be gained by considering the empirical vs.~null pathway selection frequency distributions for each dataset in Figures \ref{fig:SP2_pathways_distributions} (b) and \ref{fig:SiMES_selFreq_distributions} (top).  Here we see that the separation between empirical and null selection frequencies is most clear for values of \emph{k} below around 30 for SP2, and around 15 for SiMES.

If we assume that the two pathway rankings are indeed in closest agreement when $k=25$, then one means of obtaining a consensus set of important pathways is to consider their intersection, 
\begin{equation*}
	\Psi_{25}^{path} = \{i: \tau^{-1}(i) \le 25 \} \cap \{j: \sigma^{-1}(j) \le 25 \},
\end{equation*}
from which we can obtain a set of average rankings as
\begin{equation*}
	\psi_{25}^{path} = \big \{\frac{\tau(z) + \sigma(z)} {2}: z \in \Psi_{25}^{path} \big \}.
\end{equation*}
Both the intersection set, $\Psi_{25}^{path}$, and ordered average rankings, $\psi_{25}^{path}$ for the two datasets under consideration are shown in Table \ref{tab:lipid_study_consensus_pathway_rankings}.  We additionally mark the consensus set $\Psi_{path}^{25}$ with asterisks in Tables \ref{tab:SP2_pathway_rankings} and \ref{tab:SiMES_pathway_rankings}.

\begin{table}[htdp]
\caption[Consensus set of pathways for SP2 and SiMES datasets with $k=25$]{Consensus set of pathways, $\Psi_{25}^{path}$, for SP2 and SiMES datasets with $k=25$.  Consensus pathways are ordered by their average rankings in $\psi_{25}^{path}$.}
\begin{center}
\begin{tabular}{ll}
Pathway&Average rank $(\psi_{25}^{path}$)\\
\hline
Dilated Cardiomyopathy & $4.5$\\
Hypertrophic Cardiomyopathy  & $7.5$\\
T Cell Receptor Signaling Pathway & $11.0$\\
Terpenoid Backbone Biosynthesis & $11.0$\\
Arrhythmogenic Right Ventricular Cardiomyopathy  & $12.0$\\
Ribosome & $13.0$\\
Ppar Signaling Pathway & $18.5$\\
\end{tabular}
\end{center}
\label{tab:lipid_study_consensus_pathway_rankings}
\end{table}%

\subsection*{Comparison of ranked gene and SNP lists}\label{sec:PSGLAW_comparison_of_ranked_gene_snp_lists}

A number of factors complicate the comparison of ranked gene and SNP lists across both datasets.  Firstly, sets of mapped SNPs and genes differ slightly between the two datasets (see Table \ref{tab:lipid_study}).  Secondly, even if we consider only those variables \emph{mapped} in both datasets, different, though overlapping sets of variables are \emph{ranked} in each.  Thirdly, ranked variables are not independent \citep{Jurman2008a}.  For example, genes may be grouped into pathways, so that a reordering of genes within a pathway might be considered less significant than a reordering of genes mapping to different pathways.  Similarly a reordering of SNPs mapping to a single gene might be considered less significant than a reordering of SNPs mapping to different genes.  

In order to compute a distance measure between pairs or ranked lists, we therefore make two simplifying assumptions.  First, we consider only variables ranked in one or both datasets.  This seems reasonable, since we can necessarily only compile a distance measure from variables that are ranked in one or both datasets.  Second, we assume that variables are independent.  This makes our distance measure conservative, in the sense that it will treat all reordering of SNPs or genes equally, irrespective of any potential functional relationship between them.

With these assumptions in mind, we begin by denoting the set of all $p^*$ variables (genes or SNPs) that are ranked in \emph{either} dataset by $\mathcal{L} = \{1, \ldots, p^* \}$.  We further denote the corresponding sets of ranked variables for SP2 and SiMES datasets by $\mathcal{L}_\tau$ and $\mathcal{L}_\sigma$ respectively.  We then have the following set relations: $\mathcal{L}_\tau, \mathcal{L}_\sigma \subset \mathcal{L}$; $\mathcal{L}_\tau \ne \mathcal{L}_\sigma$; and $|\mathcal{L}_\tau| \ne |\mathcal{L}_\sigma|$.

We now extend the previous Canberra distance measure to encompass the above set relations.  We begin, as before, by defining two ranked lists corresponding to the rankings of all the variables in $\mathcal{L}$ for each dataset, although this time we must account for the fact that not all variables in $\mathcal{L}$ are ranked in both.  We denote SP2 rankings by $\tau(i), i = 1, \ldots, p^*$, where $\tau(i)$ is the rank of variable $i$ if $i \in \mathcal{L}_\tau$, and $\tau(i) = p^*$ otherwise.  SiMES rankings are defined in the same way, and denoted by $\sigma(i), i = 1, \ldots, p^*$.

Applying this revised ranking scheme, we can then define a top-$k$ normalised Canberra distance \eqref{eq:Ca} as
\begin{equation}
	Ca^*(k,\tau,\sigma) = \frac{Ca(k,\tau,\sigma)}{E[Ca(k,p^*)]}.
	\label{eq:Ca_star_partial_lists}
\end{equation}
for any $k \le \min \{ |\mathcal{L}_\tau|,|\mathcal{L}_\sigma|   \}$.  The restriction on $k$ follows from the fact that we cannot distinguish between top-$k$ rankings for all $k > \min \{ |\mathcal{L}_\tau|,|\mathcal{L}_\sigma|   \}$.\\\\\\
\emph{Gene rankings}\\\\
Information summarising the relationship between the two ranked lists of genes is given in Table \ref{tab:ranked_gene_set_relations}.  

\begin{table}[htdp]
\caption{Summary of genes analysed and ranked in SP2 and SiMES datasets.}
\begin{center}
\begin{tabular}{l|cc}
&SP2&SiMES\\
\hline
number of genes mapped to pathways							&4,734	&4,751\\
number of genes mapping to both datasets						& \multicolumn{2}{c}{4,726}\\
number of ranked genes $(|\mathcal{L}_\tau|,|\mathcal{L}_\sigma |)$	&3,430	& 2,815\\
number of genes ranked in either dataset ($p^*)$					& \multicolumn{2}{c}{3,913}\\
number of genes ranked in both datasets $(|\mathcal{L}_\tau| \cap |\mathcal{L}_\sigma |)$	& \multicolumn{2}{c}{2,332}
\end{tabular}
\end{center}
\label{tab:ranked_gene_set_relations}
\end{table}%

\begin{figure}[htb]
\begin{center}
	\includegraphics[trim = 0mm 0mm 0mm 0mm, clip, scale=0.5]{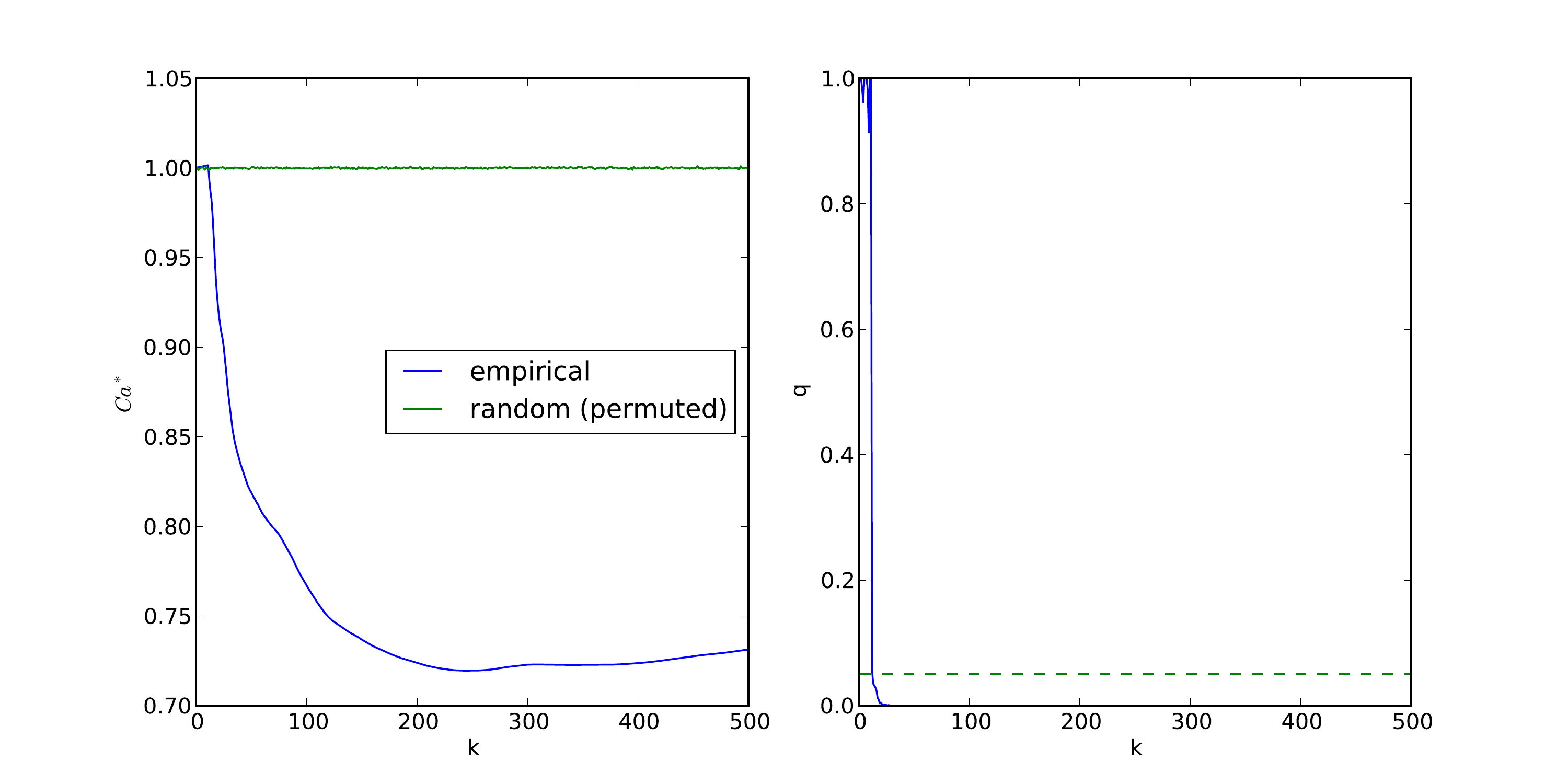}
\caption[Comparison of top-\emph{k} SP2 and SiMES gene rankings]{Comparison of top-\emph{k} SP2 and SiMES gene rankings, for $k = 1, \ldots, 500$.  \emph{Left hand plot:} Variation of normalised Canberra distance, $Ca^*$ with \emph{k} \eqref{eq:Ca_star_partial_lists} (blue curve), and corresponding mean values over 10,000 permutations of SiMES rankings \eqref{eq:Ca_star_null} (green curve).  \emph{Right hand plot:} FDR $q$-values (blue curve).  Dotted green line shows the threshold for FDR control at the 5\% level.}
\label{fig:Ca_genes}
\end{center}
\end{figure}

We consider normalised Canberra distances, $Ca^*(k,\tau,\sigma)$, for $k = 1, \ldots, 500$ only, and plot these in Figure \ref{fig:Ca_genes} (left, blue curve), along with $Ca^*_\pi (k,\tau,\sigma)$ \eqref{eq:Ca_star_null} for $Z=10,000$ permutations of the SiMES pathway rankings, $\sigma^\pi, \pi = 1, \ldots, 10,000$ (green curve).  Once again this latter curve confirms that the expected value, $E[Ca(k,p^*)]$, is indeed a good measure of $Ca$ in the random case where there is no agreement between rankings.  We also plot FDR $q$-values using the same procedure as described previously for pathways.  FDR is controlled at a nominal 5\% level for all $k > 13$ in the region tested ($1 \le k \le 500$).  The distance $Ca^*$ between SP2 and SiMES pathway rankings attains its minimum value when $k=244$, so that on this measure, the two gene rankings are in closest agreement when we consider the top 244 pathways in each ranked list only.  

Following the same strategy as implemented for pathways, we then form the consensus set, $\Psi^{gene}_{244}$, and average rankings $\psi^{gene}_{244}$.  The consensus set contains 84 genes, and we list the top 30 genes ordered by their average rank in the two datasets, in Table \ref{tab:lipid_study_consensus_gene_rankings}.\\\\
\begin{table}[htdp]
\begin{center}
\small
\caption[Consensus set of genes for SP2 and SiMES datasets with $k=244$]{Top 30 consensus genes ordered by their average rank, $\psi_{244}^{gene}$.}
\begin{tabular}{ccc}
Rank&Gene&Average rank $(\psi_{244}^{gene}$)\\
\hline
1&\emph{LAMA2} &	9.0\\
2&\emph{ADCY2} &	11.0\\
3&\emph{CACNA1C} &	11.5\\
4&\emph{PRKCB} &	11.5\\
5&\emph{PRKCA} &	21.0\\
6&\emph{EGFR} &	21.5\\
7&\emph{ITGA1} &	24.5\\
8&\emph{CACNA2D3} &	25.5\\
9&\emph{RYR2} &	26.5\\
10&\emph{IGF1R} &	30.5\\
11&\emph{PAK7} &	36.5\\
12&\emph{ADCY8} &	37.5\\
13&\emph{VAV2} &	41.0\\
14&\emph{SLC8A1} &	41.5\\
15&\emph{CACNB2} &	42.5\\
16&\emph{CACNA2D1} &	43.0\\
17&\emph{ITGA9} &	44.0\\
18&\emph{KRAS} &	47.5\\
19&\emph{MAPK10} &	50.5\\
20&\emph{CACNA1S} &	51.0\\
21&\emph{VAV3} &	54.0\\
22&\emph{PLCG2} &	55.5\\
23&\emph{BCL2} &	57.0\\
24&\emph{CD80} &	60.0\\
25&\emph{ITGA11} &	60.5\\
26&\emph{CTNNA2} &	61.0\\
27&\emph{ALDH1B1} &	61.5\\
28&\emph{MGST3} &	63.0\\
29&\emph{NEDD4L} &	63.0\\
30&\emph{PRKAG2} &	66.0
\end{tabular}
\label{tab:lipid_study_consensus_gene_rankings}
\end{center}
\end{table}\\\\
\newpage
\noindent
\emph{SNP rankings}\\\\
Information summarising the relationship between the two ranked lists of SNPs is given in Table \ref{tab:ranked_SNP_set_relations}.  In contrast to both pathway and gene rankings, it is apparent that relatively few ranked SNPs overlap both datasets -- 8,151 out of 41,452 SNPs that are ranked in either dataset.  This results in values for $Ca^*(k)$ that are close to 1, corresponding to the random list case, over a wide range of possible values for $k$ (data not shown).  

For this reason, we compute a simple summary measure
\begin{equation}
	\psi^{SNP} = \big \{\frac{\tau(j) + \sigma(j)} {2}: j \in \mathcal{L}_\tau \cap \mathcal{L}_\sigma \big \}
	\label{eq:psi_snp}
\end{equation}
and report only the top ranking SNPs using this measure in Table \ref{tab:consensus_snp_rankings}.

\begin{table}[htdp]
\caption{Summary of SNPs analysed and ranked in SP2 and SiMES datasets.}
\begin{center}
\begin{tabular}{l|cc}
&SP2&SiMES\\
\hline
number of SNPs mapped to pathways							&75,389	&78,933\\
number of SNPs mapping to both datasets						& \multicolumn{2}{c}{74,864}\\
number of ranked SNPs $(|\mathcal{L}_\tau|,|\mathcal{L}_\sigma |)$	&30,027	& 20,006\\
number of SNPs ranked in either dataset ($p^*)$					& \multicolumn{2}{c}{41,452}\\
number of SNPs ranked in both datasets $(|\mathcal{L}_\tau| \cap |\mathcal{L}_\sigma |)$	& \multicolumn{2}{c}{8,581}
\end{tabular}
\end{center}
\label{tab:ranked_SNP_set_relations}
\end{table}
\begin{table}[htdp]
\small
\caption{Top 30 SNPs ranked in both SP2 and SiMES datasets, ranked in order of mean ranking,} 
\begin{center}
\begin{tabular}{ccclll}
rank&SNP&$\psi^{SNP}$&\multicolumn{3}{c}{mapped gene(s)}\\
\hline
1 & rs897799 & 133.0 & COX6B2 & IL11 \\
2 & rs2126953 & 203.0 & ITGA1 \\
3 & rs7714110 & 213.0 & ADCY2 \\
4 & rs6924886 & 274.5 & PDSS2 \\
5 & rs2447867 & 275.5 & ITGA1 \\
6 & rs9386622 & 283.0 & PDSS2 \\
7 & rs6568474 & 283.5 & PDSS2 \\
8 & rs10446497 & 349.5 & PAK2 \\
9 & rs6583177 & 385.5 & PAK2 \\
10 & rs4765961 & 429.0 & CACNA1C \\
11 & rs759440 & 457.0 & PDSS2 \\
12 & rs10457161 & 465.0 & PDSS2 \\
13 & rs10462842 & 479.5 & ADCY2 \\
14 & rs9373932 & 529.0 & PDSS2 \\
15 & rs12206487 & 532.0 & LAMA2 \\
16 & rs743567 & 543.0 & MYH7 \\
17 & rs12472674 & 543.0 & CTNNA2 \\
18 & rs11759792 & 557.5 & PDSS2 \\
19 & rs9373924 & 566.0 & PDSS2 \\
20 & rs319070 & 623.0 & PDSS2 \\
21 & rs751877 & 630.0 & ADCY4 & LTB4R & RIPK3 \\
22 & rs2047698 & 714.0 & PDGFD \\
23 & rs4804505 & 727.5 & PDE4A & KEAP1 \\
24 & rs12672417 & 764.0 & SMURF1 \\
25 & rs7766689 & 800.5 & LAMA2 \\
26 & rs157694 & 860.0 & MAP3K7 \\
27 & rs554192 & 878.0 & NEDD4L \\
28 & rs2746543 & 896.5 & SDHB \\
29 & rs742257 & 942.0 & LAMB3 \\
30 & rs1798619 & 944.5 & PAK2 \\
\end{tabular}
\end{center}
\label{tab:consensus_snp_rankings}
\end{table}%

\section*{Discussion}\label{sec:Discussion}

We have outlined a method for the detection of pathways, SNPs and genes associated with a quantitative trait.  Our method uses a sparse regression model, the sparse group lasso, that enforces sparsity at the pathway and SNP level.  As well as identifying important pathways, this approach is designed to maximise the power to detect causal SNPs, possibly of low effect size, that might otherwise be missed if pathways information is ignored.  In a simulation study we demonstrated that where causal SNPs are enriched within a single causal pathway, SGL does indeed have greater SNP selection power, compared to an alternative sparse regression model, the lasso, that disregards pathways information.  These results mirror previous findings that support the intuition that a sparse selection penalty that promotes dual-level sparsity is better able to recover the true model in these circumstances \citep{Friedman2010,Simon2012}.

We then argued from a theoretical standpoint that where individual SNPs can map to multiple pathways, a modification (SGL-CGD) of the standard SGL-BCGD estimation algorithm that treats pathways as independent, may offer greater sensitivity for the detection of causal SNPs and pathways.  A potential concern is that this gain in power may be accompanied by an inflated number of false positives.  However, in a simulation study with overlapping pathways we found relative gains in both sensitivity and specificity, under the independence assumption.  This gain in specificity was unexpected, and appears to arise directly from treating pathways as independent in the model estimation.  As with the group lasso, the ability of SGL to recover the true model is likely to be affected by the complexity of the pathway overlap structure \citep{Percival2012}, although we expect that the gains in power and sensitivity achieved with the independence assumption will also be apparent with real data.  

Our method combines the SGL model and SGL-CGD estimation algorithm with a weight-tuning algorithm to reduce selection bias, and a resampling technique designed to provide a robust measure of SNP, gene and pathway importance in a finite sample.  As such, the latter is expected to confer advantages, in terms of the down ranking of unimportant predictors, previously observed for the lasso \citep{Meinshausen2010,Chatterjee2011}.  Once again it would be interesting to explore this further using simulations derived from real pathway and genotype data.  

We do not explore the issue of determining a selection frequency threshold for the control of false positives here.  In principal such a threshold could be determined by comparing empirical selection frequency distributions with those obtained under the `null', through permutations, although this is not a trivial exercise \citep{Valdar2012}.  An alternative method for error control has been investigated in the context of lasso selection \citep{Meinshausen2010}, but the direct application of this approach to the present case is not feasible, since overlapping pathways make clear distinctions between causal and noise variables problematic.  We instead develop a heuristic measure of ranking performance in our application study identifying SNPs and pathways associated with serum high-density lipoprotein cholesterol levels (HDLC).  Firstly, by comparing empirical and null pathway and SNP rankings for each dataset, we gain some confidence that pathway and SNP signals captured in the top rankings can be distinguished from those arising from noise or spurious associations.  Secondly, we take advantage of the fact that we are able to compare results from two independent GWAS datasets.  On the assumption that similar patterns of genetic variation are likely to impact HDLC levels in both cohorts, we set a ranking threshold based on computing distances between ranked lists from each dataset.

Interestingly, when a comparison between empirical and null rankings is made with a reduced value for the regularisation parameter $\alpha$, there is evidence of selection bias, in the sense that pathways and SNPs tend to be highly ranked both empirically and under the null (see Figures \ref{fig:SP2_pathways_distributions} and \ref{fig:SP2_SNP_distributions}).  Since a smaller $\alpha$ corresponds to a greater number of SNPs being selected at each subsample, this would seem to suggest that too many SNPs are being selected.  In this case, pathway and SNP rankings may in part reflect spurious associations, with a bias towards SNPs overlapping multiple pathways.   

There are other potentially interesting areas to explore with regard to the subsampling method used here.  For example, standard approaches consider only the set of variables selected at each subsample, and ignore potentially relevant information captured in the coefficient estimates themselves.  The use of this additional information would result in a set of ranked lists, one for each subsample, and the joint consideration of these lists has the potential to provide a more robust measure of variable importance, by taking account of the relative importance of each variable for each subsample \citep{Sculley2007,Kolde2012,Jurman2012}.  

Turning to the study results, we conduct two separate analyses on independent discovery and replication datasets.  Since subjects from both datasets are genotyped on the same platform, the large majority of SNPs mapping to pathways in one dataset do so also in the other dataset.  Thus 99.3\% of SNPs mapping to pathways in the SP2 dataset are similarly mapped in the SiMES dataset.  For the SiMES dataset, the corresponding figure is 94.8\%.  As expected, the concordance of gene coverage is even greater.  Thus 99.8\% of mapped genes in the SP2 dataset are also mapped in the SiMES dataset, and 99.5\% of mapped genes in the SiMES dataset are also mapped in SP2.  This large overlap in gene (and pathway) coverage between  datasets is likely to occur even when datasets are genotyped on different SNP arrays.  Indeed this is one advantage of methods such as the one described hereÁ that enable comparisons between pathway and gene rankings.

We obtain consensus pathway and gene rankings by considering only the top $k$ ranks in each dataset, with $k$ obtained as the value that minimises the distance between the two rankings.  We additionally derive a significance measure for each top-$k$ distance by comparing empirical distances against a null distribution obtained by permuting ranks in one list.  We note that this can only be an approximation of the true null, since in reality rankings for both datasets may be influenced by the extent to which genes and SNPs overlap multiple pathways.  However, some support for the reasonableness of this approximation can be gained from our earlier analysis, showing that the correlation between empirical and null pathway and SNP rankings is low, so that rankings under the null are indeed approximately random.

Considering the consensus pathway rankings in Table \ref{tab:lipid_study_consensus_pathway_rankings}, three out of the seven consensus pathways (ranked 1, 2 and 5), are related to cardiomyopathy.  These three pathways are the only cardiomyopathy-related pathways amongst the 185 KEGG pathways used in our analysis, so it is noteworthy that all three fall within the consensus pathway rankings.  The link between HDLC levels and cardiomyopathy is already well established \citep{Ansell2005,Gordon1989,Toth2005,Freitas2009,Gaddam2011}.  Furthermore, numerous references in the literature also describe the links between lipid metabolism and T cell receptor (consensus pathway ranking 3) and PPAR signaling (rank 7) \citep{Janes2000,Calder2007,Staels1998,Bensinger2008}.

Turning to a consideration of the top 30 consensus genes and SNPs presented in Tables \ref{tab:lipid_study_consensus_gene_rankings} and \ref{tab:consensus_snp_rankings} (and see also pathway ranking tables \ref{tab:SP2_pathway_rankings}, \ref{tab:SiMES_pathway_rankings} and \ref{tab:lipid_study_consensus_pathway_rankings}, and extended results in supplementary information).  We found that many are enriched in one of several gene families:
\begin{enumerate}
\item
	L-type calcium channel genes, including \emph{CACNA1C, CACNA1S, CACNA2D1, CACNA2D3} and \emph{CACNB2}
	\item
	Adenylate cyclase genes, including \emph{ADCY2}, \emph{ADCY4} and \emph{ADCY8}
	\item
	Integrin and laminin genes, including \emph{ITGA1}, \emph{ITGA9}, \emph{ITGA11}, \emph{LAMA2}, and \emph{LAMA3}
	\item
	MAPK signaling pathway genes, including \emph{MAPK10} and \emph{MAP3K7}
	\item
	Immunological pathway genes, including \emph{PAK2}, \emph{PAK7}, \emph{PRKCA}, \emph{PRKCB}, \emph{VAV2} and \emph{VAV3}
\end{enumerate}

These genes are highly enriched in several high ranking pathways from both datasets.  Notably, the focal adhesion pathway alone has 12 gene hits, as does the dilated cardiomyopathy pathway.  Cardiomyopathy pathways  as a whole have 30 genes hits (several of the genes overlap more than one cardiomyopathy pathway).  10 of these genes feature in the MAPK signaling pathway, while GnRH (8 genes), T and B cell receptor (8), calcium (7), ErbB (5), and Wnt signling (4) pathways also contain several genes in the list. To elucidate the biological relevance of these gene families and the connections between them, we investigated their known functional links with cardiovascular phenotypes (not restricted to HDLC) by referencing the KEGG and Genetic Association (\url{http://geneticassociationdb.nih.gov}) databases.\\\\
\emph{Voltage dependent L-type calcium channel gene family}\\
The genes in this family encode the subunits of the human voltage dependent L-type calcium channel (CaV1). The $\alpha$-1 subunit (encoded by \emph{CACNA1C}, \emph{A1S}, \emph{A2D1} and \emph{A2D3} in our study) determines channel function in various tissues. CaV1 function has significant impact on the activity of heart cells and smooth muscles.  For example, patients with malfunctioning CaV1 develop arrhythmias and shortened QT interval \citep{Splawski2004,Antzelevitch2007,Templin2011}.  Furthermore, \emph{CACNA1C} polymorphisms have been associated with variation in blood pressure in Caucasian and East Asian populations by pharmacogenetic analysis. In 120 Caucasians, 3 SNPs in this gene were significantly associated with the response to a widely applied antihypertensive CaV1 blocker \citep{Bremer2006}.  \citet{Kamide2009} also found that polymorphisms in \emph{CACNA1C} were associated with sensitivity to an antihypertensive in 161 Japanese patients.  The CaV1 $\beta$ subunit encoding \emph{CACNB2} has also been associated with blood pressure \citep{Levy2009}. 

This gene family was mapped to several pathways in our study, with the KEGG dilated cardiomyopathy pathway achieving highest rank both within individual datasets, and in the consensus pathway rankings.  Dilated cardiomyopathy is the most common form of cardiomyopathy, and  features enlarged and weakened heart muscles. Although high levels of serum HDLC lowers the risk of heart disease \citep{Castelli1988,Toth2005}, there is still no direct evidence that CaV1 is involved in HDLC metabolism.\\\\
\emph{Adenylate cyclase gene family}\\
Three adenylate cyclases genes, \emph{ADCY2}, \emph{ADCY4} and \emph{ADCY8} were highly ranked in our study. Currently, there are no reported associations of these genes with cardiovascular disease or lipid levels.  Adenylate cylcase genes catalyse the formation of cyclic adenosine monophosphate (cAMP) from adenosine triphosphate (ATP), while cAMP servers as the second messenger in cell signal transduction. Note that \emph{ADCY2} is insensitive to calcium concentration, suggesting that any association of this gene family with HDLC levels may not be due to any interactions with the CaV1 gene family.  

Among high ranking pathways, \emph{ADCY2} and \emph{ADCY8} feature in  the dilated cardiomyopathy pathway. Although ADCY4 was not in the top 30 consensus genes, rs751877 in this gene was among the top 30 consensus SNPs.\\\\
\newpage
\emph{Integrin and laminin gene families}\\
We found 3 genes encoding integrin subunits in our study. Integrins hook to the extracellular matrix (ECM) from the cell surface, and are also important signal transduction receptors which communicate aspects of the cell's physical and chemical environment \citep{Nermut1988}.  Interestingly, laminins are the major component of the ECM, and are relevant to the shape and migration of almost every type of tissue.  Both of these two families of genes are therefore highly relevant to the survival and shape of heart muscles.  A recent GWAS conducted in a Japanese population confirmed a previous association between \emph{ITGA9} and blood pressure in European populations \citep{Takeuchi2010}.  

Integrin family genes and \emph{LAMA2} were selected primarily within high-ranking cardiomyopathy, focal adhesion and ECM receptor signaling pathways, with once again the dilated cardiomyopathy pathway achieving the highest ranks.  However, evidence for \emph{LAMA3} association is weaker, since it was not in the top 30 consensus genes, although a SNP from the \emph{LAMB3} (laminin $\beta$-3) gene was ranked 29 in the consensus SNP list.\\\\
\emph{MAPK signaling pathway}\\
TAK1 (\emph{MAP3K7}) and JNK3 (\emph{MAPK10}) are kinases which regulate cell cycling. They activate or depress downstream transcription factors which mediate cell proliferation, differentiation and inflammation. 

JNK activity has been associated with obesity in a mouse model, where the absence of JNK1 (\emph{MAPK8}), a protein in the same family as \emph{MAPK10}, protects against the obesity-induced insulin resistance \citep{Hirosumi2002}.  The negative correlation between HDLC level and obesity has been well accepted \citep{Howard2003}.\\\\
\emph{Immunological pathways}\\
PAK (\emph{PAK2} and \emph{PAK7}) genes feature in the high ranking T cell signaling pathway in both SP2 and SiMES datasets. \emph{PRKC} (including \emph{PRKCA} and \emph{PRKCB}), along with \emph{VAV} (\emph{VAV2} and \emph{VAV3}) genes also feature in various high ranking immunological pathways including T cell signaling, Pathogenic Escherichia Coli Infection and Natural Killer Cell Mediated Cytotoxicity.  Genes from all 3 of these families are frequently top ranked in these pathways. 

\emph{PAK} and \emph{VAV} are activated by antigens, and regulate the T cell cytoskeleton, indicating a possible impact on T cell shape and mobility.  In a candidate gene association analysis, \emph{PRKCA} was reported to be associated with HDLC at a nominally significant level, but was not significant after adjusting for multiple testing \citep{Lu2008}.\\

In summary, genes enriched in the above gene clusters and pathways may be relevant to heart muscle cell signal transduction, shape and migration, and may thus have functional relevance to the onset of cardiovascular diseases. 
Many highly ranked genes in our study are also involved in neurological pathways.  For example polymorphisms in \emph{CACNA1C} have been associated with bipolar disorder, schizophrenia and major depression \citep{Ferreira2008,Moskvina2009,Green2010}.  This points to an interesting hypothesis that serum HDLC levels might be regulated not only by metabolism but also by neurological pathways, although the elucidation of any putative biological mechanism underlying such an association obviously exceeds the scope of this study.

Besides the gene families and associated pathways discussed above, a notable feature of the top 30 consensus SNP ranking results presented in Table \ref{tab:consensus_snp_rankings} is the inclusion of 9 SNPs mapping to the \emph{PDSS2} gene.  \emph{PDSS2} achieves its high ranking in the consensus SNP list, through its inclusion in the highly ranked terpenoid backbone biosynthesis pathway in SiMES, and this gene is in fact the second highest consensus ranking gene with this dataset.  In contrast, this gene is ranked low (216) in the SP2 dataset, which explains why it fails to make the top 30 consensus gene rankings.  \emph{PDSS2} encodes subunit 2 of prenyl diphosphate synthase, which determines the length of the isoprenoid chain of coenzyme Q10 (CoQ10) \citep{Saiki2005}. A deficiency in biosynthesis of CoQ10 has previously been associated with delayed motor development and abnormal renal function, with excess serum lipids \citep{Sobreira1997}.  

Despite the well established links between lipid metabolism and PPAR signaling noted above, no genes in this high-ranked pathway fall in the top 30 gene rankings for either dataset (see Tables \ref{tab:SP2_pathway_rankings} and \ref{tab:consensus_snp_rankings}).  This could be because the association signal in this pathway is more widely distributed, compared to other high ranking pathways, perhaps indicating heterogeneity in genetic causal factors within our sample, so that different genes and SNPs are highlighted in different subsamples.  This would result in reduced gene selection frequencies.  Also, genes that overlap multiple putative causal pathways are more likely to be selected in a given subsample, meaning that associated genes mapping to pathways with relatively few overlaps may have lower selection frequencies.  This may be the case with genes in the PPAR signaling pathway, whose 63 genes map to an average $2.7 \pm 1.8$ pathways.  As a comparison, the 84 genes in the top-ranked dilated cardiomyopathy pathway map to an average $7.2 \pm 3.8$ pathways.

Our study failed to highlight HDLC-associated SNPs identified in previous GWAS (see for example \url{www.genome.gov/gwastudies} for an up to date list).  A primary reason for this is that the large majority of SNPs identified in previous studies do not map to pathways in our study, either because they fall in intergenic regions, or because they do not feature on the Illumina arrays used here.  In addition our method is designed to highlight distributed, small SNP effects that accumulate across gene pathways, and so will likely fail to identify those SNPs with significant marginal effects targeted by GWAS.  Furthermore, where there are common mechanisms affecting phenotypes in both cohorts, we would expect to observe the most concordance between the two studies at the pathway level, followed by genes, and lastly SNPs.  Indeed this increased heterogeneity at the SNP, and to a lesser extent at the gene level is one motivation for adopting a pathways approach in the first place \citep{Holmans2009,Hirschhorn2009,Cantor2010}.   This reduced concordance at the SNP level may be due to increased heterogeneity of genetic risk factors between the two datasets.  Another important source of variation in SNP selection frequencies is LD between SNPs.  The within-pathway lasso penalty will tend to select one of a group of highly correlated SNPs at random, reducing SNP selection frequencies within LD blocks harbouring causal SNPs.  An alternative approach would be to consider a different penalty within selected pathways, for example the elastic net \citep{Zou2005}, which selects groups of correlated variables jointly, although this comes at the cost of introducing a further regularisation parameter to be tuned.

Finally, as with all pathways analyses, a number of limitations with this general approach should be noted.  Despite great efforts, pathway assembly is still in its infancy, and the relative sparsity of gene-pathway annotations reflects the fact that our understanding of how the majority of genes functionally interact is at an early stage.  As a consequence, annotations from different pathways databases often vary \citep{Soh2010}, so that the choice of pathways database will impact results \citep{Elbers2009,Cantor2010}.  Results are also subject to bias resulting from SNP to gene mapping strategies, so that for example SNP to gene mapping distances will affect the number of unmapped SNPs falling within gene `deserts' \citep{Eleftherohorinou2009}; SNPs may map to relatively large numbers of genes in gene rich areas of the genome; and the mapping of a SNP to its closest gene may obscure a true functional relationships with a more distant gene \citep{Wang2009}.  Indeed recent research from the ENCODE project indicates that functional elements may in fact be densely distributed throughout the genome \citep{Dunham2012,Sanyal2012}, and this information has the potential to radically alter future pathways analysis.  These issues, together with the fact that PGAS methods are by construction designed to highlight distributed, moderate to small SNP effects, serve to further illustrate the point that pathways analysis should be seen as complementary to studies searching for single markers \citep{Wang2010}.  

\printbibliography 


\appendix

\section*{Supplementary Information}

\section{SGL estimation algorithm} \label{SGL estimation algorithm}

For the estimation of $\boldsymbol{\beta}^{SGL}$ we proceed by noting that the optimisation \eqref{eq:SGL_objF} is convex, and (in the case of non-overlapping groups) that the penalty is block-separable, so that we can obtain a solution using block, or group-wise coordinate descent \citep{Tseng2009}.  For a single group, $l$, the corresponding minimising function is given by
\begin{equation}
	f(\boldsymbol{\beta}_l) = 
		\frac{1}{2} ||\mathbf{y}-\mathbf{X}\boldsymbol{\beta}||_2^2 + (1-\alpha) \lambda w_l||\boldsymbol{\beta}_l||_2 + \alpha 								\lambda || \boldsymbol{\beta}_l ||_1.
	\label{eq:SGL_minimising_function}
\end{equation}
An optimal solution for SNP coefficient $\beta_j$ is then derived from the subgradient equations
\begin{equation}
	-\mathbf{x}_j'(\hat{\mathbf{r}}_l - \sum_{k \ne j} \mathbf{x}_k \hat{\beta}_k - \mathbf{x}_j \beta_j) + (1-\alpha) \lambda w_l s_j + \alpha \lambda t_j = 0
	\quad j = l_1,\dots,l_{P_l},
	\label{eq:SGL_diff_all_terms}
\end{equation}
where $\hat{\beta}_k, k \ne j$ are the current estimates for other SNP coefficients in group $l$, and the group partial residual, $\mathbf{\hat{r}}_l = \mathbf{y} - \sum_{m \ne l} \mathbf{X}_m \hat{\boldsymbol{\beta}}_m $.  Here $s_j$ and $t_j$ are the respective subgradients of $|| \boldsymbol{\beta}_l ||_2$ and $| \beta_j |$, with
\begin{align}
	s_j &=
	\begin{cases}
		\frac{\beta_j}{||\boldsymbol{\beta}_l ||_2}		& \text{if }  || \boldsymbol{\beta}_l ||_2 \ne \mathbf{0}\\
		\in [-1,1] 						& \text{if }  || \boldsymbol{\beta}_l ||_2 = \mathbf{0}
	\end{cases}\notag \\
	t_j &=
	\begin{cases}
		\mbox{sign}(\beta_j) 		& \text{if }  \beta_j \ne 0	\\
		\in [-1,1] 					& \text{if }  \beta_j = 0.
	\end{cases}¥
	\label{eq:SGL_s_j_and_t_j}
\end{align}
If $\boldsymbol{\beta}_l = \mathbf{0}$, that is group $l$ is not selected by the model, then from \eqref{eq:SGL_diff_all_terms}
\begin{equation}
	-\mathbf{x}_j'\hat{\mathbf{r}}_l + (1-\alpha) \lambda w_l s_j + \alpha \lambda t_j = 0,
	\quad j = l_1,\dots,l_{P_l}.
	\label{eq:SGL_partial_deriv_bl_zero}
\end{equation}
Substituting $\mathbf{a} = \mathbf{X}_l' \mathbf{\hat{r}}_l$ gives
\begin{equation*}
	a_j =  (1-\alpha)\lambda w_l s_j + \alpha\lambda t_j, \quad j = l_1,\dots,l_{P_l}
\end{equation*}
so that 
\begin{equation*}
	s_j^2 = \frac{1}{(1-\alpha)^2\lambda^2 w_l^2} (a_j - \alpha\lambda t_j)^2, \quad j = l_1,\dots,l_{P_l}, 
\end{equation*}
and
\begin{equation*}
	\sum_j s_j^2 = \frac{1}{(1-\alpha)^2\lambda^2 w_l^2} \sum_j  (a_j - \alpha\lambda t_j)^2.
\end{equation*}	
From \eqref{eq:SGL_s_j_and_t_j}, when $\boldsymbol{\beta}_l = \mathbf{0}$, $||s||_2 = (\sum_j s_j^2)^{\frac{1}{2}} \le 1$, so that
\begin{equation}
	 \sum_j  (a_j - \alpha\lambda t_j)^2 	\le (1-\alpha)^2\lambda^2 w_l^2.
	 \label{eq:SGL_aj_inequality}
\end{equation}	
Also from \eqref{eq:SGL_s_j_and_t_j}, one further condition when $\boldsymbol{\beta}_l = \mathbf{0}$ is that $t_j \in [-1,1]$. The values, $\hat{t}_j$ that minimise the left hand size of \eqref{eq:SGL_aj_inequality} are therefore given by
\begin{equation*}
	\hat{t}_j =
	\begin{cases}
		\frac{a_j}{\alpha\lambda} 			& \text{if }  |\frac{a_j}{\alpha\lambda}| \le 1	\\
		\text{sign}(\frac{a_j}{\alpha\lambda}) 	& \text{if }  |\frac{a_j}{\alpha\lambda}| > 1.
	\end{cases}¥
\end{equation*}¥
Substituting for $a_j$, we can then write the values for $a_j - \alpha\lambda t_j$ that minimise the left hand side of \eqref{eq:SGL_aj_inequality} as
\begin{align*}
	 a_j - \alpha\lambda t_j &= 
	\begin{cases}
		0 			& \text{if }  |\mathbf{x}_j'\hat{\mathbf{r}}_l| \le \alpha \lambda	\\
		\mbox{sign}(\mathbf{x}_j'\hat{\mathbf{r}}_l)(|\mathbf{x}_j'\hat{\mathbf{r}}_l| - \alpha \lambda) 	& \text{if }  |\mathbf{x}_j'\hat{\mathbf{r}}_l| > \alpha \lambda
	\end{cases}\\
	&= S(\mathbf{x}_j'\hat{\mathbf{r}}_l, \alpha \lambda)
\end{align*}	
for $j = l_1, \ldots, l_{P_l}$, where
\begin{equation}
	S(\mathbf{x}_j'\hat{\mathbf{r}}_l, \alpha \lambda) = \mbox{sign}(\mathbf{x}_j'\hat{\mathbf{r}}_l)(|\mathbf{x}_j'\hat{\mathbf{r}}_l| - \alpha \lambda)_+
	\label{eq:SGL_soft_threshold}
\end{equation}
is the lasso soft thresholding operator.  Finally, we can now rewrite the condition for $\boldsymbol{\hat{\beta}}_l = \mathbf{0}$, \eqref{eq:SGL_aj_inequality} as
\begin{equation}
	|| S(\mathbf{X}_l'\hat{\mathbf{r}}_l, \alpha \lambda) ||_2 \le (1-\alpha) \lambda w_l, 
	\label{eq:SGL_bl_equals_zero_criterion}
\end{equation}
Where the vector $S(\mathbf{X}_l'\hat{\mathbf{r}}_l, \alpha \lambda) = [S(\mathbf{x}_{l_1}'\hat{\mathbf{r}}_l, \alpha \lambda), \ldots, S(\mathbf{x}_{l_P}'\hat{\mathbf{r}}_l, \alpha \lambda)]$.  Note that with $\alpha=0$, this reduces to the group lasso group selection criterion.

In the case that $\boldsymbol{\beta}_l \ne \mathbf{0}$, that is group $l$ is selected by the model, from \eqref{eq:SGL_diff_all_terms} and \eqref{eq:SGL_s_j_and_t_j} we see that $\beta_j = 0$ when
\begin{equation}
	-\mathbf{x}_j'(\hat{\mathbf{r}}_l - \sum_{k \ne j} \mathbf{x}_k \hat{\beta}_k) \le |\alpha \lambda|.
	\label{eq:SGL_SNP_notSel}
\end{equation}

For completeness, we rewrite the criterion for selecting pathway $l$ from \eqref{eq:SGL_bl_equals_zero_criterion} as
\begin{equation}
	|| S(\mathbf{X}_l'\hat{\mathbf{r}}_l, \alpha \lambda) ||_2 > (1-\alpha) \lambda w_l
\end{equation}
and the criterion for selecting SNP $j$ in selected pathway $l$ from \eqref{eq:SGL_SNP_notSel} as
\begin{equation}
		|\mathbf{x}_j' \mathbf{\hat{r}}_{l,j}| > \alpha \lambda
\end{equation}
where $\mathbf{\hat{r}}_{l,j} = \mathbf{\hat{r}}_l - \sum_{k \ne j} \mathbf{x}_k \hat{\beta}_k $ is the SNP partial residual, obtained by regressing out the current estimated effects of all other predictors in the model, except for predictor $j$.

A number of methods for the estimation of $\boldsymbol{\beta}_l$ in the case that $||\boldsymbol{\beta}_l||_2 \ne \mathbf{0}$ have been proposed \citep{Friedman2010,Foygel2010,Liu2010,Simon2012}.  A complicating factor is the discontinuities in the first (and second) derivatives of $s_j$  at $||\boldsymbol{\beta}_l||_2 = 0$, that is where $||\boldsymbol{\beta}_l||_2$ first moves away from zero, and of $t_j$ when $\beta_j=0$.  As with GL,  \citet{Friedman2010} describe a numerical method using coordinate descent, by combining a golden search over $\beta_j$ with parabolic interpolation.  However we find this too computationally intensive for the large datasets we wish to analyse.  \citet{Simon2012} propose an accelerated, block gradient descent method in which $\boldsymbol{\beta}_l$ is iteratively updated in a single step along the line of steepest descent of the block objective function until convergence.  We instead use a block, coordinate-wise gradient descent (BCGD) method that uses a Newton update, similar to that proposed by \citet{Zhou2010}, and we describe this below.

To update $\beta_j$ from its current estimate, $\hat{\beta}_j$, we note from \eqref{eq:SGL_diff_all_terms} and \eqref{eq:SGL_s_j_and_t_j} that if $\hat{\beta}_j \ne 0$, the subgradient equation for predictor $j$ is given by
\begin{equation}
	\partial_j = -\mathbf{x}_j'(\hat{\mathbf{r}}_l - \mathbf{X}_l \hat{\boldsymbol{\beta}_l}) + (1-\alpha) \lambda w_l \frac{\hat{\beta_j}}{|| \hat{\boldsymbol{\beta}_l} ||_2} + \alpha \lambda \cdot \mbox{sign}(\hat{\beta_j}).
		\label{eq:SGL_partial_j}
\end{equation}

We then descend along the gradient at $\hat{\beta}_j$ towards the minimum using Newton's method.  The Newton update, $\beta_j^*$, is then given by
\begin{equation}
	\begin{aligned}
		\hat{\beta}_j^* &= \hat{\beta}_j - \frac{\partial_j}{\partial_j'},\\
		\mbox{where}\quad \partial'_j &= 1 + \frac{(1-\alpha) \lambda w_l} {|| \hat{\boldsymbol{\beta}_l} ||_2} \Big(1 - \frac{\hat{\beta}_j^2}{|| \hat			{\boldsymbol{\beta}_l} ||_2^2}\Big)
	\end{aligned}
	\label{eq:SGN_Newton_update}
\end{equation}
is the derivative of \eqref{eq:SGL_partial_j} at $\hat{\beta}_j$.  The update \eqref{eq:SGN_Newton_update} is repeated until convergence.

We must also deal with the case where $\hat{\beta}_j = 0$.  Here we adopt a slightly different strategy, since the partial derivative, $t_j$ of $\beta_j$ is not continuous.  We avoid this discontinuity by testing the `directional derivatives', $\partial_j^+$ and $\partial_j^-$, respectively representing the partial derivatives at $\beta_j = 0$ in the direction of increasing and decreasing $\beta_j$.  Recalling that we are dealing with the case $||\boldsymbol{\beta}_l ||_2 \ne \mathbf{0}$, at $\beta_j = 0$ the group penalty term in \eqref{eq:SGL_partial_j} disappears.  That is, once a group is selected by model it becomes easier for each SNP coefficient to move away from zero.  The two directional derivatives are then given by
\begin{equation}
	\begin{aligned}
			\partial_j^+ &= -\mathbf{x}_j'(\hat{\mathbf{r}}_l - \mathbf{X}_l \hat{\boldsymbol{\beta}_l}) + \alpha \lambda \\
			\partial_j^- &= -\mathbf{x}_j'(\hat{\mathbf{r}}_l - \mathbf{X}_l \hat{\boldsymbol{\beta}_l}) - \alpha \lambda.
	\end{aligned}
	\label{eq:SGL_directional_derivative}
\end{equation}
Since the minimising function \eqref{eq:SGL_minimising_function} is convex, there are three possible outcomes, and we substitute for $\partial_j$ in \eqref{eq:SGN_Newton_update} accordingly:
\begin{equation}
	\partial_j \leftarrow
	\begin{cases}
		\partial_j^- \quad \text{if} \quad \partial_j^- > 0 & \text{and} \quad \partial_j^+ > 0\\
		\partial_j^+ \quad \text{if} \quad \partial_j^- < 0 & \text{and} \quad \partial_j^+ < 0 \\
		0 \quad \quad \text{if} \quad \partial_j^- > 0 & \text{and} \quad \partial_j^+ < 0
	\end{cases}¥
	\label{eq:SGL_Newton_update_betaj_zero}
\end{equation}
In the third case, $f(\boldsymbol{\beta}_l)$ is increasing either side of $\beta_j = 0$, so that $\hat{\beta}_j$ must remain at zero.  We can then proceed with the standard Newton update \eqref{eq:SGN_Newton_update}.

Finally, since the Newton update may occasionally overstep the minimum (where $\partial_j = 0$), a simple remedy proposed by \citet{Zhou2010} is to check that $f(\boldsymbol{\beta}_l)$ is decreasing at each iteration. If this is not the case, then the step size in \eqref{eq:SGN_Newton_update} is halved.  The complete algorithm for SGL estimation using BCGD is presented in Box \ref{box:SGL_estimation_algorithm}.\\

One remaining practical issue is the obtaining of a value for $\lambda_{max}$, the smallest value of $\lambda$ at which no groups are selected by the model.  Noting that $\mathbf{\hat{r}}_l = \mathbf{y}$ when no groups are selected, from \eqref{eq:SGL_bl_equals_zero_criterion} we obtain the smallest value, $\lambda_l^{min}$, for the minimum value of $\lambda$ at which group $l$ is not selected as
\begin{equation}
	\lambda_l^{min} = \frac{|| S(\mathbf{X}_l'\mathbf{y}, \alpha \lambda_l^{min}) ||_2} {(1-\alpha) w_l}
	\label{eq:SGL_lambda_min}
\end{equation} 
We can solve this in its quadratic form by first setting an upper bound for $\lambda$ at the point $\lambda_l^*$, where the soft thresholding function $S(\mathbf{X}_l' \mathbf{y}_l, \alpha \lambda) = \mathbf{0}$, that is when no SNPs are selected by the model.  We then obtain the solution by solving
\begin{equation}
	|| S(\mathbf{X}_l'\mathbf{y}, \alpha \lambda_l^{min}) ||_2^2 - (1-\alpha)^2 (\lambda_l^{min})^2 w_l^2 = 0 \quad 0 < \lambda_l^{min} < \lambda_l^*
	\label{eq:SGL_lambda_min_quadratic}
\end{equation}
for $\lambda_l^{min}$, where
\begin{equation*}
	\lambda_l^* = \max_j \frac{|\mathbf{x}_j' \mathbf{y}|}{\alpha}, \quad j = l_1,\dots,l_{P_l}.
\end{equation*}
We do this using the 1d root-finding function, $brentq$, in Python's \emph{scipy} library.
Finally, we obtain a value for $\lambda_{max}$ as
\begin{equation}
	\lambda_{max} = \max_l \lambda_l^{min}, \quad l = 1, \ldots, L.
	\label{eq:SGL_lambda_max}
\end{equation}

\section{SGL with overlaps}\label{SGL with overlaps}

We assume that $\mathbf{X}$ and $\boldsymbol{\beta}$ have been expanded to account for overlaps, but we drop the $^*$ notation for clarity.  We proceed as before by solving the block-separable optimisation \eqref{eq:SGL_overlaps_objF} for each group or pathway in turn.  However, for overlapping pathways, the assumption of pathway independence requires that each $\mathbf{X}_l, (l = 1, \ldots, L)$ is regressed against the full phenotype vector $\mathbf{y}$ rather than the partial residual, $\mathbf{\hat{r}}_l$.  With this in mind, the revised subgradient equations for group $l$ \eqref{eq:SGL_diff_all_terms} are given by
\begin{equation}
	-\mathbf{x}_j'(\mathbf{y} - \sum_{k \ne j} \mathbf{x}_k \hat{\beta}_k - \mathbf{x}_j \beta_j) + (1-\alpha) \lambda w_l s_j + \alpha \lambda t_j = 0
	\quad j = l_1,\dots,l_{P_l} .
	\label{eq:SGL_overlaps_diff_all_terms}
\end{equation}

The estimation for group $l$ then proceeds as described in the previous section, but with the partial residual $\mathbf{\hat{r}}_l$ replaced by $\mathbf{y}$, so that the group sparsity condition \eqref{eq:SGL_bl_equals_zero_criterion} for $|| \hat{\boldsymbol{\beta}_l} ||_2 = \mathbf{0}$ becomes
\begin{equation}
	|| S(\mathbf{X}_l'\mathbf{y}, \alpha \lambda) ||_2 \le (1-\alpha) \lambda w_l .
\end{equation}
As before, where group $l$ is selected by the model, the update for $\beta_j$, with current estimate $\hat{\beta}_j$, is derived from the partial derivative \eqref{eq:SGL_partial_j}, which under the independence assumption is given by
\begin{equation}
	\partial_j = 
		-\mathbf{x}_j'(\mathbf{y} - \mathbf{X}_l \hat{\boldsymbol{\beta}_l}) + (1-\alpha) \lambda w_l \frac{\hat{\beta_j}}{|| \hat{\boldsymbol{\beta}_l} ||_2} + \alpha \lambda \cdot \mbox{sign}(\hat{\beta_j}),
		\label{eq:SGL_partial_j_revised}
\end{equation}
for $j = l_1, \ldots, l_{P_l}$.  The Newton update \eqref{eq:SGN_Newton_update} remains the same.  When $\hat{\beta}_j = 0$, the revised directional derivatives \eqref{eq:SGL_directional_derivative} are given by
\begin{equation}
	\begin{aligned}
			\partial_j^+ &= -\mathbf{x}_j'(\mathbf{y} - \mathbf{X}_l \hat{\boldsymbol{\beta}_l}) + \alpha \lambda \\
			\partial_j^- &= -\mathbf{x}_j'(\mathbf{y} - \mathbf{X}_l \hat{\boldsymbol{\beta}_l}) - \alpha \lambda.
	\end{aligned}
	\label{eq:SGL_directional_derivative_revised}
\end{equation}
As before the conditions for SNP sparsity within a selected group are determined by \eqref{eq:SGL_Newton_update_betaj_zero}.

The value of $\lambda_{max}$, the smallest $\lambda$ value at which no group is selected by the model, is determined in the same way as before, since this procedure (described in \eqref{eq:SGL_lambda_min}, \eqref{eq:SGL_lambda_min_quadratic} and \eqref{eq:SGL_lambda_max}) does not depend on $\mathbf{\hat{r}}_l$.

Importantly, since each group is regressed independently against the phenotype vector $\mathbf{y}$, there is no block coordinate descent stage in the estimation, that is the revised algorithm utilises only coordinate gradient descent within each selected pathway.  For this reason we use the acronym SGL-CGD for the revised algorithm.  The new algorithm is described in Box \ref{box:SGL-CGD_estimation_algorithm}.  Note that since the block coordinate descent stage is avoided, the new algorithm has the added benefit of being much faster than would otherwise be the case.

\section{Simulation study 1}\label{sim_study_1}

A baseline phenotype, $y$ is sampled from $\mathcal{N}(10, 1)$.  To generate SNP effects, we first select a single pathway, $\mathcal{G}_l$, at random.  From this pathway we randomly select 5 SNPs to from the set $\mathcal{S} \subset \mathcal{G}_l$ of causal SNPs.  At each MC simulation we generate a genetic effect and adjust $y$ so that
\begin{equation*}
	y^* = y + w
\end{equation*}
where 
\begin{equation*}
	w = \delta \sum_{k \in \mathcal{S} } \zeta_k x_k .
\end{equation*}
Here $\delta$ controls the overall additive genetic effect on phenotype $y$ due to all casual SNPs in $\mathcal{S}$, and $\zeta_k$ determines the contribution from causal SNP $k$, with $\sum_{k \in \mathcal{S}} \zeta_k = 1$.  In our simulations we maintain a constant overall genetic effect size, 
\begin{equation*}
	\gamma = \text{E}(w)/ \text{E}(y)
\end{equation*}
across all affected phenotypes, so that $\gamma$ represents the proportionate increase in the mean value of $y$ due to all genetic effects.  We also set $\zeta_k = 1/5$, for $k \in \mathcal{S}$, so that the contribution from each causal SNP allele is equal.  This enables us to determine $\delta$ for a given $\gamma$ as
\begin{equation*}
	\delta = \frac{5 \gamma  \text{E}(y)}{2 \sum_{k \in \mathcal{S} }  m_k}.
\end{equation*}
Note that for constant $\gamma$, the proportionate effect on the mean value of $y$ due to SNP $k$ is MAF dependent, and is given by $2 \delta m_k / \text{E}(y)$.

\section{Weight tuning for bias reduction}\label{weight_tuning}

For fixed $\alpha$, and with $\lambda$ tuned to select a single pathway, we need to establish which pathway enters the model first, as $\lambda$ is reduced from its maximal value, $\lambda_{max}$.  From \eqref{eq:SGL_lambda_max}, at phenotype permutation $r$, the pathway $\hat{\mathcal{C}}_r$ selected with permuted phenotype $\mathbf{y}_r$ is given by
\begin{equation*}
	\hat{\mathcal{C}}_r = \argmax_l \lambda_l^{min}, \quad, l = 1, \ldots, L.
\end{equation*}
$\lambda_l^{min}$ is obtained by solving
\begin{equation*}
	\lambda_l^{min} = \frac{|| S(\mathbf{X}_l'\mathbf{y}_r, \alpha \lambda_l^{min}) ||_2} {(1-\alpha) w_l},
\end{equation*}
using the procedure described at the end of Section \ref{SGL estimation algorithm}.  For $R$ permutations of the phenotype vector, $\mathbf{y}$, the empirical pathway selection frequency distribution is then given by
\begin{equation*}
	\Pi^*(\mathbf{w}) = \frac{1}{R} \sum_{r=1}^{R} \{ \hat{\mathcal{C}}_r = l \}, \quad l = 1, \ldots, L.
\end{equation*}

\end{document}